\documentclass[12pt]{article}

\usepackage{amssymb}

\input epsf.tex

\newcommand{\J}{{\sf J}}

\newcommand{\cD}{{\cal D}}

\newcommand{\ii}{{\rm i}}

\newcommand{\dd}{{\rm d}}
\newcommand{\spinbox}{\left|\!\overline{\underline{{~}^{{~}^{{~}^{{~}^{{~}^{{~}^{{}^{}}}}}}}_{{~}^{{~}^{{~}^{{~}^{{~}^{{}^{{}^{}}}}}}}}}}\!\right|}

\newcommand{\spin}[2]{{#1}\stackrel{\displaystyle#2}{\spinbox}}

\def\Dirac{{D\!\!\!\!/\,}} 

\newcommand{\eq}{\begin{equation}}
\newcommand{\eqend}{\end{equation}}
\newcommand{\eqa}{\begin{eqnarray}}
\newcommand{\nonueqa}{\begin{eqnarray*}}
\newcommand{\eqaend}{\end{eqnarray}}
\newcommand{\nonueqaend}{\end{eqnarray*}}

\newcommand{\bma}[1]{\begin{array}{#1}}
\newcommand{\ema}{\end{array}}
\newcommand{\bc}{\begin{center}}
\newcommand{\ec}{\end{center}}

\renewcommand{\thefootnote}{\fnsymbol{footnote}}

\def\appendix#1{\addtocounter{section}{1}\setcounter{equation}{0}
\renewcommand{\thesection}{\Alph{section}}
\section*{Appendix \thesection\protect\indent \parbox[t]{11.715cm} {#1}}
\addcontentsline{toc}{section}{Appendix \thesection\ \ \ #1} }

\newcommand{\complex}{{\bb C}} 
\newcommand{\zed}{{\bb Z}} 
\newcommand{\real}{{\bb R}} 
\newcommand{\zedm}{{\bbm Z}} 
\newcommand{\NO}{\,\mbox{$\circ\atop\circ$}\,} 
\newcommand{\id}{{1\!\!1}} 
\def\Dirac{{D\!\!\!\!/\,}} 

\newif\ifold             \oldtrue            \def\new{\oldfalse}

\font\mybb=msbm10 at 12pt
\def\bb#1{\hbox{\mybb#1}}
\font\mybbm=msbm10 at 11pt
\def\bbm#1{\hbox{\mybbm#1}}

\def\nn{\nonumber}
\newcommand{\tr}[1]{\:{\rm tr}\,#1}
\newcommand{\Tr}[1]{\:{\rm Tr}\,#1}

\def\e{{\,\rm e}\,}

\def\beq{\begin{equation}}
\def\eeq{\end{equation}}
\def\bea{\begin{eqnarray}}
\def\eea{\end{eqnarray}}
\def\bd{\begin{displaymath}}
\def\ed{\end{displaymath}}

\newcommand{\DD}{{\cal D}}

\makeatletter
\newdimen\normalarrayskip              
\newdimen\minarrayskip                 
\normalarrayskip\baselineskip
\minarrayskip\jot
\newif\ifold             \oldtrue            \def\new{\oldfalse}
\def\arraymode{\ifold\relax\else\displaystyle\fi} 
\def\@arrayskip{\ifold\baselineskip\z@\lineskip\z@
     \else
     \baselineskip\minarrayskip\lineskip2\minarrayskip\fi}
\def\@arrayclassz{\ifcase \@lastchclass \@acolampacol \or
\@ampacol \or \or \or \@addamp \or
   \@acolampacol \or \@firstampfalse \@acol \fi
\edef\@preamble{\@preamble
  \ifcase \@chnum
     \hfil$\relax\arraymode\@sharp$\hfil
     \or $\relax\arraymode\@sharp$\hfil
     \or \hfil$\relax\arraymode\@sharp$\fi}}
\def\@array[#1]#2{\setbox\@arstrutbox=\hbox{\vrule
     height\arraystretch \ht\strutbox
     depth\arraystretch \dp\strutbox
     width\z@}\@mkpream{#2}\edef\@preamble{\halign \noexpand\@halignto
\bgroup \tabskip\z@ \@arstrut \@preamble \tabskip\z@ \cr}%
\let\@startpbox\@@startpbox \let\@endpbox\@@endpbox
  \if #1t\vtop \else \if#1b\vbox \else \vcenter \fi\fi
  \bgroup \let\par\relax
  \let\@sharp##\let\protect\relax
  \@arrayskip\@preamble}
\makeatother

\begin{document}
\begin{flushright}

\baselineskip=12pt

HWM--02--18\\
EMPG--02--11\\
hep--th/0207142\\
\hfill{ }\\
July 2002
\end{flushright}

\vspace{0.25cm}

\begin{center}

{\Large\bf BUSSTEPP LECTURES ON STRING THEORY}

\vspace{0.25cm}

{\large\bf An Introduction to String Theory and D-Brane Dynamics}

\baselineskip=14pt

\vspace{1cm}

{\large\sc Richard J. Szabo}
\\[3mm]
{\it Department of Mathematics\\ Heriot-Watt University\\ Riccarton, Edinburgh
EH14 4AS, U.K.}
\\{\tt R.J.Szabo@ma.hw.ac.uk}
\\[10mm]

\end{center}

\begin{abstract}

\baselineskip=12pt

This paper comprises the written version of the lectures on string theory
delivered at the 31st British Universities Summer School on Theoretical
Elementary Particle Physics which was held in Manchester, England, August 28 --
September~12~2001.

\end{abstract}


\tableofcontents

\renewcommand{\thefootnote}{\arabic{footnote}} \setcounter{footnote}{0}

\section{Introduction \label{Intro}}

These notes comprise an expanded version of the string theory lectures
given by the author at the 31st British Universities Summer School on
Theoretical Elementary Particle Physics (BUSSTEPP) which was held in
Manchester, England in 2001. The school is attended mostly by
Ph.D. students in theoretical high-energy physics who have just
completed their first year of graduate studies at a British
university. The lectures were thereby appropriately geared for this
level. No prior knowledge of string theory was assumed, but a good
background in quantum field theory, introductory level particle
physics and group theory was. An acquaintance with the basic
ideas of general relativity is helpful but not absolutely
essential. Some familiarity with supersymmetry was also assumed
because the supersymmetry lectures preceeded the string theory
lectures at the school, although the full-blown machinery and
techniques of supersymmetry were not exploited to any large extent.

The main references for string theory used during the course were the
standard books on the subject~\cite{GSW,PolBook} and the more recent
review article~\cite{JohnsonRev}. The prerequisite supersymmetry
lectures can be found
in~\cite{BUSSTEPPJMF}.\footnote{\baselineskip=12pt Further references
  are cited in the text, but are mostly included for historical
  reasons and are by no means exhaustive. Complete sets of references
  may be found in the various cited books and review articles.} The
lectures were delivered in the morning and exercises were assigned for
the tutorial sessions which took place in the afternoons. These
problems are also included in these notes. Many of them are intended
to fill in the technical gaps which due to time constraints were not
covered in the lectures. Others are intended to give the student a
better grasp of some ``stringy'' topics. The present paper has
expanded on many aspects of string theory that were addressed during
the school, mainly to make the presentation clearer.

There were six one-hour lectures in total. Since string theory is
nowadays such a vast and extensive subject, some focus in the subject
material was of course required. The lectures differ perhaps from most
introductory approaches since the intent was to provide the student
not only with the rudiments of perturbative string theory, but also
with an introduction to the more recently discovered non-perturbative
degrees of freedom known as ``D-branes'', which in the past few years
have revolutionalized the thinking about string theory and have
brought the subject to the forefront of modern theoretical particle
physics once again. This means that much of the standard introductory
presentation was streamlined in order to allow for an introduction to
these more current developments. The hope was that the student will
have been provided with enough background to feel comfortable in
starting to read current research articles, in addition to being
exposed to some of the standard computational techniques in the
field. The basic perturbative material was covered in roughly the
first three lectures and comprises
sections~\ref{History}--\ref{Superstrings}. Lecture~4
(section~\ref{RRCharge}) then started to rapidly move towards
explaining what D-branes are, and at the same time introducing some
more novel stringy physics. Lectures~5 and 6
(sections~\ref{DBraneGauge} and \ref{DBraneDyn}) then dealt with
D-branes in detail, studied their dynamics, and provided a brief
account of the gauge theory/string theory correspondence which has
been such an active area of research over the past few years.

\section{Overview: A Brief History of String Theory \label{History}}

To help introduce the topics which follow and their significance in
high energy physics, in this section we will briefly give a
non-technical historical account of the development of string theory
to date, focusing on its achievements, failures and prospects. This
will also help to motivate the vast interest in string theory within
the particle theory community. It will further give an overview of the
material which will follow.

\begin{figure}[htb]
\epsfxsize=2 in
\bigskip
\centerline{\epsffile{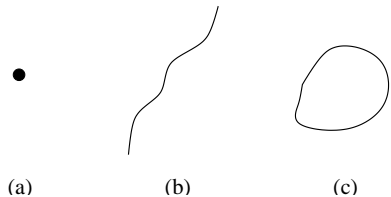}}
\caption{\baselineskip=12pt (a) {\it A point particle.} (b) {\it An open
string.} (c) {\it A closed string.}}
\bigskip
\label{pointlineloop}\end{figure}

In conventional quantum field theory, the fundamental objects are
mathematical points in spacetime, modeling the elementary point
particles of nature. String theory is a rather radical generalization
of quantum field theory whereby the fundamental objects are extended,
one-dimensional lines or loops~(fig.~\ref{pointlineloop}). The various
elementary particles observed in nature correspond to different
vibrational modes of the string. While we cannot see a string (yet) in
nature, if we are very far away from it we will be able to see its
point-like oscillations, and hence measure the elementary particles
that it produces. The main advantage of this description is that while
there are many particles, there is only one string. This indicates
that strings could serve as a good starting point for a unified field
theory of the fundamental interactions.

This is the idea that emerged by the end of the 1960's from several
years of intensive studies of dual models of hadron
resonances~\cite{Ven1}. In this setting, string theory attempts to
describe the strong nuclear force. The excitement over this formalism
arose from the fact that string S-matrix scattering amplitudes agreed
with those found in meson scattering experiments at the time. The
inclusion of fermions into the model led to the notion of a
supersymmetric string, or ``superstring'' for
short~\cite{NS1,Ramond1}. The massive particles sit on ``Regge
trajectories'' in this setting.

However, around 1973 the interest in string theory quickly began to fade,
mainly because quantum chromodynamics became recognized as the correct quantum
field theory of the strong interactions. In addition, string theories possessed
various undesirable features which made them inappropriate for a theory of
hadrons. Among these were the large number of extra spacetime dimensions
demanded by string theory, and the existence of massless particles other than
the spin~1 gluon in the spectrum of string states.

\begin{figure}[htb]
\epsfxsize=2 in
\bigskip
\centerline{\epsffile{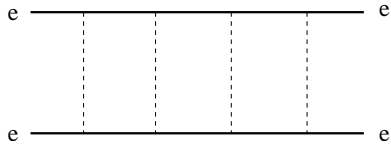}}
\caption{\baselineskip=12pt {\it A non-renormalizable ultraviolet
    divergent Feynman diagram in quantum gravity. The dashed lines
    depict graviton exchanges.}}
\bigskip
\label{UVQG}\end{figure}

In 1974 the interest in string theory was revived for another
reason~\cite{SchSchw1,Yoneya1}. It was found that, among the massless
string states, there is a spin~2 particle that interacts like a
graviton. In fact, the only consistent interactions of massless spin~2
particles are gravitational interactions. Thus string theory naturally
includes general relativity, and it was thereby proposed as a unified
theory of the fundamental forces of nature, including gravity, rather
than a theory of hadrons. This situation is in marked contrast to that
in ordinary quantum field theory, which does not allow gravity to
exist because its scattering amplitudes that involve graviton
exchanges are severely plagued by non-renormalizable ultraviolet
diveregences (fig.~\ref{UVQG}). On the other hand, string theory is a
consistent quantum theory, free from ultraviolet divergences, which
necessarily {\it requires} gravitation for its overall consistency.

With these facts it is possible to estimate the energy or length scale
at which strings should be observed in nature. Since string theory is
a relativistic quantum theory that includes gravity, it must involve
the corresponding three fundamental constants, namely the speed of
light $c$, the reduced Planck constant $\hbar$, and the Newtonian
gravitational constant $G$. These three constants may combined into a
constant with dimensions of length. The characteristic length scale of
strings may thereby be estimated by the {\it Planck length} of quantum
gravity:
\beq
\begin{tabular}{|c|}\hline\\
$\displaystyle
\ell_{\rm P}=\left(\frac{\hbar\,G}{c^3}\right)^{3/2}=1.6\times10^{-33}~
{\rm cm} \ .
$\\\\
\hline\end{tabular}
\label{Plancklength}\eeq
This is to be compared with the typical size of hadrons, which is of
the order of $10^{-13}~{\rm cm}$. The corresponding energy scale is
known as the {\it Planck mass}:
\beq
\begin{tabular}{|c|}\hline\\
$\displaystyle
m_{\rm P}=\left(\frac{\hbar\,c}G\right)^{1/2}=1.2\times10^{19}~
{\rm GeV}/c^2 \ .
$\\\\
\hline\end{tabular}
\label{Planckmass}\eeq
These scales indicate the reasons why strings have not been observed
in nature thus far. Present day particle accelerators run at energies
$\ll m_{\rm P}c^2$ and thus cannot resolve distances as short as the
Planck length. At such energies, strings look like point particles,
because at very large distance scales compared to the Planck length
all one can observe is the string's center of mass motion, which is
point-like. Thus at these present day scales, strings are accurately
described by quantum field theory.

For many of the subsequent years superstring theory began showing
great promise as a unified quantum theory of all the fundamental
forces including gravity. Some of the general features which were
discovered are:
\begin{itemize}
\item General relativity gets modified at very short distances/high
  energies (below the Planck scale), but at ordinary distances and
  energies it is present in string theory in exactly the same form as
  Einstein's theory.
\item ``Standard model type'' Yang-Mills gauge theories arise very
  naturally in string theory. However, the reasons why the gauge group
  $SU(3)\times SU(2)\times U(1)$ of the standard model should be
  singled out is not yet fully understood.
\item String theory predicts supersymmetry, because its mathematical
  consistency depends crucially on it. This is a generic feature of
  string theory that has not yet been discovered experimentally.
\end{itemize}
This was the situation for some years, and again the interest in
string theory within the high energy physics community began to
fade. Different versions of superstring theory existed, but none of
them resembled very closely the structure of the standard model.

Things took a sharp turn in 1985 with the birth of what is known as
the ``first superstring revolution''. The dramatic achievement at this
time was the realization of how to cancel certain mathematical
inconsistencies in quantum string theory. This is known as
Green-Schwarz anomaly cancellation~\cite{GrSchw1} and its main
consequence is that it leaves us with only five consistent superstring
theories, each living in ten spacetime dimensions. These five theories
are called Type~I, Type~IIA, Type~IIB, $SO(32)$ heterotic, and
$E_8\times E_8$ heterotic. The terminology will be explained later
on. For now, we simply note the supersymmetric Yang-Mills gauge groups
that arise in these theories. The Type~I theories have gauge group
$SO(32)$, both Type~II theories have $U(1)$, and the heterotic
theories have gauge groups as in their names. Of particular
phenomenological interest was the $E_8\times E_8$ heterotic string,
because from it one could construct grand unified field theories
starting from the exceptional gauge group $E_6$.

The spacetime dimensionality problem is reconciled through the notion
of ``compactification''. Putting six of the spatial directions on a
``small'' six-dimensional compact space, smaller than the resolution
of the most powerful microscope, makes the 9+1-dimensional spacetime
look 3+1-dimensional, as in our observable world. The six-dimensional
manifolds are restricted by string dynamics to be ``Calabi-Yau
spaces''~\cite{CHSW1}. These compactifications have tantalizingly
similar features to the standard model. However, no complete
quantitative agreement has been found yet between the two theories,
such as the masses of the various elementary particles. This reason,
and others, once again led to the demise of string theory towards to
the end of the 1980's. Furthermore, at that stage one only understood
how to formulate superstring theories in terms of divergent
perturbation series analogous to quantum field theory. Like in quantum
chromodynamics, it is unlikely that a realistic vacuum can be
accurately analysed within perturbation theory. Without a good
understanding of nonperturbative effects (such as the analogs of QCD
instantons), superstring theory cannot give explicit, quantitative
predictions for a grand unified model.

\begin{figure}[htb]
\epsfxsize=5 in
\bigskip
\centerline{\epsffile{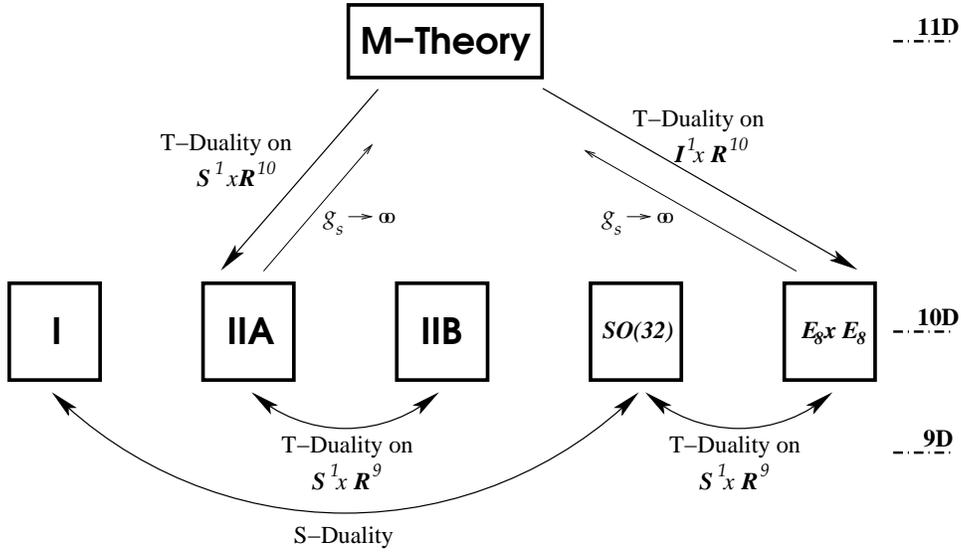}}
\caption{\baselineskip=12pt {\it The various duality transformations
    that relate the superstring theories in nine and ten
    dimensions. T-Duality inverts the radius $R$ of the circle ${\bf
      S}^1$, or the length of the finite interval ${\bf I}^1$, along
    which a single direction of the spacetime is compactified,
    i.e. $R\mapsto\ell_{\rm P}^2/R$. S-duality inverts the
    (dimensionless) string coupling constant $g_s$, $g_s\mapsto1/g_s$,
    and is the analog of electric-magnetic duality (or strong-weak
    coupling duality) in four-dimensional gauge theories. M-Theory
    originates as the strong coupling limit of either the Type~IIA or
    $E_8\times E_8$ heterotic string theories.}}
\bigskip
\label{MTheory}\end{figure}

\begin{figure}[htb]
\epsfxsize=3 in
\bigskip
\centerline{\epsffile{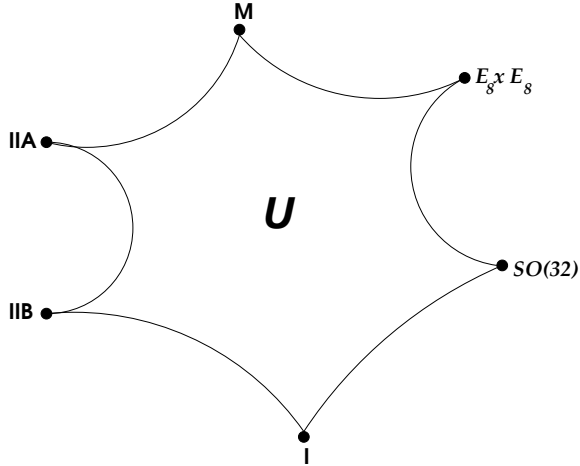}}
\caption{\baselineskip=12pt {\it The space $\cal U$ of quantum string
    vacua. At each node a weakly-coupled string description is
    possible.}}
\bigskip
\label{calU}\end{figure}

\begin{figure}[htb]
\epsfxsize=1 in
\bigskip
\centerline{\epsffile{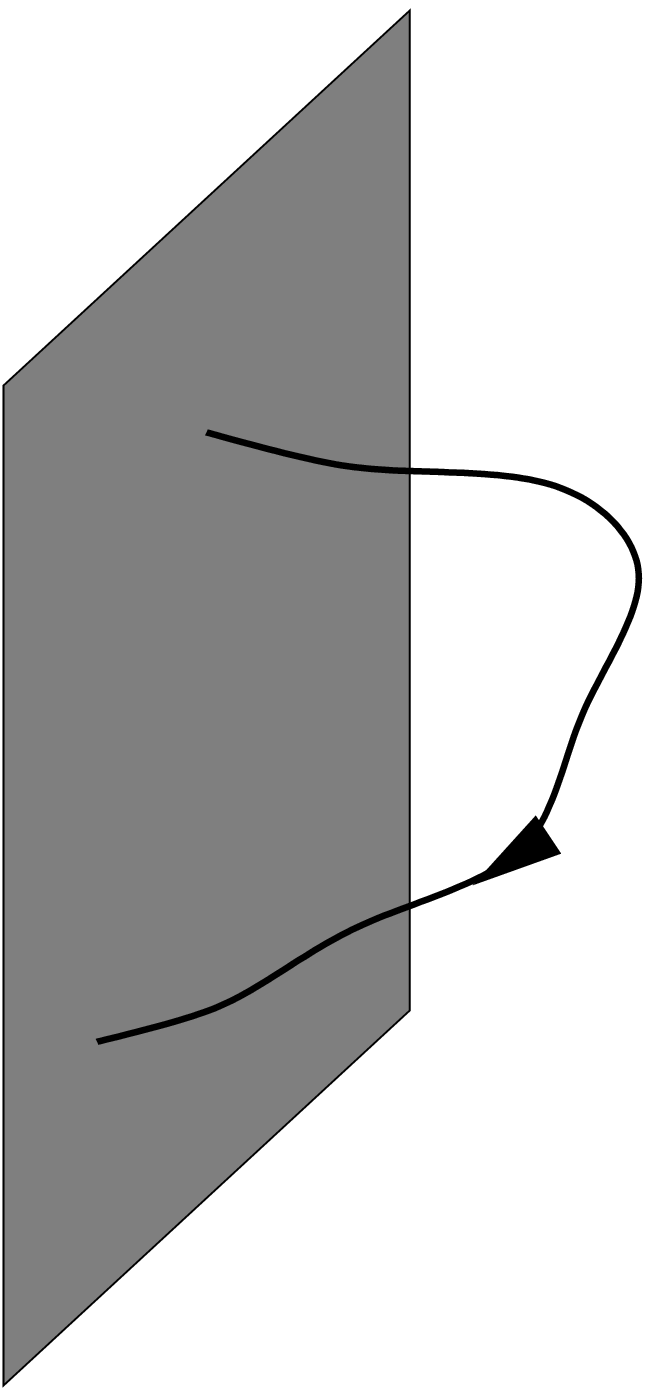}}
\caption{\baselineskip=12pt {\it A fundamental open string (wavy line)
    starting and ending (with Dirichlet boundary conditions) on a
    D$p$-brane (shaded region) which is a $p+1$-dimensional hyperplane
    in spacetime.}}
\bigskip
\label{Dpbrane}\end{figure}

This was the state of affairs until around 1995 when the ``second
superstring revolution'' set in. For the first time, it was understood
how to go beyond the perturbation expansion of string theory via
``dualities'' which probe nonperturbative features of string
theory~\cite{FILQ1,HullTown1,KachVafa1,Schw1,Sen1}. The three major
implications of these discoveries were:
\begin{itemize}
\item{\it Dualities relate all five superstring theories in ten
    dimensions to one another.}
\end{itemize}
The different theories are just perturbative expansions of a unique
underlying theory $\cal U$ about five different, consistent quantum
vacua~\cite{Schw2,Schw3}. Thus there is a completely unique theory of
nature, whose equation of motion admits many vacua. This is of course
a most desirable property of a unified theory.
\begin{itemize}
\item{\it The theory $\cal U$ also has a solution called ``M-Theory''
    which lives in 11 spacetime dimensions~\cite{Duff1,Town1,Witten1}.}
\end{itemize}
The low-energy limit of M-Theory is 11-dimensional
supergravity~\cite{CJS1}. All five superstring theories can be thought
of as originating from M-Theory~\cite{Duff1,Schw3}
(fig.~\ref{MTheory}). The underlying theory $\cal U$ is depicted in
fig.~\ref{calU}.
\begin{itemize}
\item{\it In addition to the fundamental strings, the theory $\cal U$
    admits a variety of extended nonperturbative excitations called
    ``$p$-branes''~\cite{HorStrom1}, where $p$ is the number of
    spatial extensions of the objects.}
\end{itemize}
Especially important in this regard are the ``Dirichlet
$p$-branes''~\cite{DLP1,Horava1,Pol1}, or ``D-branes'' for short,
which are $p$-dimensional soliton-like hyperplanes in spacetime whose
quantum dynamics are governed by the theory of {\it open strings}
whose ends are constrained to move on them~(fig.~\ref{Dpbrane}).

We will not attempt any description of the theory $\cal U$, which at
present is not very well understood. Rather, we wish to focus on the
remarkable impact in high-energy physics that the discovery of
D-branes has provided. Amongst other things, they have led to:
\begin{itemize}
\item Explicit realizations of nonperturbative string
  dualities~\cite{Pol1}. For example, an elementary closed string
  state in Theory~A (which is perturbative because its amplitudes are
  functions of the string coupling $g_s$) gets mapped under an
  S-duality transformation to a D-brane state in the dual Theory~B
  (which depends on $1/g_s$ and is therefore nonperturbative).
\item A microscopic explanation of black hole entropy and the rate of
  emission of thermal (Hawking) radiation for black holes in string
  theory~\cite{CalMald1,StromVafa1}.
\item The gauge theory/gravity (or AdS/CFT)
  correspondence~\cite{AdSCFTRev,Mald1}. D-branes carry gauge fields,
  while on the other hand they admit a dual description as solutions
  of the classical equations of motion of string theory and
  supergravity. Demanding that these two descriptions be equivalent
  implies, for some special cases, that string theory is equivalent to
  a gauge field theory. This is an explicit realization of the old
  ideas that Yang-Mills theory may be represented as some sort of
  string theory.
\item Probes of short-distances in spacetime~\cite{DKPS1}, where
  quantum gravitational fluctuations become important and classical
  general relativity breaks down.
\item Large radius compactifications, whereby extra compact dimensions
  of size $\gg({\rm TeV})^{-1}$ occur~\cite{IADD1,ADD1}. This is the
  distance scale probed in present-day accelerator experiments, which
  has led to the hope that the extra dimensions required by string
  theory may actually be observable.
\item Brane world scenarios, in which we model our world as a
  D-brane~\cite{RS1,RS2}. This may be used to explain why gravity
  couples so weakly to matter, i.e. why the effective Planck mass in
  our 3+1-dimensional world is so large, and hence gives a potential
  explanation of the hierarchy problem $m_{\rm P}\gg m_{\rm weak}$.
\end{itemize}
In what follows we will give the necessary background into the
description of D-brane dynamics which leads to these exciting
developments in theoretical high-energy physics.

\setcounter{equation}{0}

\section{Classical String Theory \label{Classical}}

In this section we will start making the notions of the previous
section quantitative. We will treat the classical dynamics of strings,
before moving on in the next section to the quantum theory. Usually,
at least at introductory levels, one is introduced to quantum field
theory by using ``second quantization'' which is based on field
operators that create or destroy quanta. Here we will describe string
dynamics in ``first quantization'' instead, which is a
sum-over-histories or Feynman path integral approach closely tied to
perturbation theory. To familiarize ourselves with this geometrical
description, we will start by explaining how it can be used to
reproduce the well-known dynamics of a massive, relativistic point
particle. This will easily introduce the technique that will readily
generalize to the case of extended objects such as strings and
D-branes. We will then describe the bosonic string within this
approach and analyse its classical equations of motion. For the
remainder of this paper we will work in natural units whereby
$c=\hbar=G=1$.

\subsection{The Relativistic Particle \label{Particle}}

\begin{figure}[htb]
\epsfxsize=5 in
\bigskip
\centerline{\epsffile{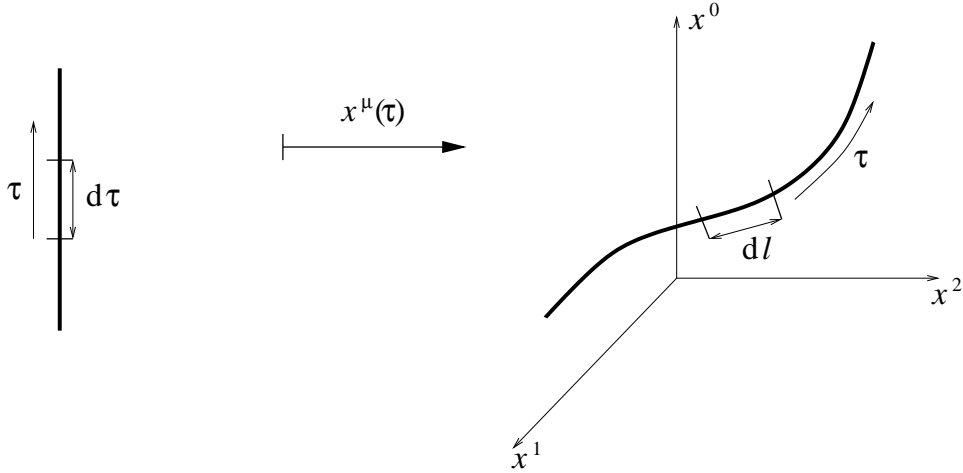}}
\caption{\baselineskip=12pt {\it The embedding $\tau\mapsto x^\mu(\tau)$ of a
particle trajectory into $d$-dimensional spacetime. As $\tau$ increases the
particle propagates along its one-dimensional worldline in the target space.}}
\bigskip
\label{worldline}\end{figure}

Consider a particle which propagates in $d$-dimensional ``target
spacetime'' $\real^{1,d-1}$ with coordinates
\beq
(t,\vec x\,)=x^\mu=(x^0,x^1,\dots,x^{d-1}) \ .
\label{spacetimecoords}\eeq
It sweeps out a path $x^\mu(\tau)$ in spacetime, called a
``worldline'' of the particle, which is parametrized by a proper time
coordinate $\tau\in\real$ (fig.~\ref{worldline}). The infinitesimal,
Lorentz-invariant path length swept out by the particle is
\beq
\dd l=(-\dd s^2)^{1/2}=(-\eta_{\mu\nu}~\dd x^\mu~\dd x^\nu)^{1/2}\equiv(-
\dd x^\mu~\dd x_\mu)^{1/2} \ ,
\label{infpathlength}\eeq
where $l$ is the proper-time of the particle and
\beq
(\eta_{\mu\nu})=\pmatrix{-1& & &0\cr &1& & \cr & &\ddots& \cr0& & &1\cr}
\label{etamunu}\eeq
is the flat Minkowski metric. The action for a particle of mass $m$ is
then given by the total length of the trajectory swept out by the
particle in spacetime:
\beq
\begin{tabular}{|c|}\hline\\
$\displaystyle
S[x]=-m\int\dd l(\tau)=-m\int\dd\tau~\sqrt{-\dot x^\mu\,\dot x_\mu}
$\\\\
\hline\end{tabular}
\label{particleaction}\eeq
where
\beq
\dot x^\mu\equiv\frac{\dd x^\mu}{\dd\tau} \ .
\label{dotxparticle}\eeq
The minima of this action determine the trajectories of the particle
with the smallest path length, and therefore the solutions to the
classical equations of motion are the geodesics of the free particle
in spacetime.

\bigskip

\begin{center}
\begin{minipage}{15cm}

\small

{\bf Exercise~3.1.} {\sl\baselineskip=12pt Show that the Euler-Lagrange
equations resulting from the action (\ref{particleaction}) give the usual
equations of relativistic particle kinematics:
$$
\dot p^\nu=0 \ , ~~ p^\nu=\frac{m\,\dot x^\nu}{\sqrt{-\dot x^\mu\,\dot
x_\mu}} \ .
$$
}

\end{minipage}
\end{center}

\bigskip

\subsubsection*{Reparametrization Invariance}

The Einstein constraint $p^2\equiv p^\mu\,p_\mu=-m^2$ on the classical
trajectories of the particle (Exercise~3.1) is related to the fact that the
action (\ref{particleaction}) is invariant under arbitrary, local
reparametrizations of the worldline, i.e.
\beq
\tau~\longmapsto~\tau(\tau') \ , ~~ \frac{\dd\tau}{\dd\tau'}>0 \ .
\label{particlerepar}\eeq
This is a kind of local worldline ``gauge invariance'' which means that the
form of $S[x]$ is unchanged under such a coordinate change, since
$\dd\tau~\sqrt{-\dot x^\mu(\tau)\dot x_\mu(\tau)}=\dd\tau'~\sqrt{-\dot
x^\mu(\tau')\dot x_\mu(\tau')}$. It is a one-dimensional version of the usual
four-dimensional general coordinate invariance in general relativity, in the
sense that it corresponds to a worldline diffeomorphism symmetry of the theory.
An application of the standard Noether procedure to the 0+1-dimensional field
theory (\ref{particleaction}) leads to a conserved Noether current whose
continuity equation is precisely the constraint $p^2=-m^2$. It tells us how to
eliminate one of the $p^\mu$'s, and in the quantum theory it becomes the
requirement that physical states and observables must be gauge invariant. It
enables us to select a suitable gauge. Let us look at a couple of simple
examples, as this point will be crucial for our later generalizations.

\subsubsection*{Examples}

\noindent
${\underline{\rm (1)~Static~Gauge:}}$ Here we take the gauge choice
\beq
x^0=\tau\equiv t
\label{staticgauge}\eeq
in which the action assumes the simple form
\beq
S[x]=-m\int\dd t~\sqrt{1-v^2}
\label{actionstatic}\eeq
where
\beq
\vec v=\frac{\dd\vec x}{\dd t}
\label{velocityparticle}\eeq
is the velocity of the particle. The equations of motion in this gauge take the
standard form of those for a free, massive relativistic particle:
\beq
\frac{\dd\vec p}{\dd t}=\vec0 \ , ~~ \vec p=\frac{m\,\vec v}{\sqrt{1-v^2}} \ .
\label{staticeomparticle}\eeq

\noindent
${\underline{\rm (2)~Galilean~Gauge:}}$ An even simpler gauge choice results
from selecting
\beq
\dot x^\mu\,\dot x_\mu=-1 \ .
\label{Galgauge}\eeq
The momentum of the particle in this gauge is given by the non-relativistic
form (c.f. exercise~3.1)
\beq
p^\mu=m\,\dot x^\mu \ ,
\label{momGalgauge}\eeq
and the equations of motion are therefore
\beq
\ddot x^\mu=0
\label{Galeom}\eeq
whose solutions are given by the Galilean trajectories
\beq
x^\mu(\tau)=x^\mu(0)+p^\mu\,\tau \ .
\label{Galtraj}\eeq

\subsection{The Bosonic String \label{Bosonic}}

\begin{figure}[htb]
\epsfxsize=5 in
\bigskip
\centerline{\epsffile{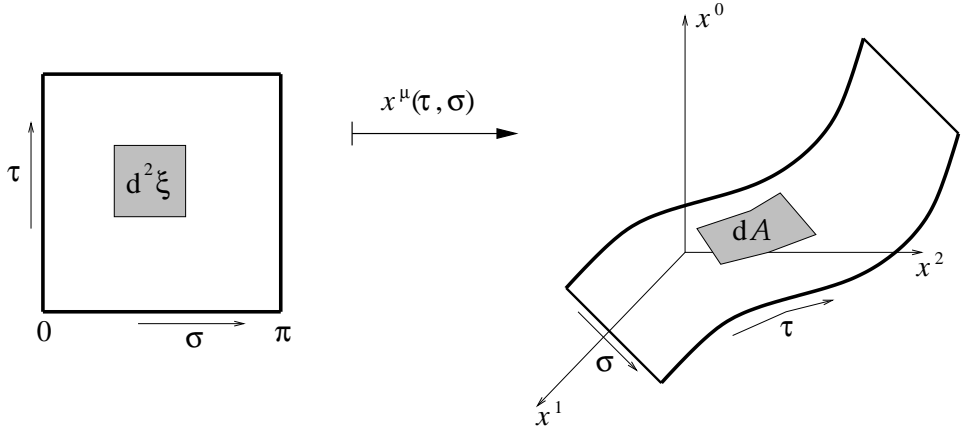}}
\caption{\baselineskip=12pt {\it The embedding $(\tau,\sigma)\mapsto
x^\mu(\tau,\sigma)$ of a string trajectory into $d$-dimensional spacetime. As
$\tau$ increases the string sweeps out its two-dimensional worldsheet in the
target space, with $\sigma$ giving the position along the string.}}
\bigskip
\label{worldsheet}\end{figure}

In the previous subsection we analysed a point particle, which is a
zero-dimensional object described by a one-dimensional worldline in spacetime.
We can easily generalize this construction to a {\it string}, which is a
one-dimensional object described by a two-dimensional ``worldsheet'' that the
string sweeps out as it moves in time with coordinates
\beq
(\xi^0,\xi^1)=(\tau,\sigma) \ .
\label{tausigmacoords}\eeq
Here $0\leq\sigma\leq\pi$ is the spatial coordinate along the string, while
$\tau\in\real$ describes its propagation in time. The string's evolution in
time is described by functions $x^\mu(\tau,\sigma)$, $\mu=0,1,\dots,d-1$ giving
the shape of its worldsheet in the target spacetime (fig.~\ref{worldsheet}).
The ``induced metric'' $h_{ab}$ on the string worldsheet corresponding to its
embedding into spacetime is given by the ``pullback'' of the flat Minkowski
metric $\eta_{\mu\nu}$ to the surface,
\beq
h_{ab}=\eta_{\mu\nu}\,\partial_ax^\mu\,\partial_bx^\nu \ ,
\label{indmetric}\eeq
where\footnote{\baselineskip=12pt Notation: Greek letters $\mu,\nu,\dots$
denote spacetime indices, beginning Latin letters $a,b,\dots$ denote worldsheet
indices, and later Latin letters $i,j,\dots$ label spatial directions in the
target space. Unless otherwise stated, we also adhere to the standard Einstein
summation convention for summing over repeated upper and lower indices.}
\beq
\partial_a\equiv\frac\partial{\partial\xi^a} \ , ~~ a=0,1 \ .
\label{partialadef}\eeq
An elementary calculation shows that the invariant, infinitesimal area element
on the worldsheet is given by
\beq
\dd A=\sqrt{-\det_{a,b}\,(h_{ab})}~\dd^2\xi \ ,
\label{areaelt}\eeq
where the determinant is taken over the indices $a,b=0,1$ of the $2\times2$
symmetric nondegenerate matrix $(h_{ab})$.

In analogy to the point particle case, we can now write down an action whose
variational law minimizes the total area of the string worldsheet in spacetime:
\beq
\begin{tabular}{|c|}\hline\\
$\displaystyle
S[x]=-T\int\dd A=-T\int\dd^2\xi~\sqrt{-\det_{a,b}\,\left(\partial_a
x^\mu\,\partial_bx_\mu\right)} \ .
$\\\\
\hline\end{tabular}
\label{stringaction}\eeq
The quantity $T$ has dimensions of mass per unit length and is the {\it
tension} of the string. It is related to the ``intrinsic length'' $\ell_s$ of
the string by
\beq
\begin{tabular}{|c|}\hline\\
$\displaystyle
T=\frac1{2\pi\alpha'} \ , ~~ \alpha'=\ell_s^2 \ .
$\\\\
\hline\end{tabular}
\label{tension}\eeq
The parameter $\alpha'$ is called the ``universal Regge slope'', because the
string vibrational modes all lie on linear parallel Regge trajectories with
slope $\alpha'$. The action (\ref{stringaction}) defines a 1+1-dimensional
field theory on the string worldsheet with bosonic fields $x^\mu(\tau,\sigma)$.

\bigskip

\begin{center}
\begin{minipage}{15cm}

\small

{\bf Exercise~3.2.} {\sl\baselineskip=12pt Show that the action
(\ref{stringaction}) is reparametrization invariant, i.e. if
$\xi\mapsto\xi(\xi')$, then it takes the same form when expressed in terms of
the new worldsheet coordinates~$\xi'$.}

\end{minipage}
\end{center}

\bigskip

Evaluating the determinant explicitly in (\ref{stringaction}) leads to the form
\beq
\begin{tabular}{|c|}\hline\\
$\displaystyle
S[x]=-T\int\dd\tau~\dd\sigma~\sqrt{\dot x^2\,x'^2-(\dot x\cdot x')^2}
$\\\\
\hline\end{tabular}
\label{NGaction}\eeq
where
\beq
\dot x^\mu=\frac{\partial x^\mu}{\partial\tau} \ , ~~
x'^\mu=\frac{\partial x^\mu}{\partial\sigma} \ .
\label{dotxprimexdef}\eeq
This is the form that the original string action appeared in and is
known as the ``Nambu-Goto action''~\cite{Goto1,Nambu1}. However, the
square root structure of this action is somewhat ackward to work
with. It can, however, be eliminated by the fundamental observation
that the Nambu-Goto action is classically equivalent to another action
which does not have the square root:
\beq
\begin{tabular}{|c|}\hline\\
$\begin{array}{lll}
S[x,\gamma]&=&\displaystyle-T\int\dd^2\xi~\sqrt{-\gamma}\,\gamma^{ab}\,h_{ab}\\
&=&\displaystyle-T\int\dd^2\xi~\sqrt{-\gamma}\,\gamma^{ab}\,\partial_a
x^\mu\,\partial_bx^\nu\,\eta_{\mu\nu} \ . \end{array}
$\\\\
\hline\end{tabular}
\label{Polaction}\eeq
Here the auxilliary rank two symmetric tensor field
$\gamma_{ab}(\tau,\sigma)$ has a natural interpretation as a metric on
the string worldsheet, and we have defined
\beq
\gamma=\det_{a,b}\,(\gamma_{ab}) \ , ~~ \gamma^{ab}=(\gamma^{-1})^{ab} \ .
\label{worldsheetmetric}\eeq
The action (\ref{Polaction}) is called the ``Polyakov
action''~\cite{Poly1}.

\bigskip

\begin{center}
\begin{minipage}{15cm}

\small

{\bf Exercise~3.3.} {\sl\baselineskip=12pt Show that the Euler-Lagrange
equations obtained by varying the Polyakov action with respect to $\gamma^{ab}$
are
$$
T_{ab}\equiv\partial_ax^\mu\,\partial_bx_\mu-\frac12\,\gamma_{ab}\,
\gamma^{cd}\,\partial_cx^\mu\,\partial_dx_\mu=0 \ , ~~ a,b=0,1 \ .
$$
Then show that this equation can be used to eliminate the worldsheet metric
$\gamma_{ab}$ from the action, and as such recovers the Nambu-Goto action.}

\end{minipage}
\end{center}

\bigskip

\subsubsection*{Worldsheet Symmetries}

The quantity $T_{ab}$ appearing in exercise~3.3 is the energy-momentum
tensor of the 1+1-dimensional worldsheet field theory defined by the
action (\ref{Polaction}). The conditions $T_{ab}=0$ are often refered
to as ``Virasoro constraints''~\cite{Vir1} and they are equivalent to
two local ``gauge symmetries'' of the Polyakov action, namely the
``reparametrization invariance''
\beq
(\tau,\sigma)~\longmapsto~\Bigl(\tau(\tau',\sigma')\,,\,\sigma(\tau',\sigma')
\Bigr) \ ,
\label{2Dreparinv}\eeq
and the ``Weyl invariance'' (or ``conformal invariance'')
\beq
\gamma_{ab}~\longmapsto~\e^{2\rho(\tau,\sigma)}\,\gamma_{ab}
\label{Weylinv}\eeq
where $\rho(\tau,\sigma)$ is an arbitrary function on the worldsheet. These two
local symmetries of $S[x,\gamma]$ allow us to select a gauge in which the three
functions residing in the symmetric $2\times2$ matrix $(\gamma_{ab})$ are
expressed in terms of just a {\it single} function. A particularly convenient
choice is the ``conformal gauge''
\beq
(\gamma_{ab})=\e^{\phi(\tau,\sigma)}\,(\eta_{ab})=\e^{\phi(\tau,\sigma)}\,
\pmatrix{-1&0\cr0&1\cr} \ .
\label{confgauge}\eeq
In this gauge, the metric $\gamma_{ab}$ is said to be ``conformally flat'',
because it agrees with the Minkowski metric $\eta_{ab}$ of a flat worldsheet,
but only up to the scaling function $\e^\phi$. Then, at the classical level,
the conformal factor $\e^\phi$ drops out of everything and we are left with the
simple gauge-fixed action $S[x,\e^\phi\,\eta]$, i.e. the Polyakov action in the
conformal gauge, and the constraints $T_{ab}=0$:
\beq
\begin{tabular}{|c|}\hline\\
$\begin{array}{rll}
S[x,\e^\phi\,\eta]&=&\displaystyle T\int\dd\tau~\dd\sigma~\left(\dot x^2-x'^2
\right) \ , \\T_{01}&=&T_{10}~=~\dot x\cdot x'~=~0 \ , \\
T_{00}&=&\displaystyle T_{11}~=~\frac12\,\Bigl(\dot x^2+x'^2\Bigr)~=~0 \ .
\end{array}
$\\\\
\hline\end{tabular}
\label{Polconfgauge}\eeq
Note that apart from the constraints, (\ref{Polconfgauge}) defines a {\it free}
(noninteracting) field theory.

\subsection{String Equations of Motion \label{StringEOM}}

The equations of motion for the bosonic string can be derived by applying the
variational principle to the 1+1-dimensional field theory (\ref{Polconfgauge}).
Varying the Polyakov action in the conformal gauge with respect to the $x^\mu$
gives
\beq
\delta S[x,\e^\phi\,\eta]=T\int\dd\tau~\dd\sigma~\left(\eta^{ab}\,
\partial_a\,\partial_bx_\mu\right)\,\delta x^\mu-T\left.\int\dd\tau~x'_\mu\,
\delta x^\mu\right|_{\sigma=0}^{\sigma=\pi} \ .
\label{Polvar}\eeq
The first term in (\ref{Polvar}) yields the usual bulk equations of motion
which here correspond to the two-dimensional wave equation
\beq
\left(\frac{\partial^2}{\partial\sigma^2}-\frac{\partial^2}{\partial\tau^2}
\right)x^\mu(\tau,\sigma)=0 \ .
\label{2dwaveeqn}\eeq
The second term comes from the integration by parts required to arrive at the
bulk differential equation, which involves a total derivative over the spatial
interval $0\leq\sigma\leq\pi$ of the string. In order that the total variation
of the action be zero, these boundary terms must vanish as well. The manner in
which we choose them to vanish depends crucially on whether we are dealing with
{\it closed} or {\it open} strings. The solutions of the classical equations of
motion then correspond to solutions of the wave equation (\ref{2dwaveeqn}) with
the appropriate boundary conditions.

\begin{figure}[htb]
\epsfxsize=2 in
\bigskip
\centerline{\epsffile{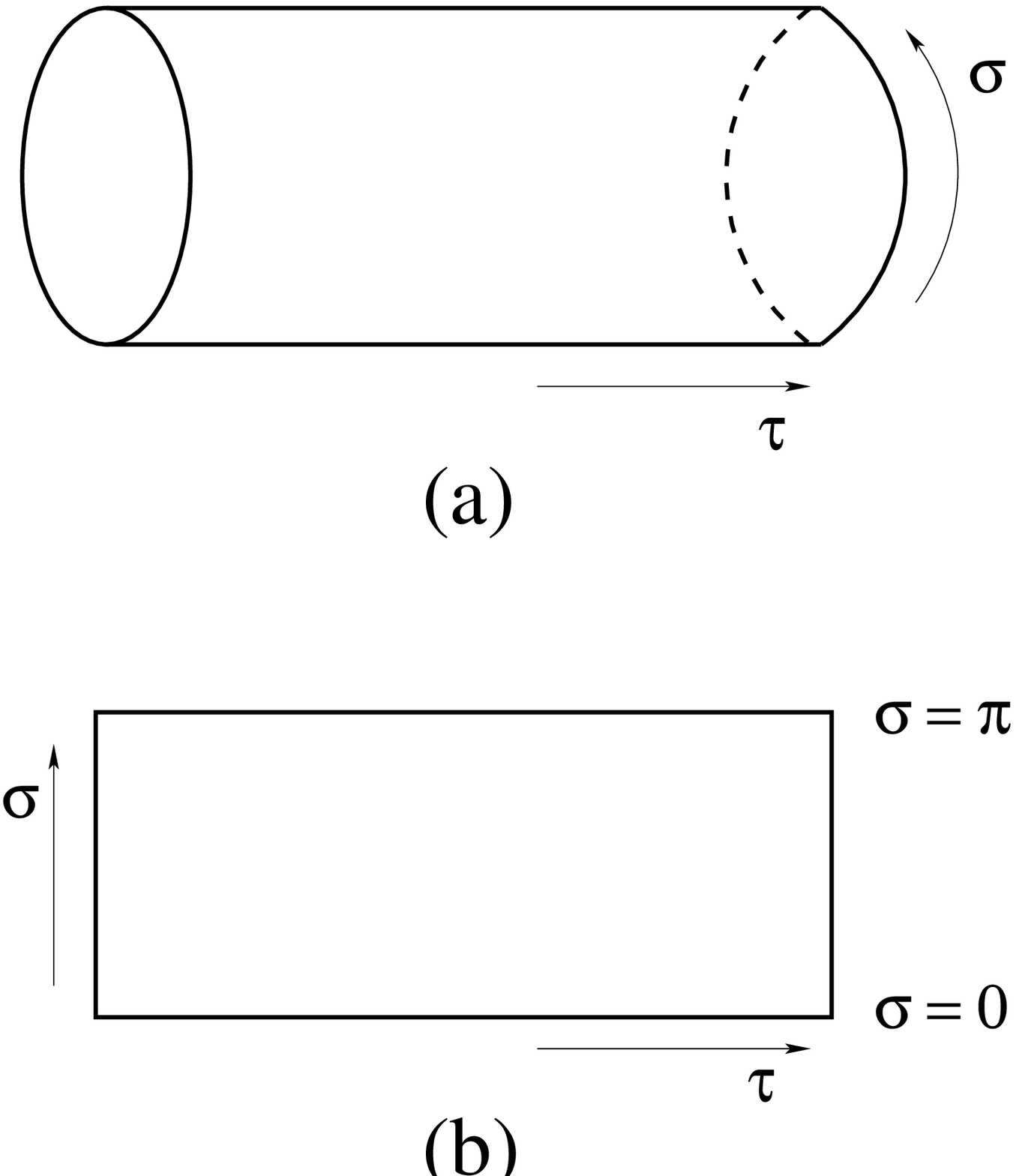}}
\caption{\baselineskip=12pt {\it The worldsheet of} (a) {\it a closed string is
an infinite cylinder $\real\times{\bf S}^1$, and of} (b) {\it an open string is
an infinite strip $\real\times{\bf I}^1$.}}
\bigskip
\label{closedopen}\end{figure}

\noindent
$\underline{\rm Closed~Strings:}$ Here we tie the two ends of the string at
$\sigma=0$ and $\sigma=\pi$ together by imposing periodic boundary conditions
on the string embedding fields (fig.~\ref{closedopen}):
\bea
x^\mu(\tau,0)&=&x^\mu(\tau,\pi) \ , \nn\\x'^\mu(\tau,0)&=&x'^\mu(\tau,\pi) \ .
\label{closedbcs}\eea

\noindent
$\underline{\rm Open~Strings:}$ Here there are two canonical choices of
boundary conditions.\footnote{\baselineskip=12pt These are by no means the only
boundary conditions possible, but they are the only ones of direct physical
relevance to the topics covered in these notes.} {\it Neumann} boundary
conditions are defined by
\beq
x'^\mu(\tau,\sigma)\Bigm|_{\sigma=0,\pi}=0 \ .
\label{Neumannbcs}\eeq
In this case the ends of the string can sit anywhere in spacetime. {\it
Dirichlet} boundary conditions, on the other hand, are defined by
\beq
\dot x^\mu(\tau,\sigma)\Bigm|_{\sigma=0,\pi}= 0 \ .
\label{Dirichletbcs}\eeq
Integrating the condition (\ref{Dirichletbcs}) over $\tau$ specifies a
spacetime location on which the string ends, and so Dirichlet boundary
conditions are equivalent to fixing the endpoints of the string:
\beq
\delta x^\mu(\tau,\sigma)\Bigm|_{\sigma=0,\pi}=0 \ .
\label{Dirichletdelta}\eeq
We will see later on that this spacetime point corresponds to a physical object
called a ``D-brane''. For the time being, however, we shall focus our attention
on Neumann boundary conditions for open strings.

\subsubsection*{Mode Expansions}

To solve the equations of motion, we write the two-dimensional wave equation
(\ref{2dwaveeqn}) in terms of worldsheet light-cone coordinates:
\beq
\partial_+\partial_-x^\mu=0 \ ,
\label{2dwavelightcone}\eeq
where
\beq
\xi^\pm=\tau\pm\sigma \ , ~~ \partial_\pm=\frac\partial{\partial\xi^\pm} \ .
\label{lightconecoords}\eeq
The general solution of (\ref{2dwavelightcone}) is then the sum of an analytic
function of $\xi^+$ alone, which we will call the ``left-moving'' solution, and
an analytic function of $\xi^-$ alone, which we call the ``right-moving''
solution, $x^\mu(\tau,\sigma)=x_{\rm L}^\mu(\xi^+)+x_{\rm R}^\mu(\xi^-)$. The
precise form of the solutions now depends on the type of boundary conditions.

\noindent
$\underline{\rm Closed~Strings:}$ The periodic boundary conditions
(\ref{closedbcs}) accordingly restrict the Taylor series expansions of the
analytic functions which solve (\ref{2dwavelightcone}), and we arrive at the
solution
\beq
\begin{tabular}{|c|}\hline\\
$\begin{array}{rll}
x^\mu(\tau,\sigma)&=&x_{\rm L}^\mu(\xi^+)+x_{\rm R}^\mu(\xi^-) \ , \\
x_{\rm L}^\mu(\xi^+)&=&\displaystyle\frac12\,x_0^\mu+\alpha'\,p_0^\mu\,\xi^+
+\ii\,\sqrt{\frac{\alpha'}2}\,\sum_{n\neq0}\frac{\tilde\alpha_n^\mu}n~
\e^{-2\,\ii\,n\xi^+} \ , \\x_{\rm R}^\mu(\xi^-)&=&\displaystyle\frac12\,x_0^\mu
+\alpha'\,p_0^\mu\,\xi^-+\ii\,\sqrt{\frac{\alpha'}2}\,\sum_{n\neq0}
\frac{\alpha_n^\mu}n~\e^{-2\,\ii\,n\xi^-} \ .
\end{array}
$\\\\
\hline\end{tabular}
\label{closedmodeexp}\eeq
We have appropriately normalized the terms in these Fourier-type series
expansions, which we will refer to as ``mode expansions'', according to
physical dimension. Reality of the string embedding function $x^\mu$ requires
the integration constants $x_0^\mu$ and $p_0^\mu$ to be real, and
\beq
(\tilde\alpha_n^\mu)^*=\tilde\alpha_{-n}^\mu \ , ~~ (\alpha_n^\mu)^*=
\alpha_{-n}^\mu \ .
\label{alphaconj}\eeq
By integrating $x^\mu$ and $\dot x^\mu$ over $\sigma\in[0,\pi]$ we see that
$x_0^\mu$ and $p_0^\mu$ represent the center of mass position and momentum of
the string, respectively. The $\tilde\alpha_n^\mu$ and $\alpha_n^\mu$ represent
the oscillatory modes of the string. The mode expansions (\ref{closedmodeexp})
correspond to those of left and right moving travelling waves circulating
around the string in opposite directions.

\begin{figure}[htb]
\epsfxsize=4 in
\bigskip
\centerline{\epsffile{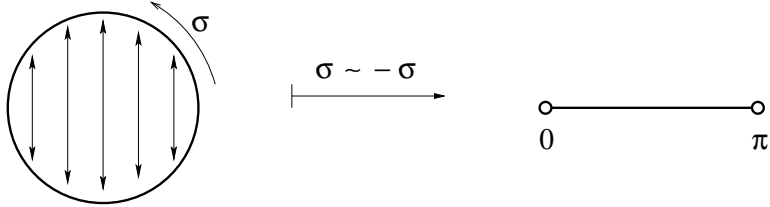}}
\caption{\baselineskip=12pt {\it The doubling trick identifies opposite points
on the circle and maps it onto a finite interval.}}
\bigskip
\label{doublingtrick}\end{figure}

\noindent
$\underline{\rm Open~Strings:}$ For open strings, the spatial worldsheet
coordinate $\sigma$ lives on a finite interval rather than a circle. The open
string mode expansion may be obtained from that of the closed string through
the ``doubling trick'', which identifies $\sigma\sim-\sigma$ on the circle
${\bf S}^1$ and thereby maps it onto the finite interval $[0,\pi]$
(fig.~\ref{doublingtrick}). The open string solution to the equations of motion
may thereby be obtained from (\ref{closedmodeexp}) by imposing the extra
condition $x^\mu(\tau,\sigma)=x^\mu(\tau,-\sigma)$. This is of course still
compatible with the wave equation (\ref{2dwaveeqn}) and it immediately implies
the Neumann boundary conditions (\ref{Neumannbcs}). We therefore find
\beq
\begin{tabular}{|c|}\hline\\
$\displaystyle
x^\mu(\tau,\sigma)=x_0^\mu+2\alpha'\,p_0^\mu\,\tau
+\ii\,\sqrt{2\alpha'}\,\sum_{n\neq0}\frac{\alpha_n^\mu}n~
\e^{-\ii\,n\tau}\,\cos(n\sigma) \ .
$\\\\
\hline\end{tabular}
\label{openmodeexp}\eeq
The open string mode expansion has a standing wave for its solution,
representing the left and right moving sectors reflected into one another by
the Neumann boundary condition~(\ref{Neumannbcs}).

\subsubsection*{Mass-Shell Constraints}

The final ingredients to go into the classical solution are the physical
constraints $T_{ab}=0$. In the light-cone coordinates (\ref{lightconecoords}),
the components $T_{-+}$ and $T_{+-}$ are identically zero, while in the case of
the closed string the remaining components are given by

\vbox{\bea
T_{++}(\xi^+)&=&\frac12\,\Bigl(\partial_+x_{\rm L}\Bigr)^2~=~
\sum_{n=-\infty}^\infty\tilde L_n~\e^{2\,\ii\,n\xi^+}~=~0 \ , \nn\\
T_{--}(\xi^-)&=&\frac12\,\Bigl(\partial_-x_{\rm R}\Bigr)^2~=~
\sum_{n=-\infty}^\infty L_n~\e^{2\,\ii\,n\xi^-}~=~0 \ ,
\label{Tlightcone}\eea}
\noindent
where we have defined
\beq
\begin{tabular}{|c|}\hline\\
$\displaystyle
\tilde L_n=\frac12\,\sum_{m=-\infty}^\infty\tilde\alpha_{n-m}\cdot\tilde
\alpha_m \ , ~~ L_n=\frac12\,\sum_{m=-\infty}^\infty\alpha_{n-m}\cdot\alpha_m
$\\\\
\hline\end{tabular}
\label{Lndef}\eeq
with
\beq
\tilde\alpha_0^\mu=\alpha_0^\mu=\sqrt{\frac{\alpha'}2}\,p_0^\mu \ .
\label{alpha0def}\eeq
For open strings, we have only the constraint involving untilded quantities,
and the definition of the zero modes (\ref{alpha0def}) changes to
$\alpha_0^\mu=\sqrt{2\alpha'}\,p_0^\mu$.

This gives an infinite number of constraints corresponding to an infinite
number of conserved currents of the 1+1-dimensional field theory. They are
associated with the residual, local, infinite-dimensional ``conformal
symmetry'' of the theory which preserves the conformal gauge condition
(\ref{confgauge}):
\bea
\xi^+&\longmapsto&\xi'^{\,+}~=~f(\xi^+) \ , \nn\\\xi^-&\longmapsto&\xi'^{\,-}
{}~=~g(\xi^-) \ ,
\label{confsym}\eea
where $f$ and $g$ are arbitrary analytic functions. Only the conformal
factor $\phi$ in (\ref{confgauge}) is affected by such coordinate
transformations, and so the entire classical theory is invariant under
them. They are known as ``conformal transformations'' and they rescale
the induced worldsheet metric while leaving preserved all angles in
two-dimensions. This ``conformal invariance'' of the worldsheet field
theory makes it a ``conformal field theory''~\cite{BPZ1,Ginsparg1},
and represents one of the most powerful results and techniques of
perturbative string theory.

\setcounter{equation}{0}

\section{Quantization of the Bosonic String \label{BosQuant}}

We will now proceed to quantize the classical theory of the previous section.
We will proceed by applying the standard canonical quantization techniques of
quantum field theory to the 1+1-dimensional worldsheet field theory. Imposing
the mass-shell constraints will thereby lead to the construction of physical
states and the appearence of elementary particles in the quantum string
spectrum. We will then turn to some more formal aspects of the quantum theory,
namely a description of the operators which create particles from the string
vacuum, a heuristic explanation of the structure of the string perturbation
expansion, and how to incorporate non-abelian gauge degrees of freedom into the
spectrum of string states.

\subsection{Canonical Quantization \label{CanQuant}}

Recall from the previous section that the string worldsheet field theory is
essentially a free field theory, and hence its quantization is particularly
simple. The basic set-up for canonical quantization follows the standard
practise of quantum field theory and is left as an exercise.

\bigskip

\begin{center}
\begin{minipage}{15cm}

\small

{\bf Exercise~4.1.} {\sl\baselineskip=12pt {\bf (a)} Starting from the Polyakov
action in the conformal gauge, show that the total classical Hamiltonian is
given by
$$
H=\left\{\begin{array}{l}\displaystyle
\frac12\,\sum_{n=-\infty}^\infty\alpha_{-n}\cdot
\alpha_n=L_0~~~~{\rm for~open~strings} \ , \\\displaystyle \frac12\,
\sum_{n=-\infty}^\infty\left(\tilde\alpha_{-n}\cdot\tilde\alpha_n+
\alpha_{-n}\cdot\alpha_n\right)
=\tilde L_0+L_0~~~~{\rm for~closed~strings} \ .
\end{array}\right.
$$

\noindent
{\bf (b)} Calculate the canonical momentum conjugate to the string embedding
field $x^\mu$, and hence show that the oscillator modes have the quantum
commutators
$$
\begin{array}{rll}
\displaystyle\left[x_0^\mu\,,\,p_0^\nu\right]&=&\ii\,\eta^{\mu\nu} \ , \\
\displaystyle\left[x_0^\mu\,,x_0^\nu\right]&=&\displaystyle
\left[p_0^\mu\,,p_0^\nu\right]~=~0 \ , \\
\displaystyle\left[\alpha_m^\mu\,,\,\alpha_n^\nu\right]&=&\displaystyle
\left[\tilde\alpha_m^\mu\,,\,\tilde\alpha_n^\nu\right]~=~m\,\delta_{m+n,0}\,
\eta^{\mu\nu} \ , \\\displaystyle
\left[\alpha_m^\mu\,,\,\tilde\alpha_n^\nu\right]&=&0 \ .
\end{array}
$$}

\end{minipage}
\end{center}

\bigskip

{}From exercise~4.1 it follows that
$(a_m^\mu,a_m^{\mu\,\dag})=(\frac1{\sqrt m}\,\alpha_m^\mu,\frac1{\sqrt
  m}\,\alpha_{-m}^\mu)$ define quantum mechanical raising and lowering
operators for the simple harmonic oscillator, i.e.
\beq
\left[a_m^\mu\,,\,a_m^{\mu\,\dag}\right]=1 \ .
\label{ammualg}\eeq
This structure is of course anticipated, because free quantum fields are simply
composed of infinitely many harmonic oscillators. The corresponding Hilbert
space is therefore a Fock space spanned by products of states $|n\rangle$,
$n=0,1,2,\dots$, which are built on a normalized ground state $|0\rangle$
annihilated by the lowering operators $a=a_m^\mu$:
\bea
\langle0|0\rangle&=&1 \ , \nn\\ a|0\rangle&=&0 \ , \nn\\
|n\rangle&=&\frac{(a^\dag)^n}{\sqrt{n!}}\,|0\rangle \ .
\label{Fockstates}\eea
By repeated use of the commutation relation (\ref{ammualg}), in the usual way
one can arrive at the relations
\bea
\langle m|n\rangle&=&\delta_{nm} \ , \nn\\a^\dag a|n\rangle&=&n|n\rangle \ .
\label{Fockstatesrels}\eea
In each sector of the theory, i.e. open, and closed left-moving and
right-moving, we get $d$ independent families of such infinite sets of
oscillators, one for each spacetime dimension $\mu=0,1,\dots,d-1$. The only
subtlety in this case is that, because $\eta^{00}=-1$, the time components are
proportional to oscillators with the wrong sign, i.e.
$[a_m^0,a_m^{0\,\dag}]=-1$. Such oscillators are potentially dangerous, because
they create states of negative norm which can lead to an inconsistent,
non-unitary quantum theory (with negative probabilities and the like). However,
as we will see, the Virasoro constraints $T_{ab}=0$ eliminate the negative norm
states from the physical spectrum of the string.

{}From exercise~4.1 it also follows that the zero mode operators $x_0^\mu$ and
$p_0^\nu$ obey the standard Heisenberg commutation relations. They may thereby
be represented on the Hilbert space spanned by the usual plane wave basis
$|k\rangle=\e^{\ii\,k\cdot x}$ of eigenstates of $p_0^\mu$. Thus the Hilbert
space of the string is built on the states $|k;0\rangle$ of center of mass
momentum $k^\mu$ with
\bea
p_0^\mu|k;0\rangle&=&k^\mu|k;0\rangle \ , \nn\\\alpha_m^\mu|k;0\rangle&=&0
\label{groundstate}\eea
for $m>0$ and $\mu=0,1,\dots,d-1$. For closed strings, there is also an
independent left-moving copy of the Fock space.

\subsubsection*{Normal Ordering}

For the quantum versions of the Virasoro operators defined in (\ref{Lndef}), we
use the usual ``normal ordering'' prescription that places all lowering
operators to the right. Then the $L_n$ for $n\neq0$ are all fine when promoted
to quantum operators. However, the Hamiltonian $L_0$ needs more careful
definition, because $\alpha_n^\mu$ and $\alpha_{-n}^\mu$ do not commute. As a
quantum operator we have
\beq
L_0=\frac12\,\alpha_0^2+\sum_{n=1}^\infty\alpha_{-n}\cdot\alpha_n+
\varepsilon_0 \ ,
\label{L0op}\eeq
where
\beq
\varepsilon_0=\frac{d-2}2\,\sum_{n=1}^\infty n
\label{Casimirenergy}\eeq
is the total zero-point energy of the families of infinite field oscillators
(coming from the usual zero point energy $\frac12$ of the quantum mechanical
harmonic oscillator). The factor of $d-2$ appears, rather than $d$, because
after imposition of the physical constraints there remain only $d-2$
independent polarizations of the string embedding fields $x^\mu$. Explicitly,
the constraints can be satisfied by imposing a spacetime light-cone gauge
condition $p^+=0$ which retains only transverse degrees of freedom.

The quantity $\varepsilon_0$ is simply the Casimir energy arising from the fact
that the 1+1-dimensional quantum field theory here is defined on a box, which
is an infinite strip for open strings and an infinite cylinder for closed
strings. Of course $\varepsilon_0=\infty$ formally, but, as usual in quantum
field theory, it can be regulated to give a finite answer corresponding to the
total zero-point energy of all harmonic oscillators in the system. For this, we
write it as
\beq
\varepsilon_0=\frac{d-2}2\,\zeta(-1) \ ,
\label{Casimirzeta}\eeq
where
\beq
\zeta(z)=\sum_{n=1}^\infty\frac1{n^z}
\label{zetafn}\eeq
for $z\in\complex$ is called the ``Riemann zeta-function''. The function
$\zeta(z)$ can be regulated and has a well-defined analytical continuation to a
finite function for ${\rm Re}(z)\leq1$. In this paper we will only require its
values at $z=0$ and $z=-1$~\cite{GradRhy1}:
\bea
\zeta(0)&=&-\frac12 \ , \nn\\\zeta(-1)&=&-\frac1{12} \ .
\label{zeta01}\eea
The vacuum energy is thereby found to be
\beq
\begin{tabular}{|c|}\hline\\
$\displaystyle
\varepsilon_0=-\frac{d-2}{24} \ .
$\\\\
\hline\end{tabular}
\label{Casimirreg}\eeq

\bigskip

\begin{center}
\begin{minipage}{15cm}

\small

{\bf Exercise~4.2.} {\sl\baselineskip=12pt {\bf (a)} Show that the angular
momentum operators of the worldsheet field theory are given by
$$\displaystyle
J^{\mu\nu}=x_0^\mu\,p_0^\nu-x_0^\nu\,p_0^\mu-\ii\,\sum_{n=1}^\infty
\frac1n\,\Bigl(\alpha_{-n}^\mu\,\alpha_n^\nu-\alpha_{-n}^\nu\,
\alpha_n^\mu\Bigr) \ .
$$

\noindent
{\bf (b)} Use the canonical commutation relations to verify that the Poincar\'e
algebra
$$
\begin{array}{rll}
\left[p_0^\mu\,,\,p_0^\nu\right]&=&0 \ , \\\left[p_0^\mu\,,\,J^{\nu\rho}
\right]&=&-\ii\,\eta^{\mu\nu}\,p_0^\rho+\ii\,\eta^{\mu\rho}\,p_0^\nu \ , \\
\left[J^{\mu\nu}\,,\,J^{\rho\lambda}\right]&=&-\ii\,\eta^{\nu\rho}\,
J^{\mu\lambda}+\ii\,\eta^{\mu\rho}\,J^{\nu\lambda}+\ii\,\eta^{\nu\lambda}
\,J^{\mu\rho}-\ii\,\eta^{\mu\lambda}\,J^{\nu\rho}
\end{array}
$$
is satisfied.

\noindent
{\bf (c)} Show that for all $n$,
$$
\left[L_n\,,\,J^{\mu\nu}\right]=0 \ .
$$
This will guarantee later on that the string states are Lorentz multiplets.}

\end{minipage}
\end{center}

\bigskip

\subsection{The Physical String Spectrum \label{Spectrum}}

Our construction of quantum states of the bosonic string will rely heavily on a
fundamental result that is at the heart of the conformal symmetry described at
the end of section~\ref{StringEOM}.

\bigskip

\begin{center}
\begin{minipage}{15cm}

\small

{\bf Exercise~4.3.} {\sl\baselineskip=12pt Use the oscillator algebra
  to show that the operators $L_n$ generate the infinite-dimensional
  ``Virasoro algebra of central charge $c=d$''~\cite{Vir1},
$$
[L_n,L_m]=(n-m)\,L_{n+m}+\frac c{12}\,\Bigl(n^3-n\Bigr)\,\delta_{n+m,0} \ .
$$
The constant term on the right-hand side of these commutation
relations is often called the ``conformal anomaly'', as it represents
a quantum breaking of the classical conformal symmetry
algebra~\cite{Poly1}.}

\end{minipage}
\end{center}

\bigskip

We define the ``physical states'' $|{\rm phys}\rangle$ of the full Hilbert
space to be those which obey the Virasoro constraints $T_{ab}\equiv0$:
\bea
(L_0-a)|{\rm phys}\rangle&=&0 \ , ~~ a\equiv-\varepsilon_0>0 \ , \nn\\
L_n|{\rm phys}\rangle&=&0
\label{GBprescr}\eea
for $n>0$. These constraints are just the analogs of the
``Gupta-Bleuler prescription'' for imposing mass-shell constraints in
quantum electrodynamics. The $L_0$ constraint in (\ref{GBprescr}) is a
generalization of the Klein-Gordon equation, as it contains
$p_0^2=-\partial_\mu\,\partial^\mu$ plus oscillator terms. Note that
because of the central term in the Virasoro algebra, it is
inconsistent to impose these constraints on both $L_n$ and $L_{-n}$.

\subsubsection*{The Open String Spectrum}

We will begin by considering open strings as they are somewhat easier to
describe. Mathematically, their spectrum is the same as the that of the
right-moving sector of closed strings. The closed string spectrum will thereby
be straightforward to obtain afterwards. The constraint $L_0=a$ in
(\ref{GBprescr}) is then equivalent to the ``mass-shell condition''
\beq
\begin{tabular}{|c|}\hline\\
$\displaystyle
m^2=-p_0^2=-\frac1{2\alpha'}\,\alpha_0^2=\frac1{\alpha'}\,\Bigl(N-a\Bigr) \ ,
$\\\\
\hline\end{tabular}
\label{openmassshell}\eeq
where $N$ is the ``level number'' which is defined to be the oscillator number
operator
\beq
N=\sum_{n=1}^\infty\alpha_{-n}\cdot\alpha_n=\sum_{n=1}^\infty n\,a_n^\dag
\cdot a_n=0,1,2,\dots \ ,
\label{levelnumber}\eeq
and $N_n\equiv a_n^\dag\cdot a_n=0,1,2,\dots$ is the usual number operator
associated with the oscillator algebra (\ref{ammualg}).

\noindent
$\underline{{\rm Ground~State}~N=0:}$ The ground state has a unique realization
whereby all oscillators are in the Fock vacuum, and is therefore given by
$|k;0\rangle$. The momentum $k$ of this state is constrained by the Virasoro
constraints to have mass-squared given by
\beq
-k^2=m^2=-\frac a{\alpha'}<0 \ .
\label{groundmass}\eeq
Since the vector $k^\mu$ is space-like, this state therefore describes a
``tachyon'', i.e. a particle which travels faster than the speed of light. So
the bosonic string theory is a {\it not} a consistent quantum theory, because
its vacuum has imaginary energy and hence is {\it unstable}. As in quantum
field theory, the presence of a tachyon indicates that one is perturbing around
a local maximum of the potential energy, and we are sitting in the wrong
vacuum. However, in perturbation theory, which is the framework in which we are
implicitly working here, this instability is not visible. Since we will
eventually remedy the situation by studying tachyon-free superstring theories,
let us just plug along without worrying for now about the tachyonic state.

\noindent
$\underline{{\rm First~Excited~Level}~N=1:}$ The only way to get $N=1$ is to
excite the first oscillator modes once, $\alpha_{-1}^\mu|k;0\rangle$. We are
also free to specify a ``polarization vector'' $\zeta_\mu$ for the state. So
the most general level 1 state is given by
\beq
|k;\zeta\rangle=\zeta\cdot\alpha_{-1}|k;0\rangle \ .
\label{level1state}\eeq
The Virasoro constraints give the energy
\beq
m^2=\frac1{\alpha'}\,\Bigl(1-a\Bigr) \ .
\label{level1mass}\eeq
Furthermore, using the commutation relations and (\ref{Lndef}) we may compute
\beq
L_1|k;\zeta\rangle=\sqrt{2\alpha'}\,(k\cdot\alpha_1)(\zeta\cdot\alpha_{-1})
|k;0\rangle=\sqrt{2\alpha'}\,(k\cdot\zeta)|k;0\rangle \ ,
\label{L1level1action}\eeq
and thus the physical state condition $L_1|k;\zeta\rangle=0$ implies that the
polarization and momentum of the state must obey
\beq
k\cdot\zeta=0 \ .
\label{kzeta0}\eeq
The cases $a\neq1$ turn out to be unphysical, as they contain tachyons and
ghost states of negative norm. So we shall take $a=1$, which upon comparison
with (\ref{Casimirreg}) fixes the ``bosonic critical dimension of spacetime'':
\beq
\begin{tabular}{|c|}\hline\\
$\displaystyle
d=26 \ .
$\\\\
\hline\end{tabular}
\label{boscritdim}\eeq
The condition (\ref{boscritdim}) can be regarded as the requirement of
cancellation of the conformal anomaly in the quantum
theory~\cite{Poly1}, obtained by demanding the equivalence of the
quantizations in both the light-cone and conformal gauges. The latter
approach relies on the introduction of worldsheet Faddeev-Popov ghost
fields for the gauge-fixing of the conformal gauge in the quantum theory.

\bigskip

\begin{center}
\begin{minipage}{15cm}

\small

{\bf Exercise~4.4.} {\sl\baselineskip=12pt Consider the ``spurious state''
defined by $|\psi\rangle=L_{-1}|k;0\rangle$. Show that:

\noindent
{\bf (a)} It can be written in the form
$$\displaystyle
|\psi\rangle=2\,\sqrt{2\alpha'}\,|k;k\rangle \ .
$$

\noindent
{\bf (b)} It is orthogonal to any physical state.

\noindent
{\bf (c)}
$$
L_1|\psi\rangle=\alpha'\,k^2|k;0\rangle \ .
$$}

\end{minipage}
\end{center}

\bigskip

The $N=1$ state $|k;\zeta\rangle$ constructed above with $k^2=m^2=0$ and
$k\cdot\zeta=0$ has $d-2=24$ independent polarization states, as the physical
constraints remove two of the initial $d$ vector degrees of freedom. It
therefore describes a massless spin 1 (vector) particle with polarization
vector $\zeta_\mu$, which agrees with what one finds for a massless Maxwell or
Yang-Mills field. Indeed, there is a natural way to describe an associated
gauge invariance in this picture.

\noindent
$\underline{\rm Gauge~Invariance:}$ The corresponding spurious state
$|\psi\rangle$ is both physical and null, and so we can add it to {\it any}
physical state with no physical consequences. We should therefore impose an
equivalence relation
\beq
|{\rm phys}\rangle\sim|{\rm phys}\rangle+\lambda\,|\psi\rangle
\label{physequivrel}\eeq
where $\lambda$ is any constant. For the physical state $|{\rm
phys}\rangle=|k;\zeta\rangle$, exercise~4.4~(a) implies that
(\ref{physequivrel}) is the same as an equivalence relation on the polarization
vectors,
\beq
\zeta^\mu\sim\zeta^\mu+2\lambda\,\sqrt{2\alpha'}\,k^\mu \ .
\label{zetaequivrel}\eeq
This is a $U(1)$ gauge symmetry (in momentum space), and so at level
$N=1$ we have obtained the 24 physical states of a photon field
$A_\mu(x)$ in 26-dimensional spacetime.

\noindent
$\underline{{\rm Higher~Levels}~N\geq2:}$ The higher level string states with
$N\geq2$ are all massive and will not be dealt with here. We simply note that
there is an infinite tower of them, thereby making the string theory suited to
describe all of the elementary particles of nature.

\subsubsection*{The Closed String Spectrum}

The case of closed strings is similar to that of open strings. We now
have to also incorporate the left-moving sector Fock states. Thus we
can easily extend the analysis to the closed string case by simply
taking the tensor product of two copies of the open string
result and changing the definition of the zero-modes as in
(\ref{alpha0def}). However, a new condition now arises. Adding and subtracting
the two physical state conditions $(L_0-1)|{\rm phys}\rangle=(\tilde
  L_0-1)|{\rm phys}\rangle=0$ yields, respectively, the quantum
  constraints
\bea
\left(L_0+\tilde L_0-2\right)|{\rm phys}\rangle&=&0 \ , \nn\\
\left(L_0-\tilde L_0\right)|{\rm phys}\rangle&=&0 \ ,
\label{closedopconstrs}\eea
where we have again fixed the value $a=1$. The first constraint yields
the usual mass-shell relation, since
$H=L_0+\tilde L_0-2$ is the worldsheet Hamiltonian which generates time
translations on the string worldsheet. The second constraint can be
understood by noting that the operator $P=L_0-\tilde L_0$ is the
worldsheet momentum, and so it generates translations in the string
position coordinate $\sigma$. This constraint therefore simply
reflects the fact that there is no physical significance as to where
on the string we are, and hence that the physics is invariant under
translations in $\sigma$. It amounts to equating the number of right-moving and
left-moving oscillator modes. We thereby arrive at the new mass-shell
relation
\beq
\begin{tabular}{|c|}\hline\\
$\displaystyle
m^2=\frac4{\alpha'}\,\Bigl(N-1\Bigr) \ ,
$\\\\
\hline\end{tabular}
\label{closedmassshell}\eeq
and the additional ``level-matching condition''
\beq
\begin{tabular}{|c|}\hline\\
$\displaystyle
N=\tilde N \ .
$\\\\
\hline\end{tabular}
\label{levelmatching}\eeq

\noindent
$\underline{{\rm Ground~State}~N=0:}$ The ground state is
$|k;0,0\rangle$ and it has mass-squared
\beq
m^2=-\frac4{\alpha'}<0 \ .
\label{groundmassclosed}\eeq
It again represents a spin 0 tachyon, and the closed string vacuum is
also unstable.

\noindent
$\underline{{\rm First~Excited~Level}~N=1:}$ The first excited state
is generically of the form
\beq
|k;\zeta\rangle=\zeta_{\mu\nu}\,\Bigl(\alpha_{-1}^\mu|k;0\rangle\otimes
\tilde\alpha_{-1}^\nu|k;0\rangle\Bigr)
\label{kzetaclosed}\eeq
and it has mass-squared
\beq
m^2=0 \ .
\label{m20closed}\eeq
The Virasoro constraints in addition give
\beq
L_1|k;\zeta\rangle=\tilde L_1|k;\zeta\rangle=0
\label{L1tildeL10}\eeq
which are equivalent to
\beq
k^\mu\,\zeta_{\mu\nu}=0 \ .
\label{kzeta0closed}\eeq

A polarization tensor $\zeta_{\mu\nu}$ obeying (\ref{kzeta0closed})
encodes three distinct spin states according to the decomposition of
$\zeta_{\mu\nu}$ into irreducible representations of the spacetime
``Little group'' $SO(24)$, which classifies massless fields in this
case~\cite{BUSSTEPPJMF}. This is the residual Lorentz symmetry group
that remains after the Virasoro constraints have been taken into
account, so that the spectrum is exactly what one would get from 24
vectors in each sector which are transversely polarized to the
light-cone. From a group theoretical perspective, this decomposition
comes from writing down the Clebsch-Gordan decomposition of the tensor
product of two vector representations ${\mathbf{24}}$ of $SO(24)$ into
the irreducible symmetric, antisymmetric, and trivial representations,
\beq
{\mathbf{24}}\otimes{\mathbf{24}}={\mathbf{S}}\oplus{\mathbf{A}}
\oplus{\mathbf{1}} \ .
\label{24decomp}\eeq
More concretely, we decompose the rank 2 tensor $\zeta_{\mu\nu}$
according to (\ref{24decomp}) as
\bea
\zeta_{\mu\nu}&=&\left[\frac12\,\Bigl(\zeta_{\mu\nu}+\zeta_{\nu\mu}\Bigr)-
\frac1{25}\,\tr(\zeta)\right]+\left[\frac12\,\Bigl(\zeta_{\mu\nu}-
\zeta_{\nu\mu}\Bigr)\right]+\left[\frac1{25}\,\eta_{\mu\nu}\,\tr(\zeta)
\right]\nn\\&\equiv&[g_{\mu\nu}]+[B_{\mu\nu}]+[\eta_{\mu\nu}\,\Phi] \ .
\label{zetadecomp}\eea
The symmetric, traceless tensor $g_{\mu\nu}$ corresponds to the spin 2
``graviton field'' and it yields the spacetime metric. The antisymmetric
spin 2 tensor $B_{\mu\nu}$ is called the ``Neveu-Schwarz $B$-field'', while
the scalar field $\Phi$ is the spin 0 ``dilaton''.

\noindent
$\underline{\rm Gauge~Invariance:}$ The equivalence relations
generated by spurious states built from $L_{-1}$ and $\tilde L_{-1}$
give the ``gauge transformations''
\bea
g_{\mu\nu}&\longmapsto&g_{\mu\nu}+\partial_\mu\Lambda_\nu+\partial_\nu
\Lambda_\mu \ , \nn\\B_{\mu\nu}&\longmapsto&B_{\mu\nu}+\partial_\mu
\Lambda_\nu-\partial_\nu\Lambda_\mu \ .
\label{gBgaugetransf}\eea
In this sense the tensor $g_{\mu\nu}$ admits a natural interpretation
as a graviton field, as it has the correct diffeomorphism gauge
invariance. Its presence accounts for the fact that string theory
contains gravity, which is a good approximation at energies
$\ll1/\ell_s$. Its vacuum expectation value $\langle
g_{\mu\nu}\rangle$ determines the spacetime geometry. Similarly, the
vacuum expectation value of the dilaton $\Phi$ determines the ``string
coupling constant'' $g_s$ through
\beq
g_s=\left\langle\e^\Phi\,\right\rangle \ .
\label{gsPhi}\eeq
This relationship can be derived using vertex operators, which will be
discussed in the next subsection.

The gauge transformation rule for $B_{\mu\nu}$, on the other hand, is
a generalization of that for the Maxwell field:
\beq
A_\mu~\longmapsto~A_\mu+\partial_\mu\Lambda \ .
\label{Maxwellgaugetransf}\eeq
The importance of the $B$-field resides in the fact that a fundamental
string is a source for it, just like a charged particle is a source for an
electromagnetic vector potential $A_\mu$ through the coupling
\beq
q\int\dd\tau~\dot x^\mu(\tau)\,A_\mu \ .
\label{Amincoupling}\eeq
In an analogous way, the $B$-field couples to strings via
\beq
q\int\dd^2\xi~\epsilon^{ab}\,\partial_ax^\mu\,\partial_bx^\nu\,
B_{\mu\nu} \ ,
\label{Bmincoupling}\eeq
where $\epsilon^{ab}$ is the antisymmetric tensor with
$\epsilon^{01}=-\epsilon^{10}=1$.

\subsubsection*{Worldsheet-Spacetime Interplay}

The upshot of the physical string spectrum can be summarized through
the interplay between worldsheet and spacetime quantities. At the
lowest level of massless states ($N=1$), open strings correspond to gauge
theory while closed strings correspond to gravity. This interplay will
be a recurring theme in these notes. The higher levels $N\geq2$ give
an infinite tower of massive particle excitations. Notice that the
massless states are picked out in the limit $\alpha'\to0$
($\ell_s\to0$) in which the strings look like point-like objects. We
shall refer to this low-energy limit as the ``field theory limit'' of
the string theory.

\subsection{Vertex Operators \label{Vertex}}

\begin{figure}[htb]
\epsfxsize=5 in
\bigskip
\centerline{\epsffile{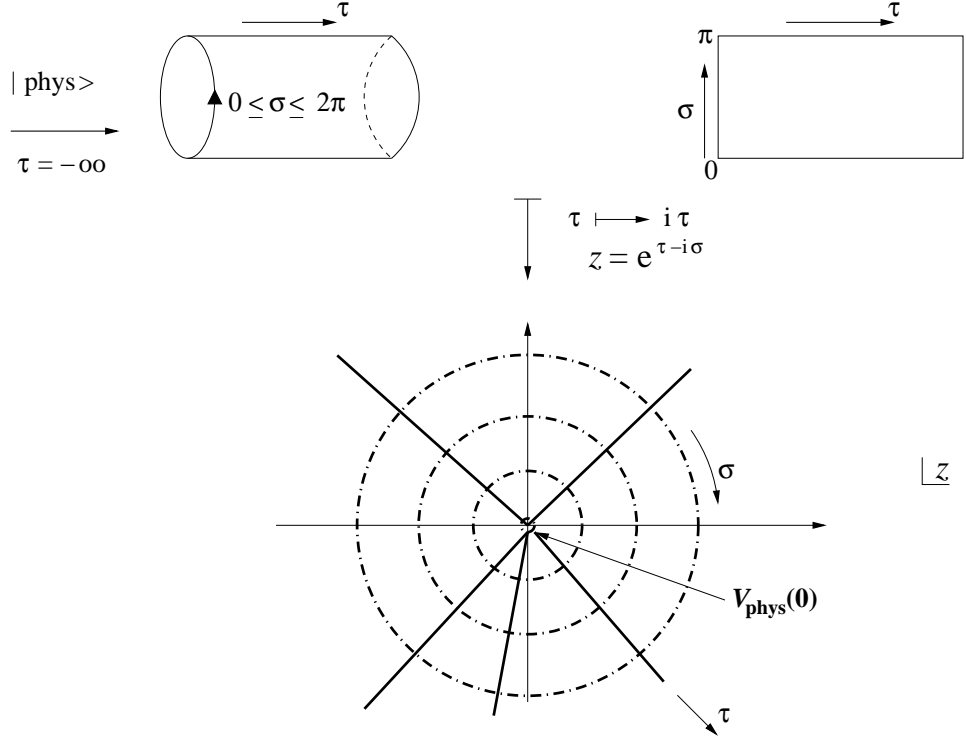}}
\caption{\baselineskip=12pt {\it The mapping of the worldsheet
    cylinder (resp. strip) onto the complex plane (resp. upper complex
  half-plane). The lines of constant $\tau$ form concentric circles
  about the origin in the complex $z$-plane, while the lines of
  constant $\sigma$ correspond to lines emerging in the radial
  directions from the origin. An incoming physical state $|{\rm
  phys}\rangle$ in the infinite worldsheet past ($\tau=-\infty$)
  corresponds to the insertion of a vertex operator $V_{\rm phys}(z)$
  at the origin $z=0$.}}
\bigskip
\label{complexmap}\end{figure}

Any local and unitary quantum field theory has an appropriate operator-state
correspondence which allows one to associate quantum fields to quantum
states in a one-to-one manner. The states may then be regarded as being created
from the vacuum by the quantum fields. We will now describe how to formulate
this correspondence in the context of string theory. For this, we map
the closed string cylinder and the open string strip to the complex
plane and the upper complex half-plane, respectively, by first a Wick
rotation $\tau\mapsto\ii\,\tau$ to Euclidean worldsheet signature,
followed by the coordinate transformation $z=\e^{\tau-\ii\,\sigma}$
(fig.~\ref{complexmap}). The advantage of this coordinate
transformation is that it allows us to reinterpret the mode expansions
of section~\ref{StringEOM} as Laurent series in the complex plane. In
particular, for closed strings the coordinate transformation
$\xi^\pm\mapsto z,\overline{z}$ allows us to write (\ref{closedmodeexp})
as
\bea
\partial_zx_{\rm L}^\mu(z)&=&-\ii\,\sqrt{\frac{\alpha'}2}~
\sum_{n=-\infty}^\infty\alpha_n^\mu~z^{-n-1} \ , \nn\\
\partial_{\overline{z}}x_{\rm R}^\mu(\overline{z})&=&-\ii\,
\sqrt{\frac{\alpha'}2}~\sum_{n=-\infty}^\infty\tilde
\alpha_n^\mu~\overline{z}^{\,-n-1} \ .
\label{modeLaurent}\eea
These relations can be inverted by using the Cauchy integral formula
to give
\bea
\alpha_{-n}^\mu&=&\sqrt{\frac2{\alpha'}}\,\oint\frac{\dd z}{2\pi}~
z^{-n}~\partial_zx_{\rm L}^\mu(z) \ , \nn\\\tilde
\alpha_{-n}^\mu&=&\sqrt{\frac2{\alpha'}}\,\oint\frac{\dd\overline{z}}
{2\pi}~\overline{z}^{\,-n}~\partial_{\overline{z}}
x_{\rm R}^\mu(\overline{z}) \ ,
\label{Cauchyalphas}\eea
where the contour integrations encircle the origin $z=0$ of the
complex plane with counterclockwise orientation and are non-vanishing
for $n\geq0$. It follows that the states built from the
$\alpha^\mu_{-n}$'s are related to the residues of $\partial_z^nx_{\rm
  L}^\mu(z)$ at the origin, where $\partial_z^nx_{\rm L}^\mu(0)$
corresponds to an insertion of a pointlike operator at $z=0$. These
operators are called ``vertex operators''. Let us consider some
elementary examples of this operator-state correspondence.

\subsubsection*{Examples}

\noindent
${\underline{\rm (1)~Closed~String~Tachyon~Vertex~Operator:}}$
Geometrically, this is the spacetime Fourier transform of the operator
\beq
V(x)=\int\dd^2z~\delta\Bigl(x-x(z,\overline{z})\Bigr)
\label{Vxpin}\eeq
which pins a string at the spacetime point $x$. In other words, the
correspondence between tachyon ground states $|k;0,0\rangle=\e^{\ii\,
  k\cdot x_0}$ and vertex operators is given by
\beq
|k;0,0\rangle~\longleftrightarrow~\int\dd^2z~\NO\e^{\ii\, k\cdot
x(z,\overline{z})}\NO \ ,
\label{tachyoncorr}\eeq
where $\NO\cdot\NO$ denotes normal ordering of quantum operators. This
correspondence agrees with the anticipated behaviour under
translations $x^\mu\mapsto x^\mu+c^\mu$ by constant vectors $c^\mu$
generated by the target space momentum, under which the operator (and
state) pick up a phase factor $\e^{\ii\, k\cdot c}$. Note that the state here
is obtained by averaging over the absorption or emission point on the string
worldsheet, as it should be independent of the particular insertion point.

\noindent
${\underline{\rm (2)~Closed~String~Level-1~Vertex~Operator:}}$ The
emission or absorption of the gravitational fields $g_{\mu\nu}$,
$B_{\mu\nu}$ and $\Phi$ are described via the operator state
correspondence
\beq
\zeta_{\mu\nu}\,\alpha_{-1}^\mu\,\tilde\alpha_{-1}^\nu|k;0,0\rangle~
\longleftrightarrow~\int\dd^2z~\zeta_{\mu\nu}\,\NO\partial_zx^\mu\,
\partial_{\overline{z}}x^\nu~\e^{\ii\,k\cdot x}\NO \ .
\label{gravitycorr}\eeq

\noindent
${\underline{\rm (3)~Photon-Emission~Vertex~Operator:}}$ The vertex
operator corresponding to the open string photon state is given by the
correspondence
\beq
\zeta_\mu\,\alpha_{-1}^\mu|k;0\rangle~\longleftrightarrow~
\int\dd l~\zeta_\mu\,\NO\partial_\parallel x^\mu~\e^{\ii\,k\cdot x}\NO \ ,
\label{photoncorr}\eeq
where $l$ is the coordinate of the real line representing the boundary
of the upper half-plane (corresponding to the $\sigma=0,\pi$ boundary
lines of the worldsheet strip), and $\partial_\parallel$ denotes the
derivative in the direction tangential to the boundary. From this
correspondence, which follows from the doubling trick $z\sim\overline{z}$ of
section~\ref{StringEOM}, it is evident that the photon is associated with the
endpoints of the open string. This is a quantum version of the
classical property that the string endpoints move at the speed of
light, which can be easily deduced from examining the
worldsheet canonical momentum of the open string theory.

\subsection{String Perturbation Theory \label{PertTheory}}

\begin{figure}[htb]
\epsfxsize=4 in
\bigskip
\centerline{\epsffile{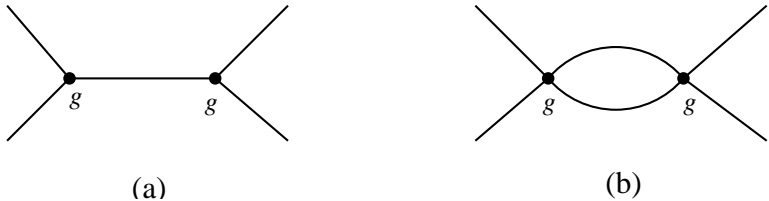}}
\caption{\baselineskip=12pt {\it Feynman graphs for four-particle} (a) {\it
tree-level and} (b) {\it one-loop scattering amplitudes in quantum field
theory. The lines denote propagators which correspond to the worldlines of
particles in spacetime. Each vertex has a coupling constant $g$ associated to
it and represents a singular worldline junction responsible for the ultraviolet
divergences in loop amplitudes.}}
\bigskip
\label{pointFeynman}\end{figure}

We will now heuristically describe the structure of the string
perturbation expansion, with emphasis on how it differs from that of
ordinary quantum field theory. In the latter case, perturbation theory
calculations are carried out by computing Feynman diagrams, which are
webs of worldlines of point particles. The particles interact at a
well-defined point in spacetime where straight lines, whose amplitudes
are given by their Feynman propagators, intersect at vertices
(fig.~\ref{pointFeynman}). A scattering amplitude is then calculated
by drawing the corresponding Feynman diagrams, and multiplying
together all the propagators and the coupling constants at each
vertex.

\begin{figure}[htb]
\epsfxsize=4 in
\bigskip
\centerline{\epsffile{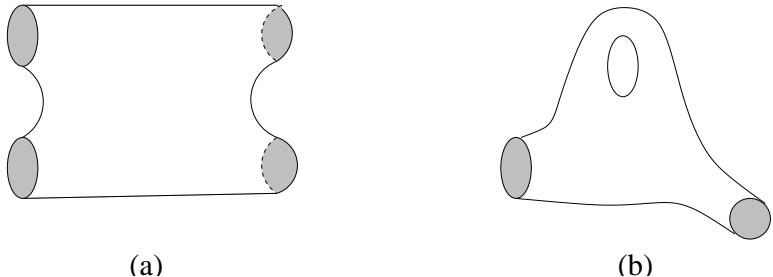}}
\caption{\baselineskip=12pt {\it Feynman diagrams in (closed) string theory
representing} (a) {\it tree-level four-string and} (b) {\it one-loop two-string
scattering. Loops are associated with handles on the string worldsheet. }}
\bigskip
\label{stringFeynman}\end{figure}

In string theory, the situation is similar, except that the Feynman diagrams
are {\it smooth} two-dimensional surfaces representing string worldsheets. The
propagation of a string is represented by a tube (fig.~\ref{stringFeynman}). It
turns out that after including the contributions from the infinite tower of
massive particles in the string spectrum, the non-renormalizable divergences in
quantum gravity loop amplitudes completely cancel each other out. The
particularly significant feature here is that string interactions are
``smoother'' than the interactions of point particles, because the worldsheets
are generically smooth. The ultraviolet divergences of quantum field theory,
which are rendered finite in loop amplitudes by strings, can be traced back to
the fact that its interactions are associated with worldline junctions at
specific spacetime points. But because the string worldsheet is smooth (with
{\it no} singular points), string theory scattering amplitudes have {\it no}
ultraviolet divergences. A profound consequence of this smoothness property is
that in string theory the structure of interactions is completely determined by
the {\it free} worldsheet field theory, and there are no arbitrary interactions
to be chosen. The interaction is a consequence of worldsheet topology (the
``handles''), rather than of local singularities.

To analyse more carefully these Feynman diagrams, it is convenient
again to Wick rotate to Euclidean signature so that the worldsheet
metric $\gamma_{ab}$ is positive definite. Then the various topologies
that arise in string perturbation theory are {\it completely}
understood. The classification of two-dimensional Euclidean surfaces
is a long-solved problem in mathematics. The schematic structure of a
generic string scattering amplitude is then given by a path integral
of the form
\beq
{\cal A}=\int\DD\gamma_{ab}(\xi)~\DD x^\mu(\xi)~\e^{-S[x,\gamma]}~
\prod_{i=1}^{n_c}~\int\limits_M\dd^2\xi_i~V_{\alpha_i}(\xi_i)~
\prod_{j=1}^{n_o}~\int\limits_{\partial M}\dd l_j~V_{\beta_j}(l_j) \ .
\label{stringscattschem}\eeq
Here $\gamma_{ab}$ is the metric on the string worldsheet $M$, $S[x,\gamma]$ is
the (ungauged) Polyakov action (\ref{Polaction}), $V_{\alpha_i}$ is the vertex
operator that describes the emission or absorption of a closed string state of
type $\alpha_i$ from the interior of $M$, and $V_{\beta_j}$ is the vertex
operator that describes the emission or absorption  of an open string state of
type $\beta_j$ from the boundary $\partial M$ of the string worldsheet $M$.

By using conformal invariance, the amplitudes $\cal A$ reduce to
integrals over conformally inequivalent worldsheets, which at a given
topology are described by a finite number of complex parameters called
``moduli''. In this representation the required momentum integrations
are already done. The amplitudes can be thereby recast as {\it finite}
dimensional integrals over the ``moduli space of $M$''. The finite
dimension ${\cal N}$ of this space is given by the number
\beq
{\cal N}=3(2h+b-2)+2n_c+n_o \ ,
\label{modulidim}\eeq
where $h$ is the number of ``handles'' of $M$, $b$ is its number of boundaries,
and $n_c$ and $n_o$ are respectively, as above, the number of closed and open
string state insertions on $M$. As described before, the string coupling $g_s$
is dynamically determined by the worldsheet dilaton field $\Phi$.

\subsection{Chan-Paton Factors \label{ChanPaton}}

\begin{figure}[htb]
\epsfxsize=3 in
\bigskip
\centerline{\epsffile{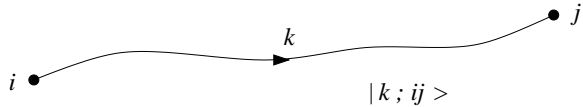}}
\caption{\baselineskip=12pt {\it The state $|k;ij\rangle$ obtained by attaching
quark labels $i,j=1,\dots,N$ to the endpoints of an open string. Only the
momentum label $k$ of the Fock states is indicated explicitly in the
notation.}}
\bigskip
\label{ChanPatonfig}\end{figure}

We will now describe how to promote the photon fields living at the
endpoints of open strings to {\it non-abelian} gauge field degrees of
freedom. For this, we attach non-dynamical ``quark'' (and
``anti-quark'') degrees of freedom\footnote{\baselineskip=12pt The
  ``quark'' terminology here is only historical, as the strings we are
  discussing here are now known not to be the long-sought QCD strings
  thought to be responsible for binding quarks together.} to the
endpoints of an open string in a way which preserves both spacetime Poincar\'e
invariance and worldsheet conformal invariance
(fig.~\ref{ChanPatonfig}). In addition to the usual Fock space labels
of a string state, we demand that each end of the string be in a state
$i$ or $j$. We further demand that the Hamiltonian of the states
$i=1,\dots,N$ is 0, so that they stay in the state that we originally
put them in for all time. In other words, these states correspond to
``background'' degrees of freedom. We may then decompose an open
string wavefunction $|k;ij\rangle$ in a basis $\lambda_{ij}^a$ of
$N\times N$ matrices as
\beq
|k;a\rangle=\sum_{i,j=1}^N|k;ij\rangle~\lambda^a_{ij} \ .
\label{ChanPatonstate}\eeq
These matrices are called ``Chan-Paton factors''~\cite{ChanPaton1}. By the
operator-state correspondence, all open string vertex operators also
carry such factors.

\begin{figure}[htb]
\epsfxsize=4 in
\bigskip
\centerline{\epsffile{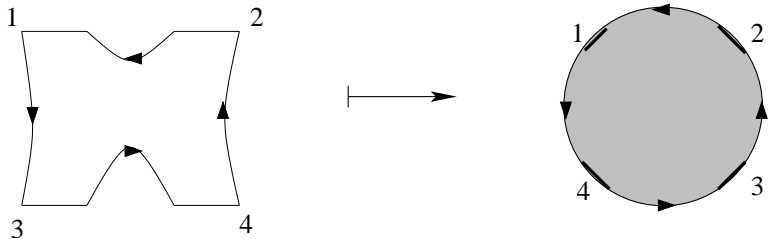}}
\caption{\baselineskip=12pt {\it A tree-level four-point open string scattering
amplitude with Chan-Paton factors. The right end of string 1 is in the same
state as the left end of string 2, and so on, because the Chan-Paton degrees of
freedom are non-dynamical. A conformal transformation can be used to map this
open string graph onto a disc diagram.}}
\bigskip
\label{4string}\end{figure}

All open string scattering amplitudes contain traces of products of Chan-Paton
factors. As an example, consider the amplitude depicted by the Feynman diagram
of fig.~\ref{4string}. Summing over all possible states involved in tying up
the ends produces the overall factor
\beq
\sum_{i,j,k,l}\lambda_{ij}^1\,\lambda_{jk}^2\,\lambda_{kl}^3\,\lambda_{li}^4
=\Tr\left(\lambda^1\,\lambda^2\,\lambda^3\,\lambda^4\right) \ .
\label{overallfactor}\eeq
The amplitude is therefore invariant under a {\it global} $U(N)$ worldsheet
symmetry
\beq
\lambda^a~\longmapsto~U\,\lambda^a\,U^{-1} \ , ~~ U\in U(N) \ ,
\label{lamdaUNsym}\eeq
under which the endpoint $i$ of the string transforms in the fundamental
$\mathbf{N}$ representation of the $N\times N$ unitary group $U(N)$, while
endpoint $j$, due to the orientation reversal between the two ends of the
string, transforms in the anti-fundamental representation
$\overline{\mathbf{N}}$. The corresponding massless vector vertex operator
\beq
V_{ij}^{a,\mu}=\int\dd l~\lambda_{ij}^a\,\NO\partial_\parallel x^\mu~
\e^{\ii\,k\cdot x}\NO
\label{ChanPatonvertex}\eeq
thereby transforms under the adjoint $\mathbf{N}\otimes\overline{\mathbf{N}}$
representation of $U(N)$,
\beq
V^{a,\mu}~\longmapsto~U\,V^{a,\mu}\,U^{-1} \ .
\label{vertexUNsym}\eeq
Thus the global $U(N)$ symmetry of the worldsheet field theory is promoted to a
local $U(N)$ {\it gauge} symmetry in spacetime, because we can make a different
$U(N)$ rotation at separate points $x^\mu(\tau,\sigma)$ in spacetime. It is in
this way that we can promote the photon fields at open string ends to {\it
non-abelian} gauge fields.

\setcounter{equation}{0}

\section{Superstrings \label{Superstrings}}

We will now generalize the bosonic string to include supersymmetry and
hence discuss superstrings. After some words of motivation for doing
this, we will simply repeat the analysis of the previous two sections
within this new setting, and the constructions will all be the same,
but with more rich features arising. After constructing the
superstring spectrum of physical states, we will discuss a consistent
truncation of it which removes the unphysical degrees of freedom
present in the bosonic string spectrum, while at the same time renders
the quantum superstring theory supersymmetric in target space. We
shall also work out in some detail a one-loop calculation which gives
an explicit example of the constructions, and also introduces some
important new concepts such as modular invariance.

\subsection{Motivation \label{Motivation}}

We will now add fermions to the bosonic string to produce quite
naturally supersymmetry, and hence a supersymmetric or spinning string
, called ``superstring'' for short. There are two main reasons why we
want to do this:
\begin{itemize}
\item Bosonic string theory is sick, because its spectrum of quantum
  states contains a tachyon, which signals an unstable vacuum.
\item Bosonic string theory contains no fermions in its quantum
  spectrum, and so it has no hope for phenomenological
  implications. As far as nature is concerned, bosonic strings cannot
  be the whole story.
\end{itemize}

There are two ways to introduce supersymmetry in string theory:
\begin{itemize}
\item{\it Ramond-Neveu-Schwarz (RNS) Formalism:} This is the original
  approach that arose within the context of dual resonance models
  around 1971~\cite{NS1,Ramond1}. It uses two-dimensional worldsheet
  supersymmetry, and it requires the ``Gliozzi-Scherck-Olive (GSO)
  projection''~\cite{GSO1} to eventually realize {\it spacetime}
  supersymmetry, a tachyon-free spectrum, and also modular invariance.
\item{\it Light-Cone Green-Schwarz Formalism:} In this approach, which
  emerged around 1981~\cite{GrSchw2}, spacetime supersymmetry is
  explicit from the onset, and it avoids having to use the GSO
  projection. However, unlike the RNS approach, this superstring
  theory cannot be quantized in a fully covariant way (as its name
  suggests, it relies heavily on light-cone gauge fixing of the
  spacetime coordinates).
\end{itemize}
Here we will only deal with the RNS formalism, which is technically
much simpler to deal with and which utilizes many of the techniques of
the previous two sections that we have now become familiar with.

\subsection{The RNS Superstring \label{RNS}}

We start with the gauge-fixed Polyakov action describing $d$ free,
massless scalar fields $x^\mu(\tau,\sigma)$. We now add $d$ free,
massless Majorana spinors $\psi^\mu(\tau,\sigma)$ on the string
worldsheet which transform as $d$-dimensional vectors under Lorentz
transformations in the target spacetime. Again, the worldsheet is the
cylinder $-\infty<\tau<\infty$, $0\leq\sigma<2\pi$ for the closed
string, and the strip $-\infty<\tau<\infty$, $0\leq\sigma<\pi$ for the
open string. The worldsheet action in the conformal gauge takes the
form~\cite{Poly2}
\beq
\begin{tabular}{|c|}\hline\\
$\displaystyle
S=-\frac T2\,\int\dd^2\xi~\Bigl(\partial_ax^\mu\,\partial^ax_\mu-\ii\,
\overline{\psi}^{\,\mu}\,\rho^a\,\partial_a\psi_\mu\Bigr) \ ,
$\\\\
\hline\end{tabular}
\label{SUSYPolaction}\eeq
where the second term in the action is the usual Dirac kinetic term
for the fermion fields.

Here $\rho^a$, $a=0,1$ are $2\times2$ Dirac matrices which in a
convenient basis for the worldsheet spinors can be taken to be
\beq
\rho^0=\pmatrix{0&-\ii\cr\ii&0\cr} \ , ~~
\rho^1=\pmatrix{0&\ii\cr\ii&0\cr} \ ,
\label{rhoexpl}\eeq
and they satisfy the worldsheet Dirac algebra
\beq
\left\{\rho^a\,,\,\rho^b\right\}=-2\,\eta^{ab} \ .
\label{2DDiracalg}\eeq
In this basis, the fermion field
\beq
\psi=\pmatrix{\psi_-\cr\psi_+\cr}
\label{psi2D}\eeq
is a two-component Majorana spinor, $\psi_\pm^*=\psi_\pm$, in order to
keep the action (\ref{SUSYPolaction}) real. Furthermore, the
two-dimensional Dirac term then becomes
\beq
\overline{\psi}\cdot\rho^a\,\partial_a\psi=
\psi_-\cdot\partial_+\psi_-+\psi_+\cdot\partial_-\psi_+ \ ,
\label{Diractermexpl}\eeq
where the derivatives $\partial_\pm$ are defined in
(\ref{lightconecoords}). The equations of motion for the fermion
fields are therefore given by the massless Dirac equation in two
dimensions,
\beq
\partial_+\psi_-^\mu=\partial_-\psi_+^\mu=0 \ .
\label{2DDiraceqn}\eeq
Thus, the Majorana-Weyl fermions $\psi_-^\mu$ describe right-movers
while $\psi_+^\mu$ describe left-movers. The equations of motion and
constraints for the $x^\mu$'s are the same as before. The
supersymmetry of the field theory defined by the action
(\ref{SUSYPolaction}) is left as an exercise.

\bigskip

\begin{center}
\begin{minipage}{15cm}

\small

{\bf Exercise~5.1.} {\sl\baselineskip=12pt {\bf (a)} Show that the
  gauge-fixed fermionic extension (\ref{SUSYPolaction}) of the
  Polyakov action is invariant under the global, infinitesimal
  worldsheet supersymmetry transformations
$$
\begin{array}{rll}
\delta_\epsilon x^\mu&=&\overline{\epsilon}\,\psi^\mu \ , \\
\delta_\epsilon\psi^\mu&=&-\ii\,\rho^a\,\partial_ax^\mu\,\epsilon \ ,
\end{array}
$$
with $\epsilon$ a constant, anticommuting two-component spinor. Since
this transformation acts on both left-moving and right-moving sectors,
the worldsheet field theory is said to have ``$(1,1)$
supersymmetry''~\cite{BUSSTEPPJMF}.

\noindent
{\bf (b)} Show that the conserved worldsheet Noether current $J_a$
associated with this symmetry is the fermionic supercurrent
$$
J_a=\frac12\,\rho^b\,\rho_a\,\psi^\mu\,\partial_bx_\mu \ .
$$

\noindent
{\bf (c)} Show that $\rho^a\,J_a=0$, so that some components of $J_a$
vanish.}

\end{minipage}
\end{center}

\bigskip

We can also easily work out the modification of the worldsheet
energy-momentum tensor. Using exercise~5.1~(c), we arrive altogether
at the mass-shell constraints
\beq
\begin{tabular}{|c|}\hline\\
$\begin{array}{rll}
T_{\pm\pm}&\equiv&\displaystyle\left(\partial_\pm x\right)^2+\frac\ii2\,
\psi_\pm\cdot\partial_\pm\psi_\pm~=~0 \ , \\J_\pm&\equiv&
\psi_\pm\cdot\partial_\pm x~=~0 \ .
\end{array}
$\\\\
\hline\end{tabular}
\label{SUSYmassshell}\eeq
The second constraint comes from the locally supersymmetric form of
the ungauged Polyakov action.

\subsubsection*{Mode Expansions}

The mode decompositions for $x^\mu(\tau,\sigma)$ are exactly the same
as before. We now need to consider boundary conditions for the free
fermionic fields $\psi^\mu(\tau,\sigma)$. Let us first consider the
case of open strings. Then the variational
principle for the Polyakov action (\ref{SUSYPolaction}) requires
\beq
\psi_+\cdot\delta\psi_+-\psi_-\cdot\delta\psi_-=0~~~~{\rm
  at}~~\sigma=0,\pi \ .
\label{fermbcs}\eeq
The Dirac equations of motion thereby admit two possible boundary
conditions consistent with Lorentz invariance, namely
$\psi_+=\pm\,\psi_-$ at $\sigma=0,\pi$ (so that also
$\delta\psi_+=\pm\,\delta\psi_-$ there). The overall relative sign
between the fields $\psi_-$ and $\psi_+$ is a matter of convention, so
by redefining $\psi_+$ if necessary, we may without loss of
generality take
\beq
\psi^\mu_+(\tau,0)=\psi_-^\mu(\tau,0) \ .
\label{psi0bc}\eeq
This still leaves two possibilities at the other endpoint
$\sigma=\pi$, which are called respectively Ramond (R) and
Neveu-Schwarz (NS) boundary conditions:
\beq
\begin{tabular}{|c|}\hline\\
$\begin{array}{rll}
\psi_+^\mu(\tau,\pi)&=&\psi_-^\mu(\tau,\pi)~~~~{\rm (R)} \ , \\
\psi_+^\mu(\tau,\pi)&=&-\psi^\mu_-(\tau,\pi)~~~~{\rm (NS)} \ .
\end{array}
$\\\\
\hline\end{tabular}
\label{RNSbcs}\eeq
We shall see later on that the R sector will give particles which are
spacetime fermions while the NS sector will yield spacetime bosons.
The mode decompositions of the Dirac equation (\ref{2DDiraceqn}) allow
us to express the general solutions as the Fourier series
\beq
\begin{tabular}{|c|}\hline\\
$\begin{array}{rll}
\psi_\pm(\tau,\sigma)&=&\displaystyle\frac1{\sqrt2}\,\sum_r
\psi_r^\mu~\e^{-\ii\,r(\tau\pm\sigma)} \ , \\r&=&0,\pm1,\pm2,
\dots~~~~{\rm (R)}\\&=&\displaystyle\pm\,\frac12,\pm\,\frac32,\dots
{}~~~~{\rm (NS)} \ ,
\end{array}
$\\\\
\hline\end{tabular}
\label{RNSmodeexpopen}\eeq
where the Majorana condition requires
\beq
\psi_{-r}^\mu=\left(\psi_r^\mu\right)^* \ .
\label{Majmode}\eeq
Note that only the R sector gives a zero mode $\psi_0^\mu$.

The closed string sector is analogous, but now we impose periodic or
anti-periodic boundary conditions on each component of $\psi^\mu$
separately to get the mode expansions
\beq
\begin{tabular}{|c|}\hline\\
$\displaystyle
\psi_+^\mu(\tau,\sigma)=\sum_r\tilde\psi_r^\mu~\e^{-2\,\ii\,r(\tau+\sigma)}
\ , ~~
\psi_-^\mu(\tau,\sigma)=\sum_r\psi_r^\mu~\e^{-2\,\ii\,r(\tau-\sigma)} \ ,
$\\\\
\hline\end{tabular}
\label{RNSmodeexpclosed}\eeq
with the mode index $r$ constrained according to
(\ref{RNSmodeexpopen}). Corresponding to the different pairings
between left-moving and right-moving modes, there are now four
distinct closed string sectors that can be grouped together according
to the spacetime boson-fermion parity of their quantum states:
\beq
\begin{array}{llrrr}
&\underline{\rm Bosons}&~~~~&\underline{\rm Fermions}&\\
&{\rm NS}-{\rm NS}&~~~~&{\rm NS}-{\rm R}&\\
&{\rm R}-{\rm R}&~~~~&{\rm R}-{\rm NS}&
\end{array}
\label{4NSRs}\eeq
Furthermore, the closed string mode expansions of the physical
constraints are given by
\bea
T_{--}(\xi^-)&=&\sum_{n=-\infty}^\infty L_n~\e^{-2\,\ii\,n\xi^-} \ ,
\nn\\J_-(\xi^-)&=&\sum_rG_r~\e^{-2\,\ii\,r\xi^-} \ ,
\label{TJclosedmodeexp}\eea
where
\beq
\begin{tabular}{|c|}\hline\\
$\begin{array}{rll}
L_n&=&\displaystyle\frac12\,\sum_{m=-\infty}^\infty\alpha_{n-m}\cdot
\alpha_m+\frac14\,\sum_r(2r-n)\,\psi_{n-r}\cdot\psi_r \ , \\
G_r&=&\displaystyle\sum_{m=-\infty}^\infty\alpha_m\cdot\psi_{r-m} \ ,
\end{array}
$\\\\
\hline\end{tabular}
\label{LnGrdefs}\eeq
and $\alpha_m^\mu$ are the bosonic oscillators for the mode expansion
of $x^\mu(\tau,\sigma)$. From the $+$ components we similarly get mode
operators $\tilde L_n$ and $\tilde G_r$.

\subsection{The Superstring Spectrum \label{SUSYSpectrum}}

As we did with the bosonic string, we can easily proceed to the
canonical quantization of the free two-dimensional field theory
(\ref{SUSYPolaction}). The fermionic modifications are straightforward
to obtain and are again left as an exercise.

\bigskip

\begin{center}
\begin{minipage}{15cm}

\small

{\bf Exercise~5.2.} {\sl\baselineskip=12pt {\bf (a)} Starting from the
  action (\ref{SUSYPolaction}), show that, in addition to the bosonic
  oscillator commutators given in exercise~4.1~(b), canonical
  quantization leads to the anti-commutators
$$
\left\{\psi_r^\mu\,,\,\psi_s^\nu\right\}=\delta_{r+s,0}\,\eta^{\mu\nu}
\ .
$$

\noindent
{\bf (b)} Show that the operators $L_n$ and $G_r$ generate the $N=1$
supersymmetric extension of the Virasoro algebra:
$$
\begin{array}{rll}
[L_n,L_m]&=&\displaystyle(n-m)\,L_{n+m}+\frac c{12}\,\Bigl(n^3-n\Bigr)
\,\delta_{n+m,0} \ ,
\\\displaystyle[L_n,G_r]&=&\displaystyle\frac12\,\Bigl(n-2r\Bigr)
\,G_{n+r} \ , \\\{G_r,G_s\}&=&\displaystyle2\,L_{r+s}+\frac c{12}\,
\Bigl(4r^2-1\Bigr)\,\delta_{r+s,0} \ ,
\end{array}
$$
where $c=d+\frac d2$ is the total contribution to the conformal
anomaly.}
\end{minipage}
\end{center}
\begin{center}
\begin{minipage}{15cm}
\small
\noindent
{\sl {\bf (c)} Show that the conserved angular momentum is given by
$J^{\mu\nu}=J_\alpha^{\mu\nu}+K^{\mu\nu}$, where $J_\alpha^{\mu\nu}$
is the contribution from the bosonic modes (exercise~4.2) and
$$
K^{\mu\nu}=-\ii\,\sum_{r\geq0}\left(\psi_{-r}^\mu\,\psi_r^\nu-\psi_{-r}^\nu
\,\psi_r^\mu\right) \ .
$$}

\end{minipage}
\end{center}

\bigskip

The canonical anticommutators obtained in exercise~5.2~(a) are just
the standard ones in the canonical quantization of free Fermi fields
$\psi^\mu(\tau,\sigma)$. The basic structure
$\{\psi_r^{\mu\,\dag},\psi_r^\mu\}=1$ is {\it very} simple, as it
describes a two-state system spanned by vectors $|0\rangle$ and
$|1\rangle$ with $\psi_r^\mu|0\rangle=0$,
$\psi_r^{\mu\,\dag}|0\rangle=|1\rangle$ for $r>0$. The $\psi_r^\mu$
with $r<0$ may then be regarded as raising operators, and as lowering
operators for $r>0$. The full state space is the free tensor product
of the bosonic and fermionic Hilbert spaces constructed in this way.

By using precisely the same techniques as for the bosonic string in
the previous section, we can again fix the spacetime dimension $d$ and
the normal ordering constant $a$. The ``superstring critical
dimension'' is now
\beq
\begin{tabular}{|c|}\hline\\
$\displaystyle
d=10 \ .
$\\\\
\hline\end{tabular}
\label{SUSYcritdim}\eeq
The bosonic oscillators $\alpha_n^\mu$ contribute, as before, a
regulated Casimir energy $-\frac{d-2}{24}$, while the fermionic modes
$\psi_r^\mu$ yield minus that value for integer $r$ and minus half
that value $\frac{d-2}{48}$ for half-integer $r$. The normal
ordering constant is thereby found to be
\beq
\begin{tabular}{|c|}\hline\\
$\displaystyle
a=\left\{\new{\begin{array}{l}0~~~~{\rm (R)}\\\frac12~~~~{\rm (NS)} \ .
\end{array}}\right.
$\\\\
\hline\end{tabular}
\label{SUSYCasimir}\eeq
The physical state conditions are given by the super-Virasoro
constraints
\bea
(L_0-a)|{\rm phys}\rangle&=&0 \ , \nn\\
L_n|{\rm phys}\rangle&=&0 \ , ~~ n>0 \ , \nn\\
G_r|{\rm phys}\rangle&=&0 \ , ~~ r>0 \ .
\label{SUSYphysstate}\eea
The $L_0=a$ constraint yields the open string mass formula
\beq
\begin{tabular}{|c|}\hline\\
$\displaystyle
m^2=\frac1{\alpha'}\,\Bigl(N-a\Bigr) \ ,
$\\\\
\hline\end{tabular}
\label{SUSYopenmass}\eeq
where the total level number is given by
\beq
N=\sum_{n=1}^\infty\alpha_{-n}\cdot\alpha_n+\sum_{r>0}r\,\psi_{-r}
\cdot\psi_r \ .
\label{SUSYlevel}\eeq
Aside from the change in values of $a$ in (\ref{SUSYCasimir}) and the
definition (\ref{SUSYlevel}), this is the same mass formula as obtained in the
previous section.

\subsubsection*{The Open Superstring Spectrum}

The open string spectrum of states has the two independent NS and R sectors,
which we will study individually.

\noindent
$\underline{\rm NS~Sector:}$ The NS ground state $|k;0\rangle^{~}_{\rm NS}$
satisfies
\bea
\alpha_n^\mu|k;0\rangle^{~}_{\rm NS}&=&\psi_r^\mu|k;0\rangle^{~}_{\rm NS}
{}~=~0 \ , ~~ n,r>0 \ , \nn\\\alpha_0^\mu|k;0\rangle^{~}_{\rm NS}&=&
\sqrt{2\alpha'}\,k^\mu|k;0\rangle^{~}_{\rm NS} \ .
\label{NSground}\eea
This sector of the Hilbert space of physical states of the RNS superstring is
then a straightforward generalization of the bosonic string spectrum. In
particular, the vacuum $|k;0\rangle^{~}_{\rm NS}$ has $m^2=-\frac1{2\alpha'}$
and is tachyonic. We'll soon see how to eliminate it from the physical
spectrum.

The first excited levels in this sector contain the massless states
$\psi^\mu_{-\frac12}|k;0\rangle^{~}_{\rm NS}$, $m^2=0$. They are vectors of the
transverse $SO(8)$ rotation group which corresponds to the Little group of
$SO(1,9)$ that leaves the light-cone momentum invariant. They describe the
eight physical polarizations of the massless, open string photon field
$A_\mu(x)$ in ten spacetime dimensions. All states in the NS sector are
spacetime bosons, because they transform in appropriate irreducible
representations of $SO(8)$ in the decompositions of tensor products of the
vectorial one.

\noindent
$\underline{\rm R~Sector:}$ In the Ramond sector there are zero modes
$\psi_0^\mu$ which satisfy the ten dimensional Dirac algebra
\beq
\left\{\psi_0^\mu\,,\,\psi_0^\nu\right\}=\eta^{\mu\nu} \ .
\label{10DDiracalg}\eeq
Thus the $\psi_0$'s should be regarded as Dirac matrices,
$\psi_0^\mu=\frac1{\sqrt2}\,\Gamma^\mu$, and in particular they are finite
dimensional operators. All states in the R sector should be ten dimensional
spinors in order to furnish representation spaces on which these operators can
act. We conclude that all states in the R sector are spacetime fermions.

The zero mode part of the fermionic constraint $J_-=0$ in (\ref{SUSYmassshell})
gives a wave equation for fermionic strings in the R sector known as the
``Dirac-Ramond equation''
\beq
G_0|{\rm phys}\rangle^{~}_{\rm R}=0 \ ,
\label{DiracRamondeq}\eeq
where
\beq
G_0=\alpha_0\cdot\psi_0+\sum_{n\neq0}\alpha_{-n}\cdot\psi_n
\label{G0expl}\eeq
obeys (exercise~5.2~(b))
\beq
G_0^2=L_0 \ .
\label{G0L0rel}\eeq
Note that the zero mode piece $\alpha_0\cdot\psi_0$ in (\ref{G0expl}) is
precisely the spacetime Dirac operator
$\partial\!\!\!/\,=\Gamma^\mu\,\partial_\mu$ in momentum space, because
$\alpha_0^\mu\propto p_0^\mu$ and $\psi_0^\mu\propto\Gamma^\mu$. Thus the R
sector fermionic ground state $|\psi^{(0)}\rangle^{~}_{\rm R}$, defined by
\beq
\alpha_n^\mu|\psi^{(0)}\rangle^{~}_{\rm R}=
\psi_n^\mu|\psi^{(0)}\rangle^{~}_{\rm R}=0 \ , ~~ n>0 \ ,
\label{Rgrounddef}\eeq
satisfies the massless Dirac wave equation
\beq
\alpha_0\cdot\psi_0|\psi^{(0)}\rangle^{~}_{\rm R}=0 \ .
\label{RgroundDiraceq}\eeq
It follows that the fermionic ground state of the superstring is a massless
Dirac spinor in ten dimensions. However, at present, it has too many components
to form a supersymmetric multiplet with the bosonic ground state of the NS
sector. In ten spacetime dimensions, it is known that one can impose both the
Majorana and Weyl conditions simultaneously on spinors, so that the ground
state $|\psi^{(0)}\rangle^{~}_{\rm R}$ can be chosen to have a definite
spacetime chirality. We will denote the two possible chiral ground states into
which it decomposes by $|a\rangle^{~}_{\rm R}$ and
$|\overline{a}\rangle^{~}_{\rm R}$, where $a,\overline{a}=1,\dots,8$ are spinor
indices labelling the two inequivalent, irreducible Majorana-Weyl spinor
representations of $SO(8)$.

\subsubsection*{The Closed Superstring Spectrum}

The spectrum of closed strings is obtained by taking tensor products of
left-movers and right-movers, each of which is very similar to the open
superstring spectrum obtained above, and by again using appropriate
level-matching conditions. Then, as mentioned before, there are four distinct
sectors (see~(\ref{4NSRs})). In the NS--NS sector, the lowest lying level is
again a closed string tachyon. The totality of {\it massless} states, which are
picked up in the field theory limit $\alpha'\to0$, can be neatly summarized
according to their transformation properties under $SO(8)$ and are given in
table~\ref{closedSUSYstates}. In this way the closed superstring sector endows
us now with spacetime ``supergravity fields''.

\begin{table}[htb]
\bigskip
\begin{center}
\begin{tabular}{|c|c|c|c|}\hline
{\sc Sector}&{\sc Boson/Fermion?}&$SO(8)$ {\sc Representation}&
{\sc Massless Fields}\\\hline\hline
NS--NS&boson&$\mathbf{8}_v\otimes\mathbf{8}_v=\mathbf{35}\oplus
\mathbf{28}\oplus\mathbf{1}$&$g_{\mu\nu}~,~B_{\mu\nu}~,~\Phi$\\\hline
NS--R&fermion&$\mathbf{8}_v\otimes\mathbf{8}_s=\mathbf{8}_s\oplus
\mathbf{56}_s$&$\Psi_\mu~,~\lambda$\\\hline R--NS&fermion&$\mathbf{8}_s
\otimes\mathbf{8}_v=\mathbf{8}_s\oplus\mathbf{56}_s$&$\Psi_\mu'~,~
\lambda'$\\\hline R--R&boson&$\mathbf{8}_s\otimes\mathbf{8}_s=p-{\rm forms}$&
Ramond-Ramond fields\\
\hline\end{tabular}
\end{center}
\caption{\baselineskip=12pt {\it The massless states of the closed superstring.
The fields in the last column are identified according to the irreducible
$SO(8)$ representations of the third column. The subscript $v$ denotes vector
and the subscript $s$ spinor representation. $\Psi_\mu$ and $\Psi_\mu'$ are
spin~$\frac32$ gravitino fields, while $\lambda$ and $\lambda'$ are
spin~$\frac12$ dilatino fields. These fermionic states all have the same
helicity. The Ramond-Ramond sector will be described in more detail in the next
section.}}
\bigskip
\label{closedSUSYstates}\end{table}

\subsection{The GSO Projection \label{GSO}}

The superstring spectrum admits a consistent truncation, called the ``GSO
projection'', which is necessary for consistency of the interacting theory. For
example, it will remedy the situation with the unwanted NS sector tachyon which
has $m^2=-\frac1{2\alpha'}$. Again we will examine this operation separately in
the NS and R sectors of the worldsheet theory.

\noindent
$\underline{\rm NS~Sector:}$ The GSO projection $P^{~}_{\rm GSO}$ here is
defined by keeping states with an odd number of $\psi$ oscillator excitations,
and removing those with an even number. This is tantamount to replacing
physical states according to
\beq
|{\rm phys}\rangle^{~}_{\rm NS}~\longmapsto~P^{~}_{\rm GSO}
|{\rm phys}\rangle^{~}_{\rm NS}
\label{physNSreplace}\eeq
with
\beq
P^{~}_{\rm GSO}=\frac12\,\Bigl(1-(-1)^F\Bigr) \ ,
\label{PGSONS}\eeq
where
\beq
F=\sum_{r>0}\psi_{-r}\cdot\psi_r
\label{fermionnumber}\eeq
is the ``fermion number operator'' which obeys
\beq
\Bigl\{(-1)^F\,,\,\psi^\mu\Bigr\}=0 \ .
\label{Kleinpsirel}\eeq
Thus only half-integer values of the level number (\ref{SUSYlevel}) are
possible, and so the spectrum of allowed physical masses are integral multiples
of $\frac1{\alpha'}$, $m^2=0,\frac1{\alpha'},\frac2{\alpha'},\dots$. In
particular, the bosonic ground state is now massless, and the spectrum no
longer contains a tachyon (which has fermion number $F=0$).

\noindent
$\underline{\rm R~Sector:}$ Here we use the same formula (\ref{PGSONS}), but
now we define the Klein operator $(-1)^F$ by
\beq
(-1)^F=\pm\,\Gamma^{11}\cdot(-1)^{\sum_{r\geq1}\psi_{-r}\cdot\psi_r} \ ,
\label{KleinR}\eeq
where
\beq
\Gamma^{11}=\Gamma^0\,\Gamma^1\cdots\Gamma^9
\label{10Dchirality}\eeq
is the ten dimensional chirality operator with
\beq
\left(\Gamma^{11}\right)^2=1 \ , ~~ \left\{\Gamma^\mu\,,\,
\Gamma^{11}\right\}=0 \ .
\label{Gamma11rels}\eeq
The operators $1\pm\,\Gamma^{11}$ project onto spinors of opposite spacetime
chirality. As we will see below, this chirality projection will guarantee
spacetime supersymmetry of the physical superstring spectrum. Although here the
choice of sign in (\ref{KleinR}), corresponding to different chirality
projections on the spinors, is merely a matter of convention, we will see in
the next section that it significantly affects the physical properties of the
{\it closed} string sector.

\subsubsection*{Spacetime Supersymmetry}

Let us now examine the massless spectrum of the GSO-projected
superstring theory. The ground state (NS) boson is described by the
state $\zeta\cdot\psi_{-\frac12}|k;0\rangle^{~}_{\rm NS}$. As before,
it is a massless vector and has $d-2=8$ physical polarizations. The
ground state (R) fermion is described by the state $P^{~}_{\rm
  GSO}|\psi^{(0)}\rangle^{~}_{\rm R}$. It is a massless Majorana-Weyl
spinor which has $\frac14\cdot2^{d/2}=8$ physical polarizations. Thus
the ground states have an equal number of bosons and fermions, as
required for supersymmetry. In particular, they form the
$\mathbf{8}_v\oplus\mathbf{8}_s$ vector supermultiplet of the $d=10$,
$N=1$ supersymmetry algebra. In fact, this state yields the pair of
fields that forms the vector supermultiplet of supersymmetric
Yang-Mills theory in ten spacetime dimensions, the extension to $U(N)$
gauge symmetry with Chan-Paton factors being straightforward. It can
also be shown that the entire interacting superstring theory (beyond
the massless states) now has spacetime
supersymmetry~\cite{GrSchw3}. Thus in addition to removing the
tachyonic instabilities of the vacuum, the GSO projection has
naturally provided us with a target space supersymmetric theory as an
additional bonus. We will see another remarkable consequence of this
projection in the next subsection.

\subsection{Example: One-Loop Vacuum Amplitude \label{1Loop}}

We will now go through an explicit calculation of a loop diagram in superstring
theory, as an illustration of some of the general features of quantum
superstring theory that we have developed thus far. This example will also
introduce a very important property of superstring perturbation theory that is
absolutely essential to the overall consistency of the model. We will need some
identities involving traces over the infinite families of oscillators that are
present in the quantum string mode expansions, which are left as an exercise
that can be straightforwardly done by carefully tabulating states of the
oscillator algebras.

\bigskip

\begin{center}
\begin{minipage}{15cm}

\small

{\bf Exercise~5.3.} {\sl\baselineskip=12pt {\bf (a)} Given a set of
one-dimensional bosonic oscillators $\alpha_n$, show that for any complex
number $q\neq1$,
$$
\tr\left(\,q^{\sum_{n=1}^\infty\alpha_{-n}\,\alpha_n}\right)=
\prod_{n=1}^\infty\frac1{1-q^n} \ .
$$

\noindent{\bf (b)} Given a set of one-dimensional fermionic oscillators
$\psi_n$, show that
$$
\tr\left(\,q^{\sum_{n=1}^\infty\psi_{-n}\,\psi_n}\right)=
\prod_{n=1}^\infty\left(1+q^n\right) \ .
$$}

\end{minipage}
\end{center}

\bigskip

\begin{figure}[htb]
\epsfxsize=3 in
\bigskip
\centerline{\epsffile{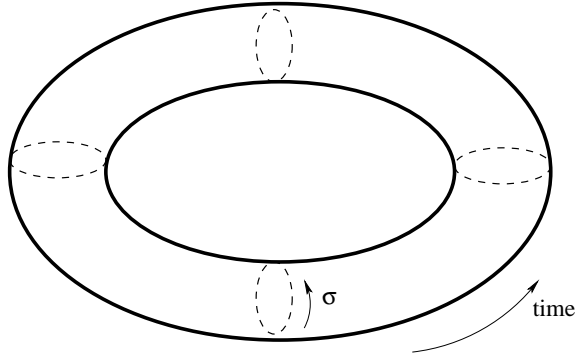}}
\caption{\baselineskip=12pt {\it The one-loop closed string vacuum diagram is a
two dimensional torus.}}
\bigskip
\label{2Dtorus}\end{figure}

\begin{figure}[htb]
\epsfxsize=3 in
\bigskip
\centerline{\epsffile{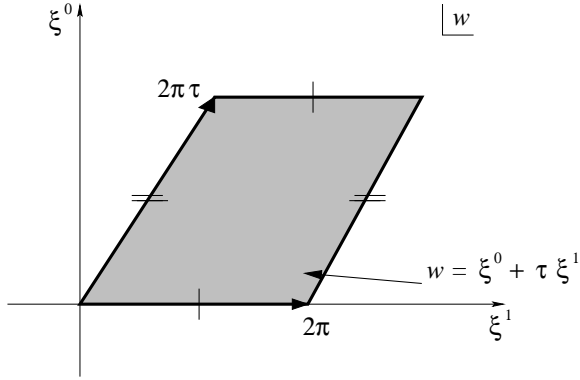}}
\caption{\baselineskip=12pt {\it The parallelogram representation of the torus.
Opposite edges of the parallelogram are periodically identified, i.e.
$0\leq\xi^0<2\pi$, $0\leq\xi^1<2\pi$. Equivalently, it can be viewed as the
region of the complex $w$-plane obtained from the identifications $w\sim w+2\pi
n$ and $w\sim w+2\pi m\tau$ for all integers $n$ and $m$.}}
\bigskip
\label{parallelogram}\end{figure}

We will now compute the one-loop closed string vacuum diagram. At
one-loop order, the Feynman graph describing a closed string state
which propagates in time $\xi^0$ and returns back to its initial state
is a donut-shaped surface, i.e. a two-dimensional torus
(fig.~\ref{2Dtorus}). The amplitude is given by the Polyakov path
integral (\ref{stringscattschem}), which, by conformal invariance,
sums over only conformally inequivalent tori. Let us first try to
understand this point geometrically. We can specify the torus by
giving a flat metric, which we will take to be $\eta_{ab}=\delta_{ab}$
in fixing the conformal gauge, along with a complex structure
$\tau\in\complex$, with ${\rm Im}(\tau)>0$. The complex number $\tau$
specifies the shape of the torus, which cannot be changed by a
conformal transformation of the metric (nor any local change of
coordinates). This is best illustrated by the ``parallelogram
representation'' of the torus in fig.~\ref{parallelogram}.

\begin{figure}[htb]
\epsfxsize=3 in
\bigskip
\centerline{\epsffile{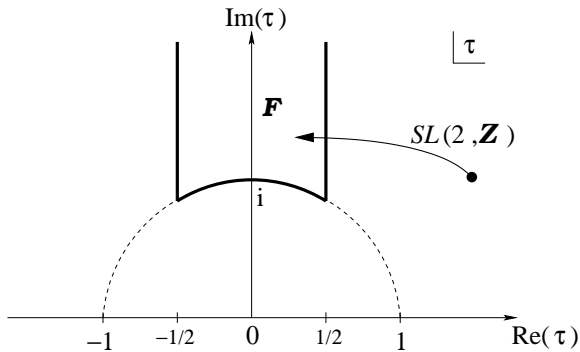}}
\caption{\baselineskip=12pt {\it A fundamental modular domain $\cal F$ for the
torus. The interior of this region is invariant under the basis transformations
(\ref{STdef}). Any point outside of $\cal F$ can be mapped into it by an
$SL(2,\zed)$ transformation (\ref{SL2Ztransfs}).}}
\bigskip
\label{modulardomain}\end{figure}

Lots of different $\tau$'s define the same torus. For example, from the
equivalence relation description of fig.~\ref{parallelogram} it is clear that
$\tau$ and $\tau+1$ give the same torus. The full family of equivalent tori can
be reached from any $\tau$ by ``modular transformations'' which are
combinations of the operations
\beq
T\,:\,\tau~\longmapsto\tau+1 \ , ~~ S\,:\,\tau~\longmapsto~-\frac1\tau \ .
\label{STdef}\eeq
The transformations $S$ and $T$ obey the relations
\beq
S^2=(ST)^3=1
\label{STrels}\eeq
and they generate the ``modular group'' $SL(2,\zed)$ of the torus acting on
$\tau$ by the discrete linear fractional transformations
\beq
\tau~\longmapsto~\frac{a\tau+b}{c\tau+d} \ , ~~ \pmatrix{a&b\cr c&d\cr}
\in SL(2,\zed) \ ,
\label{SL2Ztransfs}\eeq
where the $SL(2,\zed)$ condition restricts $a,b,c,d\in\zed$ with $ad-bc=1$. We
conclude that the ``sum'' over $\tau$ should be restricted to a ``fundamental
modular domain'' $\cal F$ in the upper complex half-plane. Any point outside
$\cal F$ can be mapped into it by a modular transformation. A convenient choice
of fundamental region is depicted in fig.~\ref{modulardomain}.

We are now ready to evaluate explicitly the vacuum amplitude. We will begin
with the bosonic string. For this, let us consider string propagation on the
torus as depicted in fig.~\ref{parallelogram}. A fixed point on the string,
which we assume is lying horizontally in the complex $w$-plane, propagates
upwards for a time $\xi^0=2\pi\,{\rm Im}(\tau)=2\pi\tau_2$, at which point it
also shifts to the right by an amount $\xi^1=2\pi\,{\rm Re}(\tau)=2\pi\tau_1$.
The time translation is affected by the worldsheet Hamiltonian $H=L_0+\tilde
L_0-2$, while the shift along the string is affected by the worldsheet momentum
$P=L_0-\tilde L_0$. So the bosonic vacuum path integral is given by
\beq
Z_{\rm bos}\equiv\Tr\left(\e^{-2\pi\tau_2H}~\e^{2\pi\,\ii\,\tau_1P}
\right)=\Tr\left(\,q^{L_0-1}\,\overline{q}^{\,\tilde L_0-1}\right) \ ,
\label{Zbosdef}\eeq
where
\beq
q\equiv\e^{2\pi\,\ii\,\tau}
\label{qtaudef}\eeq
and Tr denotes the trace taken by summing over all discrete oscillator states,
and by integrating over the continuous zero-mode momenta $k^\mu$ and
inequivalent $\tau$, along with the appropriate physical state restrictions.
Substituting in the mode expansions (\ref{Lndef},\ref{alpha0def}) for $L_0$ and
$\tilde L_0$ gives
\beq
Z_{\rm bos}=\int\limits_{\cal F}\dd^2\tau~\frac1{q\overline{q}}\,
\int\dd^{24}k~\e^{-\pi\tau_2k^2/2}~\tr\left(\,q^N\,\overline{q}^{\,
\tilde N}\right) \ ,
\label{Zbossub}\eeq
where tr denotes the trace over just the oscillator states, and $N$ and $\tilde
N$ are the number operators defined by (\ref{levelnumber}). Evaluating the
Gaussian momentum integrals and using exercise~5.3~(a) we arrive finally at
\bea
Z_{\rm bos}&=&\int\limits_{\cal F}\dd^2\tau~\frac1{\tau_2^{12}}\,
\frac1{q\overline{q}}\,\left|\,\prod_{n=1}^\infty\left(1-q^n\right)^{-24}
\,\right|^2\nn\\&=&\int\limits_{\cal F}\dd^2\tau~\frac1{\tau_2^{12}}\,
\Bigl|\eta(\tau)\Bigr|^{-48} \ ,
\label{Zbosfinal}\eea
where we have introduced the ``Dedekind function''~\cite{Mumford1}
\beq
\eta(\tau)=q^{1/24}\,\prod_{n=1}^\infty\left(1-q^n\right)
\label{Dedekindfn}\eeq
with the modular transformation properties
\beq
\eta(\tau+1)=\eta(\tau) \ , ~~ \eta\left(-\frac1\tau\right)=\sqrt{-\ii\,\tau}~
\eta(\tau) \ .
\label{Dedekindmodular}\eeq

\bigskip

\begin{center}
\begin{minipage}{15cm}

\small

{\bf Exercise~5.4.} {\sl\baselineskip=12pt Show that the integrand of the
partition function (\ref{Zbosfinal}) is ``modular invariant'', i.e. it is
unaffected by $SL(2,\zedm)$ transformations of $\tau$.}

\end{minipage}
\end{center}

\bigskip

The modular invariance property of exercise~5.4 is crucial for the overall
consistency of the quantum string theory, because it means that we are
correctly integrating over {\it all} inequivalent tori, and also that
we are counting each such torus only once. However, the modular
integral (\ref{Zbosfinal}) over $\cal F$ actually {\it diverges},
which is a consequence of the tachyonic instability of the bosonic
string theory. We shall now examine how this is remedied by the RNS
superstring. For this, we will first interpret the Ramond and
Neveu-Schwarz sectors of the theory geometrically.

\subsubsection*{Fermionic Spin Structures}

Fermions on the torus are specified by a choice of ``spin structure'',
which is simply a choice of periodic or anti-periodic boundary
conditions as $\xi^0\mapsto\xi^0+2\pi$ and $\xi^1\mapsto\xi^1+2\pi$
(see fig.~\ref{parallelogram}). There are four possible spin
structures in all, which we will denote symbolically by
\beq
\spin{\xi^0}{\xi^1}~=~\spin++~~,~~\spin+-~~,~~\spin-+~~,~~\spin-- \ ,
\label{4spinstructures}\eeq
where the squares denote the result of performing the functional
integral over fermions with the given fixed spin structure,
specified by the labelled sign change $\pm$ of the fermion fields
around the given $\xi^a$ cycle of the torus. For the moment we focus
our attention only on the right-moving sector. Looking back at the
mode expansions (\ref{RNSmodeexpopen}), we see that the NS (resp. R)
sector corresponds to anti-periodic (resp. periodic) boundary
conditions along the string in the $\xi^1$ direction. In order to
implement periodicity in the time direction $\xi^0$, we insert the
Klein operator $(-1)^F$ in the traces defining the partition
function. Because of the anticommutation property (\ref{Kleinpsirel}),
its insertion in the trace flips the boundary condition around
$\xi^0$, and without it the boundary condition is
anti-periodic, in accordance with the standard path integral
formulation of finite temperature quantum field
theory.\footnote{\baselineskip=12pt In the path integral approach to
  quantum statistical mechanics, the Euclidean ``time'' direction is taken to
  lie along a circle whose circumference is proportional to the
  inverse temperature of the system. It is known that free fermions
  must then be anti-periodic around the time direction in order to
  reproduce the correct Fermi-Dirac distribution.} The four fermionic spin
structure contributions to the path integral are thereby given as

\vbox{\bea
\spin-+&\equiv&\Tr^{~}_{\rm R}\Bigl(\e^{-2\pi\tau_2H}\Bigr) \ , \label{-+}\\
\spin++&=&\Tr^{~}_{\rm R}\Bigl((-1)^F~\e^{-2\pi\tau_2H}\Bigr) \ , \label{++}\\
\spin--&\equiv&\Tr^{~}_{\rm NS}\Bigl(\e^{-2\pi\tau_2H}\Bigr) \ , \label{--}\\
\spin+-&=&\Tr^{~}_{\rm NS}\Bigl((-1)^F~\e^{-2\pi\tau_2H}\Bigr) \ ,
\label{+-}\eea}
\noindent
where the subscripts on the traces indicate in which sector of the
fermionic Hilbert space they are taken over.

\begin{figure}[htb]
\epsfxsize=4 in
\bigskip
\centerline{\epsffile{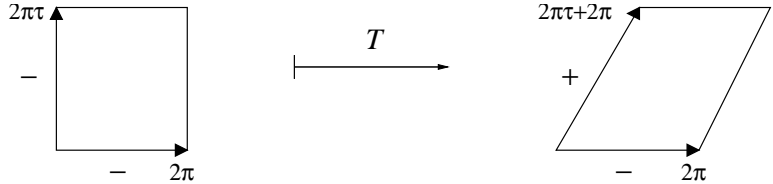}}
\caption{\baselineskip=12pt {\it The modular transformation
    $T:\tau\mapsto\tau+1$ induces an additional periodic shift from
    $\xi^1$ along the worldsheet time direction. It therefore flips the $\xi^0$
    boundary condition whenever the $\xi^1$ boundary condition is
    anti-periodic.}}
\bigskip
\label{spinmodular}\end{figure}

Let us now consider the modular properties of the fermionic
sector. The action (\ref{SL2Ztransfs}) of the modular group
$SL(2,\zed)$ in general changes the fermionic boundary conditions
(fig.~\ref{spinmodular}). It can be shown that the basis modular
transformations (\ref{STdef}) mix the various spin structures
(\ref{4spinstructures}) according to
\bea
T\,:\,&&\spin++~\longmapsto~\spin++~~,~~\spin-+~\longmapsto~\spin-+~~,~~
\nn\\&&\spin+-~\longmapsto~\spin--~~,~~\spin--~\longmapsto~\spin+- \ ,
\nn\\{~~}^{{~~}^{~~}}_{{~~}_{~~}}\nn\\
S\,:\,&&\spin++~\longmapsto~\spin++~~,~~\spin-+~\longmapsto~\spin+-~~,~~
\nn\\&&\spin+-~\longmapsto~\spin-+~~,~~\spin--~\longmapsto~\spin-- \ .
\label{STspinmix}\eea
Note that the $(+,+)$ spin structure is modular invariant. In fact,
the corresponding amplitude (\ref{++}) vanishes, because it contains a
Grassmann integral over the constant fermionic zero-modes but the
Hamiltonian $L_0$ is independent of these modes (see
(\ref{LnGrdefs})). This is the only amplitude which contains such
fermionic zero-modes. For the remaining spin structures, it follows
from (\ref{STspinmix}) that their unique modular
invariant combination is given up to an overall constant by
\beq
\spin--~~-~~\spin+-~~-~~\spin-+ \ .
\label{modinvspincomb}\eeq
The one-loop, modular invariant partition function for the
right-moving fermionic contributions is therefore given by this sum,
which from (\ref{-+})--(\ref{+-}) reads explicitly
\beq
Z_{\rm RNS}=\frac12\,\Tr^{~}_{\rm NS}\left[\Bigl(1-(-1)^F\Bigr)\,
q^{L_0-\frac12}\right]-\frac12\,\Tr^{~}_{\rm R}\left(\,q^{L_0}\right) \
{}.
\label{ZRNS}\eeq
In the first term we recognize the GSO projection operator
(\ref{PGSONS}), and so using the vanishing of (\ref{++}) we may write
the amplitude (\ref{ZRNS}) succinctly as a trace over the full
right-moving fermionic Hilbert space as
\beq
Z_{\rm RNS}=\Tr^{~}_{{\rm NS}\oplus{\rm R}}\left(P^{~}_{\rm GSO}\,
q^{L_0-a}\right) \ .
\label{ZRNSGSO}\eeq
We have thereby arrived at a beautiful interpretation of
the GSO projection, which ensures vacuum stability and spacetime
supersymmetry of the quantum string theory. Geometrically, it is simply the
{\it modular invariant} sum over spin structures. This interpretation also
generalizes to higher-loop amplitudes.

The total superstring amplitude is given by the product of $Z_{\rm RNS}$ with
the corresponding bosonic contribution $q^{1/3}\,\eta(\tau)^{-8}$ in $d=10$
spacetime dimensions (the power of 8 comes from the $d-2$ transverse
oscillators), along with their left-moving counterparts. By means of
exercise~5.3~(b) it is possible to evaluate explicitly the non-vanishing traces
in (\ref{-+})--(\ref{+-}) and we find
\bea
\spin-+&=&16\,q^{2/3}\,\prod_{n=1}^\infty\left(1+q^{2n}\right)^8 \ ,
  \nn\\\spin+-&=&q^{-1/3}\,\prod_{n=1}^\infty\left(1-q^{2n-1}\right)^8
  \ , \nn\\\spin--&=&q^{-1/3}\,\prod_{n=1}^\infty\left(1+q^{2n-1}
\right)^8 \ .
\label{spinamplexpl}\eea
The remarkable feature of the modular invariant combination
(\ref{modinvspincomb}) is that it {\it vanishes} identically, as a
consequence of the ``Jacobi abstruse identity''~\cite{Mumford1}
\beq
q^{-1/2}\,\left[\,\prod_{n=1}^\infty\left(1+q^{n-1/2}\right)^8-
\prod_{n=1}^\infty\left(1-q^{n-1/2}\right)^8\right]=16\,\prod_{n=1}^\infty
\left(1+q^n\right)^8 \ .
\label{Jacobiabstruse}\eeq
This remarkable, non-trivial mathematical identity implies that the
full superstring vacuum amplitude vanishes, which provides {\it very}
strong evidence in favour of spacetime supersymmetry in string theory
in ten dimensions. It simply reflects the fact that the spacetime NS sector
bosons and R sector fermions contribute the same way in equal numbers
(but with opposite signs due to
statistics).\footnote{\baselineskip=12pt When Jacobi discovered his
  formula (\ref{Jacobiabstruse}) in 1829, not realizing any of its immediate
  implications he refered to it as ``a very obscure formula''. Some
  150 years later string theory provided a natural explanation of it,
  namely that it is the requirement of spacetime supersymmetry at
  one-loop order in string perturbation theory.} This is of course by
no means a complete proof of spacetime supersymmetry to all orders of
string perturbation theory. For that, one needs to use the
Green-Schwarz formalism~\cite{GrSchw3}, which will not be dealt with
here.

\setcounter{equation}{0}

\section{Ramond-Ramond Charges and T-Duality \label{RRCharge}}

We will now start moving our attention away from perturbative string theory and
begin working our way towards describing D-branes. We shall begin by finishing
off our analysis of the perturbative superstring spectrum by describing in some
detail the R--R sector, which was mostly ignored in the previous section. This
will lead us to define the notion of Ramond-Ramond charge which will
immediately imply the existence of D-branes in superstring theory. To construct
them explicitly, we shall describe in detail the phenomenon of T-duality, which
is a purely stringy effect whose implications on the spacetime structure of
string theory is interesting in its own right. This duality symmetry, like
S-duality in gauge theories, is an indispensible tool in string theory. It also
yields an example of a stringy property which does not have a counterpart in
quantum field theory. We will describe individually the T-duality symmetries of
closed strings, open strings, and of superstrings.

\subsection{Ramond-Ramond Charges \label{Ramond}}

In this section we will for the most part be concerned with introducing some
deep reasons and motivation for the appearence of D-branes in superstring
theory. Our first step will be to return to the R--R sector of the closed RNS
superstring spectrum of the previous section, which we will now describe in
detail. Recall that for {\it open} strings, there are two possible Ramond
sector GSO projections given by the operators
\beq
P_{\rm GSO}^\pm=\frac12\,\Bigl(1\pm\,\Gamma^{11}\cdot(-1)^{\sum_{r\geq1}
\psi_{-n}\cdot\psi_n}\Bigr) \ .
\label{open2GSO}\eeq
The sign choice in (\ref{open2GSO}) selects a Ramond ground state spinor
$|\psi\rangle^{~}_{\rm R}$ (as well as massive fermions in the spectrum) of a
given $\pm$ spacetime chirality:
\beq
\Gamma^{11}|\psi\rangle^{~}_{\rm R}=\pm\,|\psi\rangle^{~}_{\rm R} \ .
\label{Gamma11pmpsi}\eeq
The particular chirality is simply a matter of taste, as either choice leads to
physically equivalent results.  Thus either $P_{\rm GSO}^+$ or $P_{\rm GSO}^-$
is a good GSO projection in the open string sector.

However, the distinction between $P_{\rm GSO}^\pm$ is meaningful in
the {\it closed} string sector, when left-movers and right-movers are
combined together with particular choices of chiralities. The GSO
projection is performed separately in both sectors, in each of which
there is a two-fold ambiguity. Altogether there are four possible
choices. Of these, we can identify two of them, by flipping the sign
convention of the total mod~2 fermion number operator $(-1)^{F_{\rm
    L}+F_{\rm R}}$ if necessary. Then, depending on the choice of
relative sign in defining the Klein operators $(-1)^{F_{\rm L}}$ and
$(-1)^{F_{\rm R}}$, there are two inequivalent possibilities
corresponding to the relative chirality between the surviving R-sector
Majorana-Weyl spinors $|\psi_{\rm l}\rangle^{~}_{\rm R}$ and
$|\psi_{\rm r}\rangle^{~}_{\rm R}$. Since we have to pick two copies
in order to make a closed string, there are two possible string
theories that one can construct in this way, called ``Type~IIA'' and
``Type~IIB'':

\noindent
$\underline{\rm Type~IIA:}$ In this case we take the {\it opposite}
GSO projection on both sides, so that the spinors are of opposite
chirality and hence admit the expansions
\beq
|\psi_{\rm l}\rangle^{~}_{\rm R}=\sum_a(\psi_{\rm l})_a\,|a
\rangle^{~}_{\rm R} \ , ~~ |\psi_{\rm r}\rangle^{~}_{\rm R}=
\sum_{\overline{a}}(\psi_{\rm r})_{\overline{a}}\,|\overline{a}
\rangle^{~}_{\rm R} \ .
\label{nonchiralexps}\eeq
The resulting theory is non-chiral.

\noindent
$\underline{\rm Type~IIB:}$ Here we impose the {\it same} GSO projection on
both sides, so that the spinors have the same chirality with
\beq
|\psi_{\rm l,r}\rangle^{~}_{\rm R}=\sum_a(\psi_{\rm l,r})_a\,|a
\rangle^{~}_{\rm R} \ .
\label{chiralexps}\eeq
This leads to a chiral theory.

\subsubsection*{Remarks on Superstring Types}

For completeness, we can now give brief definitions and descriptions
of the five consistent, perturbative superstring theories in ten
spacetime dimensions that were mentioned in section~\ref{History}.

\noindent
$\underline{\rm (1)~Type~II~Superstrings:}$ These are the superstrings
that we have been studying thus far. They involve strings
with oriented worldsheets (like the torus), and they possess local $N=2$
spacetime supersymmetry after implementation of the GSO
projection (this is the reason for the terminology Type~II). While the
ground state chiral structure of the Type~IIA and IIB theories differ,
their massive states are all the same. We will continue to study only
Type~II superstrings in the remainder of these notes.

\noindent
$\underline{\rm (2)~Type~I~Superstrings:}$ These can be obtained from a
projection of the Type~IIB theory that keeps only the diagonal sum of
the two gravitinos $\Psi_\mu$ and $\Psi_\mu'$ (see
table~\ref{closedSUSYstates}). This theory has only $N=1$ spacetime
supersymmetry and is a theory of unoriented string worldsheets (like
the M\"obius strip or the Klein bottle). In the open string sector,
quantum consistency (anomaly cancellation) singles out $SO(32)$ as the
only possible Chan-Paton gauge group~\cite{GrSchw1}.

\noindent
$\underline{\rm (3)~Heterotic~Strings:}$ This comprises a heterosis of
two string theories~\cite{GHMR1}, whereby we use the $d=26$ bosonic
string for the left-movers and the $d=10$ superstring for the right-movers. The
remaining 16 right-moving degrees of freedom required by $N=1$ supersymmetry
are ``internal'' ones which come from a 16-dimensional, even self-dual
lattice (these latter two constraints on the lattice are required by
modular invariance). There are only two such lattices, corresponding
to the weight lattices of the Lie groups $E_8\times E_8$ and $SO(32)$
(or more precisely its spin cover $Spin(32)/\zed_2$).\footnote{$N=1$
  supersymmetric Yang-Mills theory and supergravity in ten spacetime
  dimensions possess ``hexagon'' gauge and gravitational anomalies,
  respectively. The two types of anomalies cancel each other out if
  and only if the Chan-Paton gauge group is either $SO(32)$ or
  $E_8\times E_8$~\cite{GrSchw1}. Type~II supergravity is
  anomaly-free~\cite{AlWitten1}.}

\subsubsection*{Type II Ramond-Ramond States}

We will now study in some detail the structure of the ground states in
both the Type~IIA and IIB R--R sectors. Crucial to this analysis will
be a number of $\Gamma$-matrix identities in ten dimensions, which are given in
the following exercise.

\bigskip

\begin{center}
\begin{minipage}{15cm}

\small

{\bf Exercise~6.1.} {\sl\baselineskip=12pt Prove the following ten
  dimensional Dirac matrix identities (square brackets mean to
  antisymmetrize over the corresponding set of indices):

\noindent
{\bf (a)}
$$
\begin{array}{rll}
\Gamma^{11}\,\Gamma^{[\mu_1}\cdots\Gamma^{\mu_n]}&=&\displaystyle
\frac{(-1)^{\bigl[\frac n2\bigr]}\,n!}{\Bigl[(10-n)!\Bigr]^2}\,\epsilon^{\mu_1
\cdots\mu_n}_{~~~~~~\nu_1\cdots\nu_{10-n}}\,\Gamma^{[\nu_1}\cdots
\Gamma^{\nu_{10-n}]} \ , \\\Gamma^{[\mu_1}\cdots\Gamma^{\mu_n]}\,
\Gamma^{11}&=&\displaystyle\frac{(-1)^{\bigl[\frac{n+1}2\bigr]}\,n!}
{\Bigl[(10-n)!\Bigr]^2}\,\epsilon^{\mu_1
\cdots\mu_n}_{~~~~~~\nu_1\cdots\nu_{10-n}}\,\Gamma^{[\nu_1}\cdots
\Gamma^{\nu_{10-n}]} \ .
\end{array}
$$

\noindent
{\bf (b)}
$$
\begin{array}{rll}
\Gamma^\nu\,\Gamma^{[\mu_1}\cdots\Gamma^{\mu_n]}&=&\displaystyle
\frac{\Gamma^{[\nu}\,\Gamma^{\mu_1}\cdots\Gamma^{\mu_n]}}{n+1}-
\frac n{(n-1)!}\,\eta^{\nu[\mu_1}\,\Gamma^{\mu_2}\cdots\Gamma^{\mu_n]}
\ , \\\Gamma^{[\mu_1}\cdots\Gamma^{\mu_n]}\,\Gamma^\nu&=&\displaystyle
\frac{\Gamma^{[\mu_1}\cdots\Gamma^{\mu_n}\,\Gamma^{\nu]}}{n+1}-
\frac n{(n-1)!}\,\eta^{\nu[\mu_n}\,\Gamma^{\mu_1}\cdots\Gamma^{\mu_{n-1}]}
\ .
\end{array}
$$}

\end{minipage}
\end{center}

\bigskip

We will examine the R--R sectors {\it before} the imposition of the
physical super-Virasoro constraints on the spectrum. The R-sector
spinors are then representations of the spin cover $Spin(10)$ of the
ten-dimensonal rotation group and everything will be written in ten
dimensional notation. The two irreducible, Majorana-Weyl
representations of $Spin(10)$ are the spinor $\mathbf{16}_s$ and
conjugate spinor $\mathbf{16}_c$. From a group theoretic perspective
then, the massless states of the two R--R sectors are characterized by
their Clebsch-Gordan decompositions:

\noindent
$\underline{\rm Type~IIA:}$
\beq
\mathbf{16}_s\otimes\mathbf{16}_c=[0]\oplus[2]\oplus[4] \ .
\label{16scdecomp}\eeq

\noindent
$\underline{\rm Type~IIB:}$
\beq
\mathbf{16}_s\otimes\mathbf{16}_s=[1]\oplus[3]\oplus[5]_+ \ .
\label{16ssdecomp}\eeq

Here $[n]$ denotes the irreducible $n$-times antisymmetrized
representation of $Spin(10)$, which from a field theoretic perspective
corresponds to a completely antisymmetric tensor of rank $n$, or in
other words an ``$n$-form''. The + subscript indicates a certain
self-duality condition which will be discussed below. The $n$-forms
arise by explicitly decomposing the Ramond bi-spinor $|\psi_{\rm
  l}\rangle^{~}_{\rm R}\otimes|\psi_{\rm r}\rangle^{~}_{\rm R}$ in a
basis of antisymmetrized Dirac matrices:

\noindent
$\underline{\rm Type~IIA:}$
\beq
(\psi_{\rm l})_a\,(\psi_{\rm r})_{\overline{b}}=\sum_{n\,{\rm even}}
F_{\mu_1\cdots\mu_n}^{(n)}\,\left(\Gamma^{[\mu_1}\cdots\Gamma^{\mu_n]}
\right)_{a\overline{b}} \ .
\label{bispinorsc}\eeq

\noindent
$\underline{\rm Type~IIB:}$
\beq
(\psi_{\rm l})_a\,(\psi_{\rm r})_b=\sum_{n\,{\rm odd}}
F_{\mu_1\cdots\mu_n}^{(n)}\,\left(\Gamma^{[\mu_1}\cdots\Gamma^{\mu_n]}
\right)_{ab} \ .
\label{bispinorss}\eeq

Inverting these relations gives the $n$-forms explicitly as
\beq
F_{\mu_1\cdots\mu_n}^{(n)}={}^{~}_{\rm R}\langle\,\overline{\psi}_{\rm
  l}|\,\Gamma_{[\mu_1}\cdots\Gamma_{\mu_n]}|\psi_{\rm r}\rangle^{~}_{\rm
  R} \ .
\label{nformexpl}\eeq
Because of the GSO projection, the states $|\psi_{\rm
  l,r}\rangle^{~}_{\rm R}$ have definite $\Gamma^{11}$ eigenvalue
  $\pm\,1$. The $\Gamma$-matrix relations of exercise~6.1~(a) thereby
  generate an isomorphism
\beq
F_{\mu_1\cdots\mu_n}^{(n)}\sim\epsilon_{\mu_1\cdots\mu_n}^{~~~~~~
\nu_1\cdots\nu_{10-n}}\,F_{\nu_1\cdots\nu_{10-n}}^{(10-n)} \ .
\label{nformiso}\eeq
  This identifies the irreducible representations $[n]\cong[10-n]$,
  and in particular $[5]_+$ is ``self-dual''.

\bigskip

\begin{center}
\begin{minipage}{15cm}

\small

{\bf Exercise~6.2.} {\sl\baselineskip=12pt {\bf (a)} Verify that the
  number of independent components of the antisymmetric tensor fields
  $F_{\mu_1\cdots\mu_n}^{(n)}$ agrees with that of the tensor product
  of two Majorana-Weyl spinors in ten dimensions.

\noindent
{\bf (b)} Show that the Dirac-Ramond equations for $|\psi_{\rm
  l}\rangle^{~}_{\rm R}$ and $|\psi_{\rm r}\rangle^{~}_{\rm R}$ are
  equivalent to the field equations
$$
\partial_{[\mu}F_{\mu_1\cdots\mu_n]}^{(n)}=0 \ , ~~
\partial^\mu F_{\mu\mu_2\cdots\mu_n}^{(n)}=0 \ .
$$}

\end{minipage}
\end{center}

\bigskip

{}From exercise~6.2~(b) it follows that the physical constraints in the R--R
sector are equivalent to the ``Maxwell equation of motion'' and
``Bianchi identity'' for an antisymmetric tensor field strength:
\beq
\begin{tabular}{|c|}\hline\\
$\displaystyle
F_{\mu_1\cdots\mu_n}^{(n)}=\partial_{[\mu_1}C^{(n-1)}_{\mu_2\cdots\mu_n]}
\ ,
$\\\\
\hline\end{tabular}
\label{antisymfieldstrength}\eeq
generalizing the familiar equations of motion of electrodynamics. The
$n$-forms $F_{\mu_1\cdots\mu_n}^{(n)}$ are called ``Ramond-Ramond
fields'', while $C_{\mu_1\cdots\mu_n}^{(n)}$ are called
``Ramond-Ramond potentials''. The isomorphism (\ref{nformiso})
corresponds to an electric-magnetic duality which exchanges
equations of motion and Bianchi identities. It relates the fields
$C^{(n)}$ and $C^{(8-n)}$, which are thereby treated on equal footing
in string theory. Using (\ref{16scdecomp})--(\ref{bispinorss}) we
can look at the totality of Ramond-Ramond potentials in the Type~II
theories:

\noindent
$\underline{\rm Type~IIA:}$
\beq
C^{(1)}~~~~~~C^{(3)}~~~~~~C^{(5)}~~~~~~C^{(7)} \ .
\label{IIARRC}\eeq
Here the fields $C^{(5)}$ and $C^{(7)}$ are dual to $C^{(3)}$ and
$C^{(1)}$, respectively.

\noindent
$\underline{\rm Type~IIB:}$
\beq
C^{(0)}~~~~~~C^{(2)}~~~~~~C^{(4)}~~~~~~C^{(6)}~~~~~~C^{(8)} \ .
\label{IIBRRC}\eeq
Here the field $C^{(4)}$ is self-dual, while $C^{(6)}$ and $C^{(8)}$
are dual to $C^{(2)}$ and $C^{(0)}$, respectively.

\subsubsection*{Ramond-Ramond Charges}

We finally come to the crux of the matter here. Recall that the antisymmetric
NS--NS sector tensor $B_{\mu\nu}$ couples directly to the string worldsheet,
because as we saw in (\ref{Bmincoupling}) its vertex operator couples directly
to $B_{\mu\nu}$. In other words, the string carries (electric) ``charge'' with
respect to $B_{\mu\nu}$. However, the situation for the R--R potentials
$C^{(n)}$ above is very different, because from
(\ref{bispinorsc},\ref{bispinorss}) it follows that the vertex operators for
the R--R states involve only the R--R {\it fields} $F^{(n+1)}$ and so only the
field strengths, and not potentials, couple to the string. Thus elementary,
perturbative string states cannot carry any charge with respect to the R--R
gauge fields $C^{(p+1)}$.

We are thereby forced to search for non-perturbative degrees of freedom which
couple to these potentials. Clearly, these objects must be ``$p$-branes'',
which are defined to be $p$-dimensional extended degrees of freedom that sweep
out a $p+1$-dimensional ``worldvolume'' as they propagate in time, generalizing
the notion of a string. The minimal coupling would then involve the potential
$C^{(p+1)}$ multiplied by the induced volume element on the hypersurface, and
it takes the form
\beq
q\int\dd^{p+1}\xi~\epsilon^{a_0\cdots a_p}\,\frac{\partial x^{\mu_1}}
{\partial\xi^{a_0}}\cdots\frac{\partial x^{\mu_{p+1}}}
{\partial\xi^{a_p}}~C_{\mu_1\cdots\mu_{p+1}}^{(p+1)} \ ,
\label{pbranemincoupling}\eeq
in complete analogy with the electromagnetic coupling (\ref{Amincoupling}) and
the $B$-field minimal coupling (\ref{Bmincoupling}). Gauge transformations
shift $C^{(p+1)}$ by $p$-forms $\Lambda^{(p)}$ according to
\beq
C_{\mu_1\cdots\mu_{p+1}}^{(p+1)}~\longmapsto~C_{\mu_1\cdots\mu_{p+1}}^{(p+1)}
+\partial_{[\mu_1}\Lambda^{(p)}_{\mu_2\cdots\mu_{p+1}]} \ ,
\label{RRgaugetransf}\eeq
and they leave the field strength $F^{(p+2)}$ and all physical string states
invariant. As usual in quantum field theory, from (\ref{pbranemincoupling}) it
follows that they also leave the quantum theory invariant if at the same time
the $p$-brane wavefunction $\Psi$ transforms as
\beq
\Psi~\longmapsto~\e^{\ii\,q\int_{p-{\rm brane}}\Lambda^{(p)}}\,\Psi \ .
\label{Psigaugetransf}\eeq
This defines the ``Ramond-Ramond charge'' $q$ of the $p$-brane with respect to
the ``gauge field'' $C^{(p+1)}$.

The upshot of this analysis is that, since perturbative string states are R--R
neutral, string theory has to be complemented with {\it non-perturbative}
states which carry the R--R charges. These objects are known as ``Dirichlet
$p$-branes'', or ``D-branes'' for short, and they are dynamical objects which
are extended in $p$ spatial dimensions. In the remainder of this section we
will present a systematic derivation of how these states arise in string
theory.

\subsection{T-Duality for Closed Strings \label{TClosed}}

To systematically demonstrate the existence of D-branes in superstring theory,
i.e. to quantify their origin and explain what they are, we will first need to
describe a very important string theoretical symmetry. It is a crucial
consequence of the {\it extended} nature of strings and it has no field theory
analog. We will first consider closed bosonic strings with worldsheet the
cylinder and coordinates $-\infty<\tau<\infty$, $0\leq\sigma<2\pi$.

Let us begin by recalling the mode expansions of the string embedding fields:
\bea
x^\mu(\tau,\sigma)&=&x_{\rm L}^\mu(\tau+\sigma)+x_{\rm R}^\mu(\tau-\sigma)
\nn\\&=&x_0^\mu+\tilde x_0^\mu+\sqrt{\frac{\alpha'}2}\,\Bigl(\alpha_0^\mu+
\tilde\alpha_0^\mu\Bigr)\tau+\sqrt{\frac{\alpha'}2}\,\Bigl(\alpha_0^\mu-
\tilde\alpha_0^\mu\Bigr)\sigma+({\rm oscillators}) \ , \nn\\
\label{closedmodeexp0modes}\eea
where we focus our attention on only the zero modes as the oscillator
contributions will play no role in the present discussion. The total (center of
mass) spacetime momentum of the string is given by integrating the $\tau$
derivative of (\ref{closedmodeexp0modes}) over $\sigma$ to get
\beq
p_0^\mu=\frac1{\sqrt{2\alpha'}}\,\Bigl(\alpha_0^\mu+\tilde\alpha_0^\mu
\Bigr) \ .
\label{p0mudef}\eeq
Since the oscillator terms are periodic, a periodic shift along the string
changes the function (\ref{closedmodeexp0modes}) as
\beq
x^\mu(\tau,\sigma+2\pi)=x^\mu(\tau,\sigma)+2\pi\,\sqrt{\frac{\alpha'}2}\,
\Bigl(\alpha_0^\mu-\tilde\alpha_0^\mu\Bigr) \ ,
\label{xmuperiodic}\eeq
and so requiring that the embedding be single-valued under
$\sigma\mapsto\sigma+2\pi$ yields the constraint
\beq
\alpha_0^\mu=\tilde\alpha_0^\mu=\sqrt{\frac{\alpha'}2}\,p_0^\mu
\label{singlevaluedconstr}\eeq
with $p_0^\mu$ real. This is of course just what we worked out before in
section~\ref{StringEOM}.

All of this is true if the spatial directions in spacetime all have infinite
extent. We would now like to see how the mode expansion is modified if one of
the directions, say $x^9$, is compact. For this, we ``compactify'' $x^9$ on a
circle $\mathbf{S}^1$ of radius $R$. This means that the spacetime coordinate
is periodically identified as
\beq
x^9\sim x^9+2\pi R \ .
\label{x9periodic}\eeq
Then the basis wavefunction $\e^{\ii\,p_0^9x^9}$, which is also the generator
of translations in the $x^9$ direction, is single-valued under
(\ref{x9periodic}) {\it only} if the momentum $p_0^9$ is quantized according to
\beq
p_0^9=\frac nR
\label{p09quant}\eeq
for some integer $n$. This quantization condition applies to any
quantum system and is not particular to strings. Using (\ref{p0mudef})
we then have the zero mode constraint
\beq
\alpha_0^9+\tilde\alpha_0^9=\frac{2n}R\,\sqrt{\frac{\alpha'}2} \ .
\label{0modeconstr1}\eeq
In addition, under a periodic shift $\sigma\mapsto\sigma+2\pi$ along the
string, the string can ``wind'' around the spacetime circle. This means that,
since by (\ref{x9periodic}) the coordinate $x^9$ is no longer a periodic
function, the relation
\beq
x^9(\tau,\sigma+2\pi)=x^9(\tau,\sigma)+2\pi wR
\label{x9wchange}\eeq
is an allowed transformation for any integer $w$. For a fixed ``winding
number'' $w\in\zed$, this is represented by adding the term $wR\sigma$ to the
mode expansion (\ref{closedmodeexp0modes}) for $x^9(\tau,\sigma)$, and
comparing with (\ref{closedmodeexp0modes}) then yields a second zero mode
constraint
\beq
\alpha_0^9-\tilde\alpha_0^9=wR\,\sqrt{\frac2{\alpha'}} \ .
\label{0modeconstr2}\eeq
Solving (\ref{0modeconstr1}) and (\ref{0modeconstr2}) simultaneously then gives
\beq
\begin{tabular}{|c|}\hline\\
$\begin{array}{l}
\displaystyle
\alpha_0^9=p^{~}_{\rm L}\,\sqrt{\frac{\alpha'}2} \ , ~~ p^{~}_{\rm L}=
\frac nR+\frac{wR}{\alpha'} \ , \\\displaystyle
\tilde\alpha_0^9=\displaystyle p^{~}_{\rm R}\,\sqrt{\frac{\alpha'}2} \ ,
{}~~ p^{~}_{\rm R}=\displaystyle\frac nR-\frac{wR}{\alpha'} \ ,
\end{array}
$\\\\
\hline\end{tabular}
\label{leftrightmomenta}\eeq
where $p^{~}_{\rm L}$ and $p^{~}_{\rm R}$ are called the ``left-moving and
right-moving momenta''.

Let us now consider the mass spectrum in the remaining uncompactified 1+8
dimensions, which is given by
\bea
m^2&=&-p_\mu\,p^\mu \ , ~~ \mu=0,1,\dots,8\nn\\
&=&\frac2{\alpha'}\,\Bigl(\alpha_0^9\Bigr)^2+\frac4{\alpha'}\,
\Bigl(N-1\Bigr)\nn\\&=&\frac2{\alpha'}\,\Bigl(\tilde\alpha_0^9\Bigr)^2+
\frac4{\alpha'}\,\Bigl(\tilde N-1\Bigr) \ ,
\label{1+8mass}\eea
where the second and third equalities come from the $L_0=1$ and $\tilde L_0=1$
constraints, respectively. The sum and difference of these left-moving and
right-moving Virasoro constraints give, respectively, the Hamiltonian and
level-matching formulas. Here we find that they are modified from those of
section~\ref{Spectrum} to
\beq
\begin{tabular}{|c|}\hline\\
$\displaystyle
\begin{array}{l}
\displaystyle m^2=\frac{n^2}{R^2}+\frac{w^2R^2}{\alpha'^2}+\frac2{\alpha'}\,
\Bigl(N+\tilde N-2\Bigr) \ , \\nw+N-\tilde N=0 \ .
\end{array}
$\\\\
\hline\end{tabular}
\label{masslevelmod}\eeq
We see that there are extra terms present in both the mass formula and the
level matching condition, in addition to the usual oscillator contributions,
which come from the ``Kaluza-Klein tower of momentum states'' $n\in\zed$ and
the ``tower of winding states'' $w\in\zed$ of the string. The winding modes are
a purely {\it stringy} phenomenon, because only a string can wrap non-trivially
around a circle. The usual non-compact states are obtained by setting $n=w=0$,
and, in particular, the massless states arise from taking $n=w=0$ and $N=\tilde
N=1$.

It is interesting to compare the large and small radius limits of the
compactified string theory:

\noindent
$\underline{R\to\infty:}$ In this limit all $w\neq0$ winding states
disappear as they are energetically unfavourable. But the $w=0$ states
with all values of $n\in\zed$ go over to the usual continuum of
momentum zero modes $k^\mu$ in the mass formula. This is the same
thing that would happen in quantum field theory. In this limit, the
spacetime momenta are related by $p_{\rm R}^{~}=p_{\rm L}^{~}$.

\noindent
$\underline{R\to0:}$ In this limit all $n\neq0$ momentum states become
infinitely massive and decouple. But now the pure winding states
$n=0$, $w\neq0$ form a continuum as it costs very little energy to
wind around a small circle. So as $R\to0$ an extra uncompactified
dimension reappears. This is in marked contrast to what would occur in
quantum field theory, whereby all surviving fields in the limit would
be just independent of the compact coordinate $x^9$. In this limit the
spacetime momenta are related by $p_{\rm R}^{~}=-p_{\rm L}^{~}$.

The appearence of the extra dimension in the small radius limit requires an
explanation. This stringy behaviour is the earmark of ``T-duality'', as we
shall now explain. The mass formula (\ref{masslevelmod}) for the spectrum is
invariant under the simultaneous exchanges
\beq
\begin{tabular}{|c|}\hline\\
$\displaystyle
n~\longleftrightarrow~w \ , ~~ R~\longleftrightarrow~R'=\frac{\alpha'}R \ ,
$\\\\
\hline\end{tabular}
\label{Tdualitysym}\eeq
which by (\ref{leftrightmomenta}) is equivalent to the zero-mode
transformations
\beq
\alpha_0^9~\longmapsto~\alpha_0^9 \ , ~~ \tilde\alpha_0^9~\longmapsto~
-\tilde\alpha_0^9 \ .
\label{alpha09Tdual}\eeq
This symmetry of the compactified string theory is known as a
``T-duality symmetry''~\cite{GPR1}. The string theory compactified on
the circle of radius $R'$ (with momenta and windings interchanged) is
called the ``T-dual string theory''. The process of going from a
compactified string theory to its dual is called ``T-dualization''.

We can also realize this symmetry at the full interacting level of massive
states by a ``spacetime parity transformation'' of the worldsheet right-movers
as
\beq
\begin{tabular}{|c|}\hline\\
$\displaystyle
{\rm T}\,:\,
x_{\rm L}^9(\tau+\sigma)~\longmapsto~x_{\rm L}^9(\tau+\sigma) \ , ~~
x_{\rm R}^9(\tau-\sigma)~\longmapsto-x_{\rm R}^9(\tau-\sigma) \ .
$\\\\
\hline\end{tabular}
\label{Tdualitytransf}\eeq
This is called a ``T-duality transformation''. It holds because the entire
worldsheet quantum field theory is invariant under the rewriting of the radius
$R$ theory in terms of the dual string coordinates
\beq
x'^{\,9}=x_{\rm L}^9-x_{\rm R}^9 \ .
\label{dualstringcoords}\eeq
It is simply a matter of convention whether to add or subtract the two
worldsheet sectors, since both choices solve the two-dimensional wave equation
of section~\ref{StringEOM}. The only change which occurs is that the zero-mode
spectrum of the new variable $x'^{\,9}(\tau,\sigma)$ is that of the T-dual
$R'=\frac{\alpha'}R$ string theory. The dual theories are physically
equivalent, in the sense that all quantum correlation functions are invariant
under the rewriting (\ref{dualstringcoords}). It follows that T-duality
$R\leftrightarrow R'=\frac{\alpha'}R$ is an {\it exact} quantum symmetry of
perturbative closed string theory.

\subsubsection*{String Geometry}

\begin{figure}[htb]
\epsfxsize=3 in
\bigskip
\centerline{\epsffile{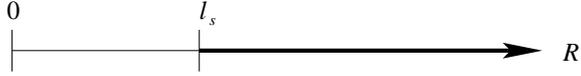}}
\caption{\baselineskip=12pt {\it The moduli space of string theory
    compactified on a circle of radius $R$. The circles seen by string
  theory correspond to the line $R\geq\ell_s$, with $\ell_s$ the
  fixed point $R'=R$ of the T-duality transformation
  $R\mapsto\frac{\alpha'}R$. Closed string probes cannot see very
  small radius circles with $0\leq R<\ell_s$ due to the finite
  intrinsic size $\ell_s$ of the string. The entire quantum
  string theory on a circle of radius $R\in[0,\ell_s)$ can be mapped
  onto a completely equivalent one with $R\geq\ell_s$ by using T-duality.}}
\bigskip
\label{Rmodulispace}\end{figure}

One of the many profound consequences of the T-duality symmetry of
closed strings is on the nature of spacetime geometry as seen by
strings~\cite{ACV1,GrossMende1,Ven2}. In particular, the ``moduli
space'' of these compactified string theories, which in this simple
case is the space of circle radii $R$ parametrizing the fields, is the
semi-infinite line $R\geq\ell_s=\sqrt{\alpha'}$, rather than the
classical bound $R\geq0$ (fig.~\ref{Rmodulispace}). By T-duality, very
small circles are equivalent to very large ones in string theory. Thus
strings see spacetime geometry very differently than ordinary point particles
do. In this sense, strings modify classical general relativity at very
short distance scales.

\subsection{T-Duality for Open Strings \label{TOpen}}

We will now turn to the case of bosonic open strings with worldsheet
the infinite strip $-\infty<\tau<\infty$, $0\leq\sigma\leq\pi$. Open
strings do not wind around the periodic direction of spacetime, and so
they have no quantum number comparable to $w$. Thus something very different
must happen as compared to the closed string case. The absence of $w$
in fact means that the open string theory looks more like a quantum
field theory in the limit $R\to0$, in that states with non-zero
Kaluza-Klein momentum $n\neq0$ become infinitely massive but no new
continuum states arise. So we have reached an apparent
paradox. Unitarity requires any fully consistent string theory to
possess both open and closed strings. The reason for this is another
worldsheet quantum duality which we will discuss briefly in
section~\ref{Forces}. However, the open strings effectively live in {\it nine}
spacetime dimensions as $R\to0$, while the closed strings live in {\it ten}
dimensions. The way out of this paradox is to note that the interior
of the open string still vibrates in ten dimensions, because there the
theory actually resembles that of a closed string. The distinguished
parts are the string {\it endpoints} which are restricted to lie on a
nine dimensional hyperplane in spacetime.

To quantify these statements, we recall the open string mode
expansions
\beq
x^\mu(\tau\pm\sigma)=\frac{x_0^\mu}2\pm\frac{x_0'^{\,\mu}}2+
\sqrt{\frac{\alpha'}2}\,\alpha_0^\mu\,(\tau\pm\sigma)+
\ii\,\sqrt{\frac{\alpha'}2}\,\sum_{n\neq0}\frac{\alpha_n^\mu}n~
\e^{-\ii\,n(\tau\pm\sigma)} \ , \nn\\
\label{openstringmodeexpnew}\eeq
with
\beq
\alpha_0^\mu=\sqrt{2\alpha'}\,p_0^\mu
\label{0modeopen}\eeq
and the total embedding coordinates
\bea
x^\mu(\tau,\sigma)&=&x^\mu(\tau+\sigma)+x^\mu(\tau-\sigma) \nn\\
&=&x_0^\mu+\alpha'p_0^\mu\,\tau+\ii\,\sqrt{2\alpha'}\,
\sum_{n\neq0}\frac{\alpha_n^\mu}n~\e^{-\ii\,n\tau}\,\cos(n\sigma) \ .
\label{xmuopennew}\eea
Note that the arbitrary integration constant $x_0'^{\,\mu}$ has
dropped out of (\ref{xmuopennew}) to give the usual open string
coordinates that we found in section~\ref{StringEOM}. Again we will
put $x^9$ on a circle of radius $R$, so that
\beq
x^9(\tau,\sigma)\sim x^9(\tau,\sigma)+2\pi R \ , ~~
p_0^9=\frac nR
\label{x9compactopen}\eeq
with $n\in\zed$.

\begin{figure}[htb]
\epsfxsize=4 in
\bigskip
\centerline{\epsffile{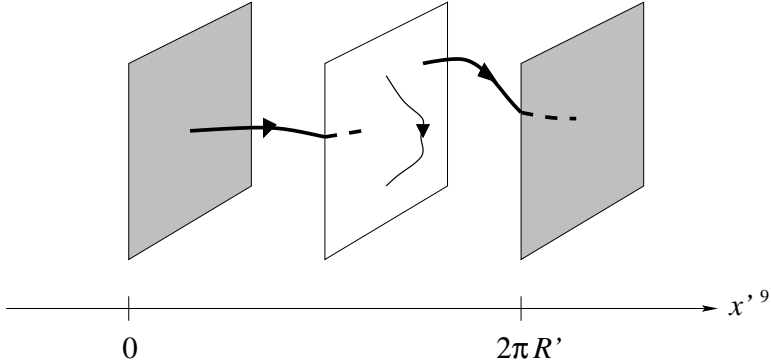}}
\caption{\baselineskip=12pt {\it The ends of the T-dual open string
    coordinate $x'^{\,9}$ attach to hyperplanes in spacetime. The
    shaded hyperplanes are periodically identified. The thick strings
    have winding number 1, while the thin string has winding number 0.}}
\bigskip
\label{openwind}\end{figure}

To get the T-dual open string coordinate, we use the ``doubling
trick'' of section~\ref{StringEOM} which may be used to generate the
open string mode expansion from the closed one (as in
(\ref{openstringmodeexpnew})). According to what we saw in the
previous subsection, this implies that we should reflect the
right-movers, i.e. set $x^9(\tau+\sigma)\mapsto x^9(\tau+\sigma)$ and
$x^9(\tau-\sigma)\mapsto-x^9(\tau-\sigma)$. The desired embedding
function is therefore given by
\bea
x'^{\,9}(\tau,\sigma)&=&x^9(\tau+\sigma)-x^9(\tau-\sigma) \nn\\
&=&x_0'^{\,9}+2\alpha'\,\frac nR\,\sigma+\sqrt{2\alpha'}\,
\sum_{n\neq0}\frac{\alpha_n^9}n~\e^{-\ii\,n\tau}\,\sin(n\sigma) \ .
\label{xprimeopen}\eea
Notice that the zero mode sector of (\ref{xprimeopen}) is independent
of the worldsheet time coordinate $\tau$, and hence the new string
embedding carries no momentum. Thus the dual string is {\it
  fixed}. Since $\sin(n\sigma)=0$ at $\sigma=0,\pi$, the endpoints do
not move in the $x^9$ direction, i.e. $\partial_\tau
x'^{\,9}|_{\sigma=0,\pi}=0$. That is to say, instead of the usual
Neumann boundary condition
\beq
\partial_\perp x^9\Bigm|_{\sigma=0,\pi}\equiv\partial_\sigma
x^9\Bigm|_{\sigma=0,\pi}=0
\label{Neumannx9}\eeq
with $\partial_\perp$ the ``normal derivative'' to the boundary of the
string worldsheet, we now have
\beq
\partial_\parallel x'^{\,9}\Bigm|_{\sigma=0,\pi}\equiv\partial_\tau
x'^{\,9}\Bigm|_{\sigma=0,\pi}=0
\label{Dirichletx9}\eeq
with $\partial_\parallel$ the ``tangential derivative'' to the
worldsheet boundary. This gives the ``Dirichlet boundary condition''
that the open string endpoints are at a fixed place in spacetime given
by the formula
\beq
\begin{tabular}{|c|}\hline\\
$\displaystyle
x'^{\,9}(\tau,\pi)-x'^{\,9}(\tau,0)=\frac{2\pi\alpha'\,n}R=2\pi nR' \ .
$\\\\
\hline\end{tabular}
\label{endpointfixed}\eeq
Thus the endpoints $x'^{\,9}|_{\sigma=0,\pi}$ are equal up to the
periodicity of the T-dual dimension. We may thereby regard this
formula as defining an ``open string of winding number $n\in\zed$''
(fig.~\ref{openwind}).

The open string ends are still free to move in the other 1+8
directions that are not T-dualized, which constitute a hyperplane
called a ``D-brane''~\cite{DLP1,Horava1}. As there are eight spatial
dimensions, we call it more specifically a ``D8-brane''. Generally,
T-dualizing $m$ directions of the spacetime gives Dirichlet boundary
conditions in the $m$ directions, and hence a hyperplane with $p=9-m$
spatial dimensions which we will call a ``D$p$-brane''. We conclude that
T-duality, as a symmetry of the fully consistent string theory
(containing both open and closed strings), necessitates {\it
  D-branes}.

\bigskip

\begin{center}
\begin{minipage}{15cm}

\small

{\bf Exercise~6.3.} {\sl\baselineskip=12pt Show explicitly that T-duality
  interchanges the definitions of normal and tangential derivatives,
  and hence it exchanges Neumann and Dirichlet boundary conditions.}

\end{minipage}
\end{center}

\bigskip

\subsection{T-Duality for Type II Superstrings \label{TTypeII}}

Finally, we will now generalize the results of the previous two
subsections to superstrings. Let us consider the effects of T-duality on
the closed, oriented Type~II theories. As we have seen, as a
right-handed parity transformation it flips the sign of the
right-mover $x^9_{\rm R}(\tau-\sigma)$. By worldsheet supersymmetry,
it must do the same on the right-moving fermion fields
$\psi_-^9(\tau-\sigma)$, so that
\beq
{\rm T}\,:\,\psi_-^9~\longmapsto~-\psi_-^9 \ .
\label{Tpsi9}\eeq
This implies that the zero-mode of $\psi_-^9$ in the Ramond sector,
which acts as the Dirac matrix $\Gamma^9$ on right-movers, changes
sign, and hence ${\rm T}:\Gamma^{11}\mapsto-\Gamma^{11}$. Thus the
relative chirality between left-movers and right-movers is flipped,
i.e. T-duality reverses the sign of the GSO projection on
right-movers:
\beq
{\rm T}\,:\,P_{\rm GSO}^\pm~\longmapsto~P_{\rm GSO}^\mp \ .
\label{TGSO}\eeq

We conclude that T-duality interchanges the Type~IIA and Type~IIB
superstring theories. It is only a symmetry of the closed string
sector, since in the open string sector it relates two different types
of theories. Furthermore, since the IIA and IIB theories have
different Ramond-Ramond fields, T-duality must transform one set into
the other. The same conclusions are reached if one T-dualizes any odd
number of spacetime dimensions, while dualizing an even number returns
the original Type~II theory.

\setcounter{equation}{0}

\section{D-Branes and Gauge Theory \label{DBraneGauge}}

In this section we will start working towards a systematic description
of D-branes, which were introduced in the previous section. We will
begin with a heuristic, qualitative description of D-branes, drawing
from the way they were introduced in the previous section, and
painting the picture for the way that they will be analysed. The
underlying theme, as we will see, is that they are intimately tied to
gauge theory. Indeed, they provide a means of embedding gauge theories into
superstring theory. In particular, we will see how to describe their
collective coordinates in terms of standard gauge theory Wilson
lines. We will then take our first step to describing the low-energy
dynamics of D-branes, which will be covered in more detail in the
next and final section. Here we shall derive the celebrated
Born-Infeld action, which will also introduce some further important
computational tools.

\subsection{D-Branes \label{DBranes}}

\begin{figure}[htb]
\epsfxsize=3 in
\bigskip
\centerline{\epsffile{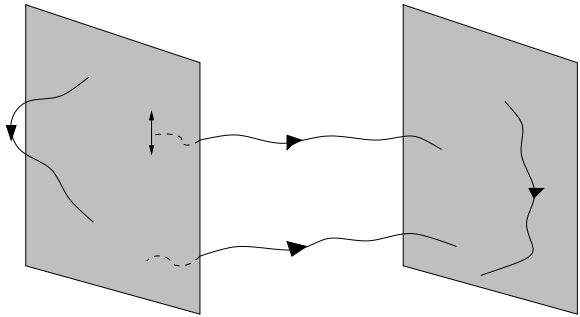}}
\caption{\baselineskip=12pt {\it A pair of D-branes (shaded regions)
    with open strings (wavy lines) attached (with Dirichlet boundary
    conditions). The string ends are free to move along the
    hyperplanes. The corresponding open string coordinates satisfy
    Neumann boundary conditions in the directions along the D-branes
    and Dirichlet boundary conditions in the directions transverse to
    the D-branes.}}
\bigskip
\label{Dbranes}\end{figure}

Let us begin with a heuristic description of the new extended degrees
of freedom that we have discovered in the previous section. By a
``D$p$-brane'' we will mean a $p+1$ dimensional hypersurface in
spacetime onto which open strings can attach
(fig.~\ref{Dbranes}). Such objects arise when we choose {\it
  Dirichlet} rather than Neumann boundary conditions for the open
strings. More precisely, the D$p$-brane is specified by choosing
Neumann boundary conditions in the directions along the hypersurface,
\beq
\partial_\sigma x^\mu\Bigm|_{\sigma=0,\pi}=0 \ , ~~ \mu=0,1,\dots,p \ ,
\label{NeumannDbrane}\eeq
and Dirichlet boundary conditions in the transverse directions,
\beq
\delta x^\mu\Bigm|_{\sigma=0,\pi}=0 \ , ~~ \mu=p+1,\dots,9 \ .
\label{DirichletDbrane}\eeq

In string perturbation theory, the position of the D$p$-brane is fixed
at the boundary coordinates $x^{p+1},\dots,x^9$ in spacetime,
corresponding to a particular string theory background (i.e. a
solution to the classical string equations of motion), while
$x^0,\dots,x^p$ are free to move along the $p+1$ dimensional
hypersurface. Using R--R vertex operators, it can be shown that they
must couple D$p$-branes and R--R potentials $C^{(p+1)}$~\cite{Pol1},
and furthermore that fundamental (perturbative) strings cannot carry the R--R
charges. Given this fact, we can readily deduce from (\ref{IIARRC})
and (\ref{IIBRRC}) what sorts of D-branes live in the different
Type~II superstring theories:

\noindent
$\underline{{\rm Type~IIA~D}p-{\rm Branes:}}$ These branes exist for
all even values of $p$,
\beq
p~=~0~~~~~~2~~~~~~4~~~~~~6~~~~~~8 \ .
\label{IIAps}\eeq
The case $p=0$ is a ``D-particle'', while $p=8$ describes a ``domain
wall'' in ten dimensional spacetime (in light of the solitonic
description of D-branes~\cite{DKL1}). The corresponding Ramond-Ramond
fields are respectively $F^{(2)}$, $F^{(4)}$, $F^{(6)}$, $F^{(8)}$,
and $F^{(10)}$. By the field equations (exercise~6.2~(b)), the latter
field admits no propagating states. The D0-brane and D6-brane are
electromagnetic duals of each other, as are the D2-brane and D4-brane.

\noindent
$\underline{{\rm Type~IIB~D}p-{\rm Branes:}}$ Here we find branes for
all odd values of $p$,
\beq
p~=~-1~~~~~~1~~~~~~3~~~~~~5~~~~~~7~~~~~~9 \ .
\label{IIBps}\eeq
The case $p=-1$ describes an object which is localized in time and corresponds
to a ``D-instanton'', while $p=1$ is a ``D-string''. The D9-branes are
spacetime filling branes, with no coupling to any R--R field strength, while
$p=3$ yields the self-dual D3-brane (which we note has worldvolume of
observable 3+1-dimensions). The D-instanton and D7-brane are electromagnetic
duals of one another, as are the D1-brane and the D5-brane.

In what follows though we will work towards a {\it non-perturbative}
description of D-branes. We will find that the massless modes of open strings
are associated with the fluctuation modes of the D-branes themselves, so that,
non-perturbatively, D-branes become dynamical $p$-dimensional objects. Let us
give a heuristic description of how the dynamical degrees of freedom of a
D-brane arise from the massless string spectrum in a {\it fixed} D-brane
background. For this, we recall that the open string spectrum contains a
massless $SO(8)$ vector $A_\mu$. In the presence of the D-brane, the gauge
field $A_\mu(x)$ decomposes into components parallel and perpendicular to the
D-brane worldvolume. Because the endpoints of the strings are tied to the
worldvolume, these massless fields can be interpreted in terms of a low-energy
field theory on the D-brane worldvolume. Precisely, a ten-dimensional gauge
field $A_\mu(x)$, $\mu=0,1,\dots,9$ will split into components $A_a$,
$a=0,1,\dots,p$ corresponding to a $U(1)$ gauge field on the D$p$-brane, and
components $\Phi^m$, $m=p+1,\dots,9$ which are scalar fields describing the
fluctuations of the D$p$-brane in the $9-p$ transverse directions. We will find
that the low-energy dynamics of such a configuration is governed by a
supersymmetric gauge theory, obtained from the dimensional reduction to the
D$p$-brane of maximally supersymmetric Yang-Mills theory in ten spacetime
dimensions. In this section and the next we will work towards making these
statements precise and explicit.

\subsection{Wilson Lines \label{Wilson}}

To make the discussion of the previous subsection more quantitative,
we will first need an elementary, but perhaps not so widely
appreciated, result from quantum mechanics, which we leave as an
exercise.\footnote{\baselineskip=12pt The free kinetic term in the
  action below is not the same as the one we studied in
  section~\ref{Particle}, but is rather a point particle version of
  the Polyakov action. The second term is the usual minimal coupling
  of a particle to a gauge potential.}

\bigskip

\begin{center}
\begin{minipage}{15cm}

\small

{\bf Exercise~7.1.} {\sl\baselineskip=12pt A relativistic particle of
  mass $m$ and charge $q$ in $d$ Euclidean spacetime dimensions
  propagates in a background electromagnetic vector potential
  $A_\mu(x)$ according to the action
$$
S=\int\dd\tau~\left(\frac m2\,\dot x^\mu\,\dot x_\mu-\ii\,q\,\dot x^\mu\,
A_\mu\right) \ .
$$
Show that if the $d$-th direction is compactified on a circle of
radius $R$, then a constant gauge field
$A_\mu=-\delta_{\mu,d}\,\frac\theta{2\pi}$ induces a fractional
canonical momentum
$$
p^d=\frac nR+\frac{q\,\theta}{2\pi R} \ .
$$
}

\end{minipage}
\end{center}

\bigskip

We will begin by again compactifying the $x^9$ direction of spacetime, $x^9\sim
x^9+2\pi R$, and introduce $U(N)$ Chan-Paton factors for a Type~II oriented
open string. Consider the open string vertex operator corresponding to a {\it
constant}, background abelian gauge field
\beq
A_\mu=\delta_{\mu,9}\,\pmatrix{\displaystyle\frac{\theta_1}{2\pi R}& &0\cr
 &\ddots& \cr0& &\displaystyle\frac{\theta_N}{2\pi R}\cr} \ ,
\label{Amubackground}\eeq
where $\theta_i$, $i=1,\dots,N$ are constants. The introduction of this
electromagnetic background generically breaks the Chan-Paton gauge symmetry as
$U(N)\to U(1)^N$, as (\ref{Amubackground}) is clearly only invariant under an
abelian subgroup of $U(N)$. {\it Locally}, it is a trivial, pure gauge
configuration, since its non-vanishing component can be written as
\beq
A_9=-\ii\,\Lambda^{-1}\,\partial_9\Lambda \ , ~~ \Lambda(x)=
\pmatrix{\e^{\ii\,\theta_1x^9/2\pi R}& &0\cr &\ddots& \cr
0& &\e^{\ii\,\theta_Nx^9/2\pi R}\cr} \ .
\label{A9puregauge}\eeq
This means that we can gauge $A_9$ away by a local gauge transformation. But
this is not true {\it globally}, because the compactness of the $x^9$ direction
still leads to non-trivial effects involving the background
(\ref{Amubackground}). This is manifested in the fact that the gauge
transformation $\Lambda(x)$ of (\ref{A9puregauge}) is not a single-valued
function on spacetime, but has the singular behaviour
\beq
\Lambda(x^9+2\pi R)=W\cdot\Lambda(x) \ , ~~ W=
\pmatrix{\e^{\ii\,\theta_1}& &0\cr &\ddots& \cr0& &\e^{\ii\,\theta_N}\cr} \ .
\label{Lambdasing}\eeq
All charged (gauge invariant) states pick up the phase factor $W$ under a
periodic tranlation $x^9\mapsto x^9+2\pi R$ due to the trivializing gauge
transformation, and the effects of the trivial gauge field
(\ref{Amubackground}) are still felt by the theory. Thus the configuration
$A_\mu$ yields no contribution to the (local) equations of motion, and its only
effects are in the holonomies as the string ends wind around the compactified
spacetime direction.

The phase factor $W$ in (\ref{Lambdasing}) is in fact just the
``Wilson line'' for the given gauge field configuration (i.e. the
group element corresponding to the gauge field), represented by the
exponential of the vertex operator
(\ref{photoncorr},\ref{ChanPatonvertex}) for the open string photon
field (\ref{Amubackground}):
\beq
W=\exp\left(\ii\,\int\dd t~\dot x^\mu(t)\,A_\mu\right)=
\exp\left(\ii\,\int\limits_0^{2\pi R}\dd x^9~A_9\right) \ .
\label{Wilsonline}\eeq
This is the observable that appears in exponentiated form in the
Polyakov path integral (\ref{stringscattschem}) for amplitudes, upon
introducing the vertex operators compatible with the gauge invariance
of the problem. It cannot be set to unity by a gauge transformation
and it is ultimately responsible for the symmetry breaking $U(N)\to
U(1)^N$. This scenario has a familiar analog in quantum mechanics, the
Aharonov-Bohm effect. Placing a solenoid of localized, point-like
magnetic flux in a region of space introduces a non-contractible loop,
representing charged particle worldlines encircling the source that cannot
be contracted to a point because of the singularity introduced by the
flux. Although everywhere outside of the source the electromagnetic
field is zero, the wavefunctions still acquire a non-trivial phase
factor $W\neq1$ which cannot be removed by the gauge symmetry of the
problem. This phase is observable through electron interference
patterns obtained from scattering off of the solenoid.

The symmetry breaking mechanism induced by the Wilson line $W$ has a very
natural interpretation in the T-dual theory in terms of {\it D-branes}. From
exercise~7.1 it follows that the string momenta along the $x^9$ direction are
{\it fractional}, and so the fields in the T-dual description have fractional
winding numbers. Recalling the analysis of section~\ref{TOpen} (see
(\ref{endpointfixed})), we see therefore that the two open string endpoints no
longer lie on the same hyperplane in spacetime.

\begin{figure}[htb]
\epsfxsize=4 in
\bigskip
\centerline{\epsffile{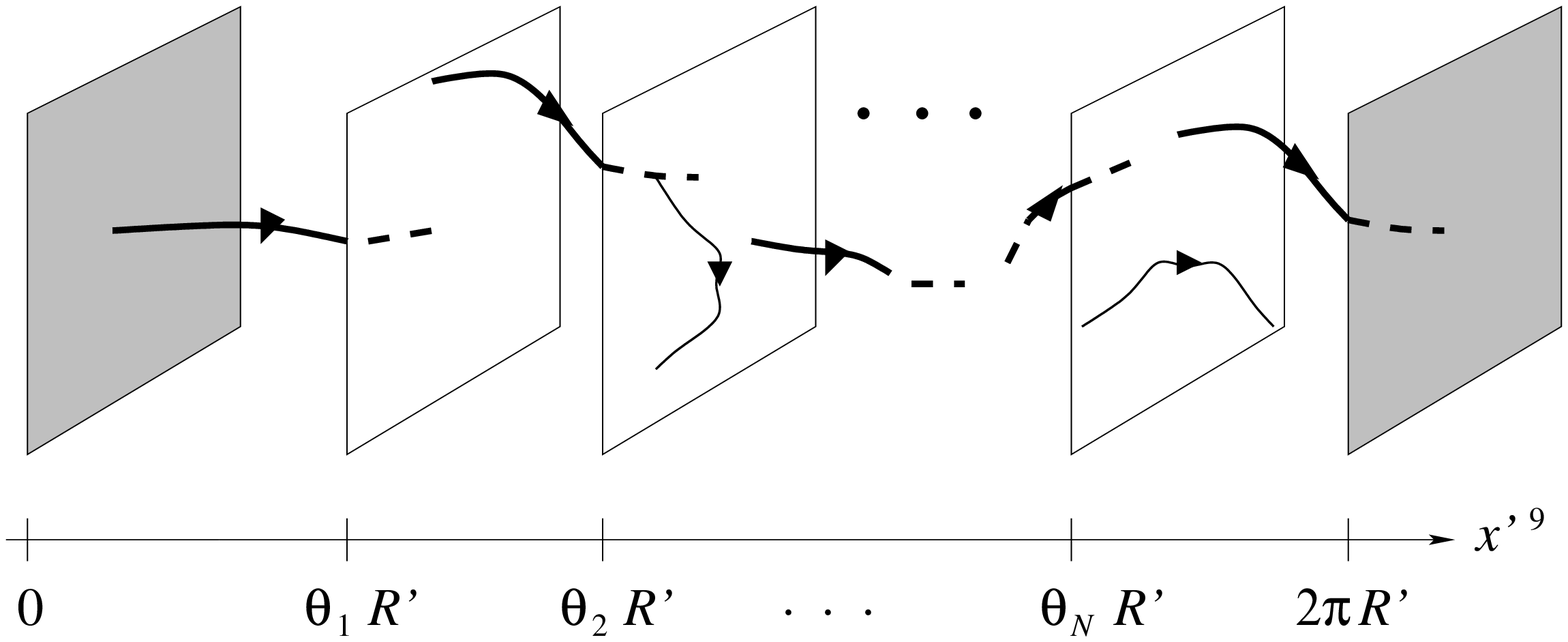}}
\caption{\baselineskip=12pt {\it An assembly of $N$ separated, parallel
D-branes with open strings attached. The shaded hyperplanes are periodically
identified.}}
\bigskip
\label{NDbranes}\end{figure}

To quantify these last remarks, let us consider a Chan-Paton wavefunction
$|k;ij\rangle$, as depicted schematically in fig.~\ref{ChanPatonfig}. The state
$i$ attached to an end of the open string will acquire a factor
$\e^{-\ii\,\theta_ix^9/2\pi R}$ due to the gauge transformation
(\ref{A9puregauge}), while state $j$ will have $\e^{\ii\,\theta_jx^9/2\pi R}$.
The total open string wavefunction will therefore gauge transform to
$|k;ij\rangle\cdot\e^{-\ii\,(\theta_i-\theta_j)x^9/2\pi R}$, and so under a
periodic translation $x^9\mapsto x^9+2\pi R$ it will acquire a Wilson line
factor given by
\beq
|k;ij\rangle~\longmapsto~\e^{\ii\,(\theta_j-\theta_i)}\,|k;ij\rangle \ .
\label{kijWilsonline}\eeq
In Fourier space this should be manifested through spacetime translations in
the plane waves $\e^{\ii\,p_9x^9}$, which means that the momentum of the state
$|k;ij\rangle$ is given by
\beq
p_{ij}^9=\frac nR+\frac{\theta_j-\theta_i}{2\pi R} \ ,
\label{pij9}\eeq
with $n\in\zed$ and $i,j=1,\dots,N$. So by appropriately modifying the mode
expansions and the endpoint calculation of section~\ref{TOpen}, we find that
the Dirichlet boundary condition corresponding to this Chan-Paton state is now
changed to
\beq
x'^{\,9}(\tau,\pi)-x'^{\,9}(\tau,0)=\left(2\pi n+\theta_j-\theta_i
\right)\,R' \ ,
\label{pij9mod}\eeq
with $R'=\frac{\alpha'}R$ the T-dual compactification radius. Thus, up to an
arbitrary additive constant, the open string endpoint in Chan-Paton state $i$
is at the spacetime position
\beq
\begin{tabular}{|c|}\hline\\
$\displaystyle
x_i'^{\,9}=\theta_i\,R'=2\pi\alpha'\,(A_9)_{ii} \ , ~~ i=1,\dots,N \ .
$\\\\
\hline\end{tabular}
\label{xi9hyperplane}\eeq
The locations (\ref{xi9hyperplane}) specify $N$ hyperplanes at different
positions corresponding to a collection of $N$ parallel, separated D-branes
(fig.~\ref{NDbranes}). In particular, from this analysis we may conclude the
remarkable fact that T-duality maps gauge fields in open string theory to the
localized positions of D-branes in spacetime.

\subsubsection*{D-Brane Terminology}

We will interpret the original ten-dimensional open strings as lying
on $N$ ``D9-branes'' which fill the spacetime. In this picture the
string endpoints can sit anywhere in spacetime and correspond to
ordinary Chan-Paton factors. Compactifying $9-p$ coordinates $x^m$,
$m=p+1,\dots,9$ confines the open string endpoints to $N$
``D$p$-brane'' hyperplanes of dimension $p+1$. This is a consequence
of the corresponding T-duality transformation which maps the Neumann
boundary conditions $\partial_\parallel x^m=0$ to the Dirichlet ones
$\partial_\perp x'^{\,m}=0$, with all other $x^a$, $a=0,1,\dots,p$
still obeying Neumann boundary conditions. Since T-duality
interchanges Neumann and Dirichlet boundary conditions (exercise~6.3),
a T-duality transformation along a direction parallel to a D$p$-brane
produces a D$(p-1)$-brane, while T-duality applied to a direction
perpendicular to a D$p$-brane yields a D$(p+1)$-brane. Comparing with
(\ref{IIAps}) and (\ref{IIBps}), we see that this is in fact the basis
of the way that the Type~IIA and IIB superstring theories are
interchanged under T-duality, as we saw in section~\ref{TTypeII}.

\subsection{Collective Coordinates for D-Branes \label{Collective}}

As we will see very soon, D-branes are in fact {\it not} rigid
hyperplanes in spacetime. They are dynamical, and can fluctuate both
in shape and position. For instance, the T-dual theory with D-branes
still contains gravity and gauge fields, whose dynamics we will begin
describing in the next subsection. To see this, let us consider, as we
did in (\ref{1+8mass}), the 1+8 dimensional massless spectrum,
interpreted in the T-dual string theory, for the case where only the
coordinate field $x^9$ is T-dualized. With $\cal N$ denoting the
occupation number of a Chan-Paton state $|k;ij\rangle$, the mass-shell
relation reads
\bea
m_{ij}^2&=&\left(p_{ij}^9\right)^2+\frac1{\alpha'}\,\Bigl({\cal N}-1
\Bigr)\nn\\&=&\frac{L_{ij}^2}{(2\pi\alpha')^2}+\frac1{\alpha'}\,
\Bigl({\cal N}-1\Bigr) \ ,
\label{mijcalN}\eea
where
\beq
L_{ij}=\Bigl|2\pi n+(\theta_i-\theta_j)\Bigr|\,R'
\label{Lijdef}\eeq
is the minimum length of an open string which winds $n$ times between
hyperplanes $i$ and $j$. To examine the massless states, we set $n=0$
(as it costs energy to wind) and ${\cal N}=1$. Then the string tension
$T$ contributes to the energy of a stretched string as
\beq
m_{ij}^{(0)}=\frac{R'}{2\pi\alpha'}\,\Bigl|\theta_i-\theta_j\Bigr|
=T\cdot L_{ij} \ ,
\label{mij0}\eeq
so that the mass is proportional to the distance $L_{ij}$ between
hyperplanes $i$ and $j$.

Thus, generically massless states {\it only} arise for non-winding
open strings whose ends lie on the same D-brane $i=j$. There are two
such types of states/vertex operators that may be characterized as
follows:

\noindent
$\bullet~\underline{\alpha_{-1}^\mu|k;ii\rangle~~,~~V=\partial_\parallel
  x^\mu:}$

\noindent
These states correspond to a gauge field $A_\mu(\xi^a)$ on the D-brane
with $p+1$ coordinates tangent to the hyperplane, where
$\mu,a=0,1,\dots,p$ and $\xi^\mu=x^\mu$ are coordinates on the D-brane
worldvolume (in the ``static gauge''). It describes the {\it shape} of
the D-brane as a ``soliton'' background, i.e. as a fixed topological
defect in spacetime~\cite{DKL1}. The quanta of $A_\mu(\xi^a)$ describe the
fluctuations of that background.

\noindent
$\bullet~\underline{\alpha_{-1}^m|k;ii\rangle~~,~~V=\partial_\parallel
  x^m=\partial_\perp x'^{\,m}:}$

\noindent
These states correspond to scalar fields $\Phi^m(\xi^a)$,
$m=p+1,\dots,9$ which originate from the gauge fields in the compact
dimensions of the original string theory, and which give the transverse
position of the D-brane in the compact dualizing directions. They
describe the {\it shape} of the D-brane as it is embedded in
spacetime, analogously to the string embedding coordinates
$x^\mu(\tau,\sigma)$.

Thus a flat hyperplane in spacetime has fluctuations described by a
certain open string state which corresponds to a {\it gauge
  field}. This gives a remarkable description of D-branes in terms of
{\it gauge theory}. These facts are the essence of the gauge
theory/geometry correspondences provided by string theory, such as the
AdS/CFT correspondence~\cite{AdSCFTRev}. We can in this way describe
D-branes using the wealth of experience we have of working with gauge
theory, which is the path we shall mostly take in the
following. Conversely, it is hoped that D-branes and the string
theories in which they live can teach us a lot about gauge theories,
such as their strong coupling expansions.

\subsubsection*{Non-Abelian Gauge Symmetry}

We have seen above that if none of the D-branes coincide,
$\theta_i\neq\theta_j$ for $i\neq j$, then there is a single massless
vector state $\alpha_{-1}^M|k;ii\rangle$ associated to each individual
D-brane $i=1,\dots,N$. Together, these states describe a gauge theory
with abelian gauge group $U(1)^N$, which is the {\it generic} unbroken
symmetry group of the problem. But suppose now that $k\leq N$ D-branes
coincide, say
\beq
\theta_1=\theta_2=\dots=\theta_k=\theta \ .
\label{thetakcoinc}\eeq
Then from (\ref{mij0}) it follows that $m_{ij}^{(0)}=0$ for $1\leq
i,j\leq k$. Thus new massless states appear in the spectrum of the
open string theory, because strings which are stretched between these
branes can now attain a vanishing length. In all there are $k^2$
massless vector states which, by the transformation properties of
Chan-Paton wavefunctions, form the adjoint representation of a $U(k)$
gauge group. In the {\it original} open string theory, the coincident
position limit (\ref{thetakcoinc}) corresponds to the Wilson line
\beq
W=\pmatrix{\e^{\ii\,\theta}\,\id_k& & &0\cr &\e^{\ii\,\theta_{k+1}}& &
  \cr & &\ddots& \cr0& & &\e^{\ii\,\theta_N}\cr} \ ,
\label{Wilsoncoinc}\eeq
where $\id_k$ denotes the $k\times k$ identity matrix. The background
field configuration leaves unbroken a $U(k)\subset U(N)$ subgroup,
acting on the upper left $k\times k$ block of
(\ref{Wilsoncoinc}). Thus the D-brane worldvolume now carries a $U(k)$
gauge field $\alpha_{-1}^\mu|k;ij\rangle\leftrightarrow
A_\mu(\xi^a)_{ij}$ and, at the same time, a set of $k^2$ massless
scalar fields
$\alpha_{-1}^m|k;ij\rangle\leftrightarrow\Phi^m(\xi^a)_{ij}$, where
$i,j=1,\dots,k$.

The geometrical implications of this non-abelian $U(k)$ symmetry are
rather profound \cite{Witten2}. The $k$ D-brane positions in spacetime
are promoted to a {\it matrix} $\Phi^m(\xi^a)$ in the adjoint
representation of the unbroken $U(k)$ gauge group. This is a curious
and exotic feature of D-brane dynamics that is difficult to visualize,
and is a consequence of the somewhat surprising aspects of string
geometry that were mentioned at the end of section~\ref{TClosed}. It
simply reflects the fact that the T-dual string theory rewrites the
$R\ll\ell_s$ limit of the original open string theory in terms of {\it
  non-commuting}, matrix-valued spacetime coordinates. These are the
variables which are more natural to use as $R\to0$, and
various puzzling features of this limit become clearer in the T-dual
picture. This illustrates once again how spacetime geometry is
significantly altered by strings at very short distances.

Generally, the gauge symmetry of the theory is broken to
the subgroup of $U(N)$ which commutes with the Wilson line $W$. But if
{\it all} $N$ D-branes coincide, then $W$ belongs to the center of
the Chan-Paton gauge group, and we recover the original $U(N)$ gauge
symmetry. Since $\Phi^m$ is a Hermitian matrix, we can diagonalize it
by a $U(N)$ gauge transformation and write
\beq
\Phi^m=U^m~\pmatrix{\phi_1^m& &0\cr &\ddots& \cr0& &\phi_N^m\cr}~
U^{m\,\dag} \ .
\label{Phimdiag}\eeq
The real-valued eigenvalue $\phi_i^m$ describes the classical position
of D-brane $i$ and corresponds to the ground state of the system of
$N$ D-branes. The $N\times N$ unitary matrices $U^m$ describe the
fluctuations $U^m_{ij}$, $i\neq j$ about classical spacetime, and they
arise from the short open strings connecting D-branes $i$ and $j$. In
this way the off-diagonal elements $\Phi^m_{ij}$, $i\neq j$ may be
thought of as ``Higgs fields'' for the symmetry breaking
mechanism~\cite{Witten2}. The $U(N)$ symmetry is broken when some (or
all) of the D-branes separate, leaving a set of {\it massive} fields, with mass
equal to that of the stretched open strings. We will see precisely how
this works dynamically in the next section.

\subsection{The Born-Infeld Action \label{BI}}

We have argued that, associated to any configuration of D-branes, there
correspond dynamical gauge fields living on the worldvolumes. At this stage it
is natural to ask what sort of gauge theory describes their
dynamics. In this subsection we will begin deriving the appropriate
actions which describe the (low-energy) dynamics of D-branes in
Type~II superstring theory. They will govern the worldvolume dynamics
of the gauge fields, the transverse scalar fields, and eventually also
the Ramond-Ramond form potentials which couple electrically to the
D-brane worldvolume as in (\ref{pbranemincoupling}). In this section
we will consider the really relevant situation that led to the gauge
theoretic description above, namely the coupling of free open strings
to a background photon field $A_\mu$ of constant field strength
\beq
F_{\mu\nu}=\partial_\mu A_\nu-\partial_\nu A_\mu \ .
\label{constfieldstrength}\eeq
Equivalently, we may regard the open strings as being attached to a
spacetime-filling D9-brane. The corresponding result for arbitrary
dimension D-branes will be derived in the next section using
T-duality.

We will work at tree-level in open string perturbation theory, and
therefore calculate the disc diagram (fig.~\ref{4string}). Using conformal
invariance we may set the radius of the disc to unity. The complex
coordinates on the disc will be written in the polar decomposition
\beq
z=r~\e^{\ii\,\theta} \ ,
\label{polardecomp}\eeq
where $0\leq\theta<2\pi$ and $0\leq r\leq1$.

\bigskip

\begin{center}
\begin{minipage}{15cm}

\small

{\bf Exercise~7.2.} {\sl\baselineskip=12pt {\bf (a)} Show that the
  solution to the Neumann problem on the disc,
$$
\begin{array}{rll}
\Delta N(z,z')~\equiv~\partial_z\,\partial_{\overline{z}}N(z,z')&=&
\delta(z-z') \ , \nn\\\displaystyle\left.\frac\partial{\partial r}N(z,z')
\right|_{r=1}&=&0 \ ,
\end{array}
$$
is given by
$$
N(z,z')=\frac1{2\pi}\,\ln\left(\,\Bigl|z-z'\Bigr|\,\Bigl|z-
\overline{z}^{\,\prime\,-1}\Bigr|\,\right) \ .
$$
[Hint: Use a conformal transformation to map the disc onto the upper
complex half-plane, and hence apply the method of images.]

\noindent
{\bf (b)} Show that on the boundary of the disc this Green's function
can be written as
$$
N(\e^{\ii\,\theta},\e^{\ii\,\theta'})=-\frac1\pi\,\sum_{n=1}^\infty
\frac{\cos\Bigl(n(\theta-\theta')\Bigr)}n \ .
$$
}

\end{minipage}
\end{center}

\bigskip

The effective bosonic string action in the conformal gauge is given by
\beq
S[x,A]=\frac1{4\pi\alpha'}\,\int\dd^2z~\partial_zx^\mu\,
\partial_{\overline{z}}x_\mu-\ii\,\int\limits_0^{2\pi}\dd\theta~
\dot x^\mu\,A_\mu\Bigm|_{r=1} \ ,
\label{SxA}\eeq
where
\beq
\dot x^\mu(\theta)\equiv\frac{\partial x^\mu}{\partial\theta}
\label{dotxtheta}\eeq
and we have used the standard minimal coupling of the point particle
at the string endpoint. We are interested in evaluating the
gauge-fixed, Euclidean Polyakov path integral
\beq
Z[F]=\frac1{g_s}\,\int\cD x^\mu~\e^{-S[x,A]} \ ,
\label{ZFpathint}\eeq
where the inverse power of string coupling indicates that we are
evaluating a tree-level diagram, and there are no moduli on the
disc. By gauge invariance, the result of this functional integration
should depend only on the field strength
(\ref{constfieldstrength}). Note that the path integral
(\ref{ZFpathint}) can be interpreted as that in
(\ref{stringscattschem}) with the appropriate vertex operator (Wilson
line) insertion. We interpret it as the effective gauge field action
induced on the D9-brane by integrating out all of the open string
modes.\footnote{Note that the Polyakov path integral computes directly
  the vacuum energy~\cite{CLNY1}. The string partition function is
  quite different from that of quantum field theory, in that it is
  more like an S-matrix.}

To compute (\ref{ZFpathint}), we will use the usual ``background field
gauge'' of quantum field theory, in which the string embedding
coordinates are expanded as
\beq
x^\mu=x_0^\mu+\xi^\mu \ ,
\label{backgauge}\eeq
where $x_0^\mu$ are the constant, worldsheet zero modes of $x^\mu$ on
the disc. We will also work in the ``radial gauge'' for the gauge
field background,
\beq
\xi^\mu\,A_\mu(x_0+\xi)=0 \ , ~~ A_\mu(x_0)=0 \ ,
\label{radialgauge}\eeq
and with slowly-varying vector potentials,
\beq
A_\mu(x_0+\xi)=\frac12\,F_{\mu\nu}(x_0)\,\xi^\nu+{\cal O}(\partial F) \ .
\label{derivexpF}\eeq
In other words, we evaluate the path integral (\ref{ZFpathint}) to
leading orders in a derivative expansion in the field strength
$F_{\mu\nu}$ (which essentially means that we work with constant~$F_{\mu\nu}$).

The path integral measure can be decomposed using (\ref{backgauge}) in
terms of bulk and boundary integrations over the disc,
i.e. schematically we have
\beq
\cD x^\mu=\prod_{z\in{\rm interior}}\cD
x^\mu(z,\overline{z})~\prod_{\theta\in{\rm boundary}}\cD\xi^\mu(\theta) \ .
\label{bulkbdryschem}\eeq
The (Gaussian) bulk integration in the
interior of the disc just produces some (functional) normalization
factor, corresponding to the closed string sector, which is independent
of $A_\mu$. This can be absorbed into an irrelevant normalization of
(\ref{ZFpathint}). Then, integrating out the bulk modes leaves a
boundary path integral
\beq
Z[F]=\frac1{g_s}\,\int\dd\vec x_0~\int\cD\xi^\mu(\theta)~\e^{-S_{\rm
    b}[\xi,A]} \ ,
\label{bdrypathint}\eeq
where the boundary action is given by
\beq
S_{\rm b}[\xi,A]=\frac12\,\int\limits_0^{2\pi}\dd\theta~\left(
\frac1{2\pi\alpha'}\,\xi^\mu\,N^{-1}\,\xi_\mu+\ii\,F_{\mu\nu}\,
\xi^\mu\,\dot\xi^\nu\right)
\label{bdryaction}\eeq
and $N^{-1}$ is the coordinate space inverse of the boundary Neumann
function given in exercise~7.2. By using the Fourier completeness relation
\beq
\frac1\pi\,\sum_{n=1}^\infty\cos\Bigl(n(\theta-\theta')\Bigr)=
\delta(\theta-\theta')-\frac1{2\pi}
\label{coscomplete}\eeq
for $0\leq\theta,\theta'\leq2\pi$, and exercise 7.2~(b), one can
easily compute
\beq
N^{-1}(\theta,\theta')=-\frac1\pi\,\sum_{n=1}^\infty n\cos
\Bigl(n(\theta-\theta')\Bigr) \ .
\label{Ninverse}\eeq
We will expand the non-constant string modes $\xi^\mu(\theta)$ in
periodic Fourier series on the circle:
\beq
\xi^\mu(\theta)=\sum_{n=1}^\infty\Bigl(a_n^\mu\,\cos(n\theta)+
b_n^\mu\,\sin(n\theta)\Bigr) \ .
\label{xiFourier}\eeq
Then the Feynman measure in (\ref{bdrypathint}) can be expressed in
terms of Fourier modes as
\beq
\cD\xi^\mu(\theta)=\prod_{n=1}^\infty\dd a_n^\mu~\dd b_n^\mu \ .
\label{FeynmanFourier}\eeq

We will now use Lorentz-invariance to simplify the form of the action
(\ref{bdryaction}). The {\it anti-symmetric} $10\times10$ matrix
$(F_{\mu\nu})$ cannot be diagonalized like a symmetric matrix, but it
can be rotated into its canonical Jordan normal form
\beq
(F_{\mu\nu})=\pmatrix{0&-f_1& & &0\cr f_1&0& & & \cr & &\ddots& & \cr
 & & &0&-f_5\cr0& & &f_5&0\cr} \ ,
\label{Fcanform}\eeq
where the ``skew-diagonal'' blocks contain the real
``skew-eigenvalues'' $f_l$, $l=1,\dots,5$ of $(F_{\mu\nu})$. The path
integral (\ref{bdrypathint},\ref{FeynmanFourier}) factorizes in this
basis into a product of five independent functional Gaussian
integrations over the pairs of coordinate modes $a_n^{2l-1},a_n^{2l}$
and $b_n^{2l-1},b_n^{2l}$, where $l=1,\dots,5$. By substituting
(\ref{Ninverse}) and (\ref{xiFourier}) into (\ref{bdryaction}), and
using standard Fourier properties of the orthogonal trigonometric
functions appearing, we find that, for each $n$ and $l$, the Gaussian
Boltzmann weight has rank~2 quadratic form
\beq
\frac12\,\frac1{2\pi n}\,\frac1{2\pi\alpha'}\,
\Bigl(a_n^{2l-1}~,~a_n^{2l}\Bigr)~\pmatrix{1&-2\pi\alpha'\,f_l\cr
2\pi\alpha'\,f_l&1\cr}~\pmatrix{a_n^{2l-1}\cr a_n^{2l}\cr} \ ,
\label{Gaussweight}\eeq
plus an analogous term for the $b_n$'s. Integrating over each of the
$a_n$'s and $b_n$'s thereby yields (up to irrelevant constants)
\beq
Z[F]=\frac1{g_s}\,\int\dd\vec x_0~\prod_{l=1}^5Z_{2l-1,2l}[f_l] \ ,
\label{ZFabint}\eeq
where
\beq
Z_{2l-1,2l}[f_l]=\prod_{n=1}^\infty\left\{\left(4\pi^2\alpha'\,n\right)^2
\,\Bigl[1+(2\pi\alpha'f_l)^2\Bigr]^{-1}\right\}
\label{functfluctdet}\eeq
is the functional fluctuation determinant arising from the two copies
of the Gaussian form~(\ref{Gaussweight}).

To deal with the infinite products in (\ref{functfluctdet}), we note
first of all that $\prod_{n=1}^\infty n^2$ diverges. However, it can
be regulated by introducing a worldsheet ultraviolet cutoff and
thereby absorbing it into an (infinite) renormalization of the string
coupling constant $g_s$. This divergence is due to the {\it tachyon}
mode of the bosonic string, and it originates in the
$\theta\to\theta'$ divergence of (\ref{Ninverse}). It can therefore be
removed by introducing worldsheet supersymmetry, and hence will simply
be dropped in what follows. The other infinite product in
(\ref{functfluctdet}) is independent of $n$ and can be evaluated by
using ``zeta-function regularization'' to write
\beq
\prod_{n=1}^\infty\,c=c^{\,\zeta(0)} \ ,
\label{zetafnreg}\eeq
where $\zeta(z)$ is the Riemann zeta-function (\ref{zetafn}). Using
(\ref{zeta01}) thereby yields the finite answer
\beq
Z_{2l-1,2l}[f_l]=\frac1{4\pi^2\alpha'}\,\sqrt{1+(2\pi\alpha'f_l)^2} \ .
\label{Z2l}\eeq
Finally, we rotate the field strength tensor $F_{\mu\nu}$ back to
general form to produce a Lorentz-invariant result. In this way we
have found that the partition function (\ref{ZFpathint}) gives the
``open string effective action''~\cite{CLNY1,FT1}
\beq
\begin{tabular}{|c|}\hline\\
$\displaystyle
Z[F]=\frac1{(4\pi^2\alpha')^5\,g_s}\,\int\dd\vec
x_0~\sqrt{\det_{\mu,\nu}
\left(\eta_{\mu\nu}+2\pi\alpha'\,F_{\mu\nu}\right)} \ ,
$\\\\
\hline\end{tabular}
\label{BornInfeld}\eeq
where we have recalled the origin of the background field dependent
terms in (\ref{Z2l}) as the determinant of the quadratic form in
(\ref{Gaussweight}).

The string theoretic action (\ref{BornInfeld}) is {\it exact} in
$\alpha'$, which is in fact the coupling constant of the original
two-dimensional worldsheet field theory defined by (\ref{SxA}). Thus
this result is non-perturbative at the level of the theory on the
disc. In other words, the action (\ref{BornInfeld}) is a truly
``stringy'' result, containing contributions from {\it all} massive
and massless string states. Remarkably, it actually dates back to
1934, and is known as the ``Born-Infeld action''~\cite{BI1}. This
model of non-linear electrodynamics was originally introduced to
smoothen out the singular electric field distributions generated by
point charges in ordinary Maxwell electrodynamics, thereby yielding a
finite total energy. This is quite unlike the situation in Maxwell
theory, where the field of a point source is singular at the origin
and its energy is infinite. Here the effective distribution of the
field has radius of order the string length $\ell_s$, and the
delta-function singularity is smeared away. It is truly remarkable how
string theory captures and revamps this model of non-linear
electrodynamics.\footnote{\baselineskip=12pt In the exercise below,
  the second equation for the fermionic Green's function defines
  anti-periodic boundary conditions. Periodic boundary conditions
  would produce a vanishing functional integral due to the Ramond zero
  modes.}

\bigskip

\begin{center}
\begin{minipage}{15cm}

\small

{\bf Exercise~7.3.} {\sl\baselineskip=12pt In this exercise
  you will generalize the above derivation to the case of the
  superstring.

\noindent
{\bf (a)} Show that by augmenting the bosonic action (\ref{SxA}) by
the fermionic action
$$
S_{\rm ferm}[\psi,A]=\frac\ii{4\pi\alpha'}\,\int\dd^2z~\left(\,
\overline{\psi}\cdot\partial_z\overline{\psi}+\psi\cdot
\partial_{\overline{z}}\psi\right)-\frac\ii2\,\int\limits_0^{2\pi}
\dd\theta~\psi^\mu\,F_{\mu\nu}\,\psi^\nu\Bigm|_{r=1}
$$
with $\overline{\psi}^{\,\mu}$ and $\psi^\mu$ independent complex fermion
fields in the bulk of the disc, the total action is invariant under
worldsheet supersymmetry transformations.}
\end{minipage}
\end{center}

\begin{center}
\begin{minipage}{15cm}
\small
\noindent
{\sl {\bf (b)} Show that the fermionic Green's function on the disc,
defined by
$$
\begin{array}{rll}
\partial_zK(z,z')&=&\delta(z-z') \ , \nn\\
K(\e^{2\pi\,\ii}\,z,z')\Bigm|_{r=1}&=&-K(z,z')\Bigm|_{r=1} \ ,
\end{array}
$$
can be written on the boundary of the disc as
$$
K(\e^{\ii\,\theta},\e^{\ii\,\theta'})=-\frac1\pi\,
\sum_r\sin\Bigl(r(\theta-\theta')\Bigr) \ , ~~
r=\frac12\,,\,\frac32\,,\,\dots \ .
$$

\noindent
{\bf (c)} Show that the superstring path integral yields the {\it
  same} Born-Infeld action (\ref{BornInfeld}), and at the same time
  removes the tachyonic divergence. You will need to use the
  ``generalized zeta-function''~\cite{GradRhy1}
$$
\zeta(z,a)=\sum_{n=0}^\infty\frac1{(n+a)^z}
$$
with
$$
\zeta(0,0)=-\frac12 \ , ~~ \zeta(0,1/2)=0 \ .
$$
}

\end{minipage}
\end{center}

\bigskip

\setcounter{equation}{0}

\section{D-Brane Dynamics \label{DBraneDyn}}

In this final section we will describe various aspects of the dynamics
of D-branes in the low-energy limit. We will start from the
Born-Infeld action (\ref{BornInfeld}) and work out its extensions to
D$p$-branes with $p<9$, some of their physical properties, and how
they couple to the spacetime supergravity fields of the closed string
sector. This will then lead us into a description of the dynamics of
D-branes in terms of supersymmetric Yang-Mills theory, a more familiar quantum
field theory which has sparked the current excitement over the D-brane/gauge
theory correspondence, and which unveils some surprising features of D-brane
physics. Finally, we will give an elementary calculation of
the interaction energy between two separated D-branes, and thereby
illustrate the role played by supersymmetry in D-brane dynamics.

\subsection{The Dirac-Born-Infeld Action \label{DBI}}

In the previous section we saw that the low-energy dynamics of a D9-brane,
induced by the quantum theory of the open strings attached to it, is governed
by the Born-Infeld action
\beq
S_{\rm BI}=\frac1{(4\pi^2\alpha')^5\,g_s}\,\int\dd^{10}x~
\sqrt{-\det_{\mu,\nu}\left(\eta_{\mu\nu}+T^{-1}\,F_{\mu\nu}\right)} \ ,
\label{SBI}\eeq
where we have Wick rotated back to Minkowski signature. Here
$T=\frac1{2\pi\alpha'}$ is the string tension and
$F_{\mu\nu}=\partial_\mu A_\nu-\partial_\nu A_\mu$ is the field
strength of the gauge fields living on the D9-brane worldvolume. This
low-energy approximation to the full dynamics is good in the static
gauge and for slowly-varying field strengths. In general there are
derivative corrections from $F_{\mu\nu}$, but (\ref{SBI}) is
nevertheless the {\it exact} result as a function of $\alpha'$. In
this subsection we will obtain the general form of the worldvolume
action for D$p$-branes (in static gauge) from (\ref{SBI}) in terms of
their low-energy field content, i.e. the gauge fields $A_\mu$ and
scalar fields $\Phi^m$, by using T-duality. Because non-linear
electrodynamics is not a very familiar subject to most, let us begin
with the following exercise to become acquainted with some of its
novel features.

\bigskip

\begin{center}
\begin{minipage}{15cm}

\small

{\bf Exercise~8.1.} {\sl\baselineskip=12pt {\bf (a)} Show that the equations of
motion which follow from the Born-Infeld action are given by
$$
\left(\frac1{\id-(T^{-1}\,F)^2}\right)^{\nu\lambda}\,
\partial_\nu F_{\lambda\mu}=0 \ .
$$
They reduce to the usual Maxwell equations in the field theory limit
$\alpha'\to0$ which decouples all massive string modes.

\noindent
{\bf (b)} Show that in $d=4$ dimensions the Born-Infeld action can be written
in the form
$$
S_{\rm BI}^{(d=4)}=\frac1{(4\pi^2\alpha')^5\,g_s}\,\int\dd^4x~
\sqrt{1+\frac1{2T^2}\,F_{\mu\nu}F^{\mu\nu}-\frac1{16T^2}\,
\Bigl(F_{\mu\nu}\tilde F^{\mu\nu}\Bigr)^2} \ ,
$$
where $\tilde
F^{\mu\nu}=\frac12\,\epsilon^{\mu\nu\lambda\rho}\,F_{\lambda\rho}$. In
this sense the Born-Infeld action interpolates between the Maxwell
form $\frac14\,F_{\mu\nu}F^{\mu\nu}$ for small $F$ and the topological
density $\frac14\,F_{\mu\nu}\tilde F^{\mu\nu}$ for large $F$.}

\end{minipage}
\end{center}

\begin{center}
\begin{minipage}{15cm}

\small

{\sl\baselineskip=12pt {\bf (c)} Show that in four spacetime dimensions the
Born-Infeld electric field generated by a point charge $Q$ at the origin is
given by
$$
E_r=F_{rt}=\frac Q{\sqrt{r^4+r_0^2}} \ , ~~ r_0^2=\frac QT \ .
$$
Thus the distribution $\rho=\frac1{4\pi}\,\nabla\cdot\vec E$ of the electric
field has an effective radius $r_0\propto\ell_s$.
}

\end{minipage}
\end{center}

\bigskip

To transform the Born-Infeld action (\ref{SBI}) to an action for a D$p$-brane
with $p<9$, we T-dualize $9-p$ of the ten spacetime directions. Then the $9-p$
directions are described by {\it Dirichlet} boundary conditions for the open
strings, which thereby sit on a $p+1$ dimensional worldvolume hyperplane. We
assume that the normal directions $x^m$, $m=p+1,\dots,9$ are circles which are
so small that we can neglect all derivatives along them. The remaining
uncompactified worldvolume directions are $x^a$, $a=0,1,\dots,p$.

\bigskip

\begin{center}
\begin{minipage}{15cm}

\small

{\bf Exercise~8.2.} {\sl\baselineskip=12pt If $\cal M$ and $\cal N$ are
invertible $p\times p$ and $q\times q$ matrices, respectively, and $\cal A$ is
$p\times q$, show that
$$
\begin{array}{rrl}
\displaystyle\det\pmatrix{{\cal N}&-{\cal A}^\top\cr{\cal A}&{\cal M}\cr}
&=&\det({\cal M})\,\det\left({\cal N}+{\cal A}^\top\,{\cal M}^{-1}\,{\cal A}
\right)\\&=&\det({\cal N})\,\det\left({\cal M}+{\cal A}\,{\cal N}^{-1}\,
{\cal A}^\top\right) \ .
\end{array}
$$
}

\end{minipage}
\end{center}

\bigskip

To expand the determinant appearing in (\ref{SBI}), we apply the determinant
formula of exercise~8.2 with
\bea
{\cal N}&=&(\eta_{ab}+2\pi\alpha'\,F_{ab}) \ , \nn\\
{\cal M}&=&(\delta_{mn}) \ , \nn\\{\cal A}&=&(2\pi\alpha'\,
\partial_aA_m) \ ,
\label{detformulawith}\eea
and use the T-duality rules to replace gauge fields in the T-dual directions by
brane coordinates according to
\beq
2\pi\alpha'\,A_m=x^m \ .
\label{Tdualreplace}\eeq
By worldvolume and spacetime reparametrization invariance of the theory, we may
choose the ``static gauge'' in which the worldvolume is aligned with the first
$p+1$ spacetime coordinates, leaving $9-p$ transverse coordinates. This amounts
to calling the $p+1$ brane coordinates $\xi^a=x^a$, $a=0,1,\dots,p$. In this
way the Born-Infeld action can be thereby written as
\beq
\begin{tabular}{|c|}\hline\\
$\displaystyle
S_{\rm DBI}=-\frac{T_p}{g_s}\,\int\dd^{p+1}\xi~\sqrt{-
\det_{0\leq a,b\leq p}\left(\eta_{ab}+\partial_ax^m\,\partial_bx_m
+2\pi\alpha'\,F_{ab}\right)} \ .
$\\\\
\hline\end{tabular}
\label{DBIaction}\eeq
This is known as the ``Dirac-Born-Infeld action'' and it describes a
model of non-linear electrodynamics on a fluctuating
$p$-brane~\cite{Leigh1}. The quantity
\beq
\begin{tabular}{|c|}\hline\\
$\displaystyle
T_p=\frac1{\sqrt{\alpha'}}\,\frac1{\left(2\pi\,\sqrt{\alpha'}\,\right)^p}
$\\\\
\hline\end{tabular}
\label{tensionp}\eeq
has dimension mass/volume and is the tension of the $p$-brane, generalizing the
$p=1$ string tension $T$. The tension formula (\ref{tensionp}) plays a pivotal
role in the dynamics of D-branes, as will be discussed in section~\ref{Forces}.

To understand the meaning of the action (\ref{DBIaction}), let us consider the
case where there is no gauge field on the D$p$-brane, so that $F_{ab}\equiv0$.
Then the Dirac-Born-Infeld action reduces to
\beq
S_{\rm DBI}(F=0)=-\frac{T_p}{g_s}\,\int\dd^{p+1}\xi~\sqrt{-\det_{a,b}\left(
-\eta_{\mu\nu}\,\partial_ax^\mu\,\partial_bx^\nu\right)} \ ,
\label{SDBIF0}\eeq
where we have rewritten the argument of the static gauge determinant argument
in (\ref{DBIaction}) in covariant form. The tensor field
$h_{ab}=-\eta_{\mu\nu}\,\partial_ax^\mu\,\partial_bx^\nu$ is the induced metric
on the worldvolume of the D$p$-brane, so that the integrand of (\ref{SDBIF0})
is the invariant, infinitesimal volume element on the D-brane hypersurface.
Thus (\ref{SDBIF0}) is just the $p$-brane generalization of the actions we
encountered in section~\ref{Classical} for a massive point particle and a
string of tension $T$. So the Dirac-Born-Infeld action is the natural geometric
extension, incorporating the worldvolume gauge fields, of the string Nambu-Goto
action to the case of D-branes.

\subsubsection*{Example}

To illustrate the utility of describing effects in string theory by
using T-duality, and to further give a nice physical origin to the
Dirac-Born-Infeld action, let us consider now the example of electric
fields in string theory, which have many exotic properties that find
their most natural dynamical explanations in the T-dual D-brane
picture~\cite{AMSS1}. A pure electric background is specified by the
field strength tensor
\beq
F_{0i}=E_i \ , ~~ F_{ij}=0 \ ,
\label{pureelectric}\eeq
where $i,j=1,\dots,9$. The Born-Infeld action (\ref{SBI}) then essentially only
involves a simple $2\times2$ determinant, and it takes the particularly simple
form
\beq
S_{\rm BI}(E)=\frac1{g_s}\,\left(\frac T{2\pi}\right)^5\,\int\dd^{10}x~
\sqrt{1-\left(T^{-1}\,\vec E\right)^2} \ .
\label{SBIE}\eeq

{}From (\ref{SBIE}) we see that, at the origin of the source for $\vec E$, the
electric field attains a {\it maximum} value
\beq
E_c=T=\frac1{2\pi\alpha'} \ .
\label{Ecrit}\eeq
This limiting value arises because for $|\vec E|>E_c$ the action
(\ref{SBIE}) becomes complex-valued and ceases to make physical
sense~\cite{FT1}. It represents an instability in the system,
reflecting the fact that the electromagnetic coupling of open strings
is not minimal and creates a divergence due to the fast rising density
of string states. Heuristically, since the string effectively carries electric
charges of equal sign at each of its endpoints, as $|\vec E|$
increases the charges start to repel each other and stretch the
string. For field strengths larger than the critical value
(\ref{Ecrit}), the string tension $T$ can no longer hold the strings
together. Note that this instability may be attributed to the
Minkowski sign factor of the time direction of the metric
$\eta_{\mu\nu}$, and hence it does not arise in a purely magnetic
background~\cite{AMSS1}.

The fact that electric fields in string theory are not completely arbitrary,
because they have a limiting value above which the system becomes unstable, is
actually {\it very} natural in the T-dual D-brane picture. For this, let us
consider the simplest case of the coupling of an open string to a time-varying
but spatially constant electric field $\vec E=\partial_0\vec A$. The worldsheet
action in the Neumann picture is
\beq
S_{\rm N}=\frac1{4\pi\alpha'}\,\int\dd^2\xi~\partial_ax^\mu\,
\partial^ax_\mu+\ii\,\int\dd l~\vec A(x^0)\cdot\partial_\parallel
\vec x \ ,
\label{SNE}\eeq
where all spacetime coordinate functions $x^\mu$ obey Neumann boundary
conditions. As we have seen, T-dualizing the nine space directions maps the
vector potential $\vec A$ onto the trajectory $\vec y$ of a D-particle and
(\ref{SNE}) into the Dirichlet picture action
\beq
S_{\rm D}=\frac1{4\pi\alpha'}\,\int\dd^2\xi~\partial_ax'^{\,\mu}\,
\partial^ax'_\mu+\frac1{2\pi\alpha'}\,\int\dd l~\vec y(x^0)\cdot
\partial_\perp\vec x^{\,\prime} \ ,
\label{SDE}\eeq
where $x'^{\,0}=x^0$ still obeys Neumann boundary conditions, while $x'^{\,i}$
obey Dirichlet boundary conditions. The boundary vertex operator in (\ref{SDE})
creates a moving D0-brane which travels with velocity
\beq
\vec v=\partial_0\vec y=2\pi\alpha'\,\vec E \ .
\label{D0velocity}\eeq

In string perturbation theory, the equivalence of the electric field and moving
D-brane problems follows from the perturbative duality between Neumann and
Dirichlet boundary conditions for the open strings. This is reflected in the
equality of the corresponding boundary propagators (see exercise~7.2)
\bea
\Bigl\langle\partial_{\sigma_1}x^\mu(\tau_1)\,\partial_{\sigma_2}
x^\nu(\tau_2)\Bigr\rangle_{\rm N}&=&-\,\Bigl\langle\partial_{\tau_1}
x'^{\,\mu}(\tau_1)\,\partial_{\tau_2}x'^{\,\nu}(\tau_2)\Bigr\rangle_{\rm D}
\nn\\&=&\frac{2\alpha'\,\eta^{\mu\nu}}{(\tau_1-\tau_2)^2} \ .
\label{NDpropsequal}\eea
This implies that the open string loop expansions are the same (modulo zero
modes). This is true on the boundary of the disc, but {\it not} in the bulk.
The Born-Infeld action (\ref{SBIE}) then simply maps onto the usual action for
a relativistic point particle (c.f.~(\ref{actionstatic})),
\beq
S_{\rm DBI}(v)=m\int\dd\tau~\sqrt{1-\vec v^{\,2}} \ ,
\label{D0action}\eeq
where
\beq
m=T_0=\frac1{g_s\,\sqrt{\alpha'}}
\label{D0mass}\eeq
is the mass of the D-particle. It follows that, in the dual picture, the
existence of a limiting electric field is merely a consequence of the laws of
relativistic particle mechanics for a 0-brane, with the ``critical'' velocity
$v_c=2\pi\alpha'\,E_c=1$ corresponding to the speed of light. We note the
string coupling dependence of the D0-brane mass (\ref{D0mass}), which reflects
the fact that D-branes are really non-perturbative degrees of freedom in
superstring theory.

At the velocity $v_c$, we can make a large Lorentz boost to bring the system to
rest, so that in the T-dual picture of the original open string theory the
existence of electric fields of strength near (\ref{Ecrit}) amounts to a boost
to large momentum. Thus string theory with electric background near the
critical limit is equivalent to string theory in the infinite momentum frame.
This illustrates the overall ease in which things may be interpreted in D-brane
language. Put differently, demanding that the D0-branes of Type~IIA superstring
theory behave as relativistic particles {\it uniquely} fixes the form of the
Born-Infeld action, which is the result of a resummation of all stringy
$\alpha'$ corrections. This demonstrates the overall consistency of the
Dirac-Born-Infeld action in superstring theory.

\subsubsection*{Supergravity Couplings}

Thus far we have been working in {\it flat} ten dimensional spacetime,
which represents a particular background of string theory, i.e. a
particular solution to the supergravity equations of motion. It is
straightforward, however, to generalize the action (\ref{DBIaction})
to {\it curved} spacetimes, which is tantamount to coupling D-branes
to supergravity fields. For this, we incorporate the massless NS--NS
spacetime fields of the closed string sector, namely the spacetime
metric $g_{\mu\nu}$, the antisymmetric tensor $B_{\mu\nu}$, and the
dilaton $\Phi$. This is done by considering a more general worldsheet
action, corresponding to the couplings of the NS--NS fields to the
fundamental strings, of the form
\beq
S=\frac1{4\pi\alpha'}\,\int\dd^2\xi~\left(g_{\mu\nu}\,\partial_ax^\mu\,
\partial^ax^\nu+2\pi\alpha'\,B_{\mu\nu}\,\epsilon^{ab}\,\partial_ax^\mu\,
\partial_bx^\nu+\alpha'\,\Phi\,R^{(2)}\right) \ ,
\label{Sworldsheetgen}\eeq
where $R^{(2)}$ is the (scalar) curvature of the two-dimensional worldsheet.

If $B_{\mu\nu}$ is constant, then the familiar $B$-field coupling in
(\ref{Sworldsheetgen}) is actually a {\it boundary} term on the disc, since it
can then be integrated by parts to give
\beq
\int\dd^2\xi~B_{\mu\nu}\,\epsilon^{ab}\,\partial_ax^\mu\,
\partial_bx^\nu=\int\limits_0^{2\pi}\dd\theta~\frac12\,B_{\mu\nu}\,
x^\nu\,\dot x^\mu\Bigm|_{r=1} \ .
\label{Btermbdry}\eeq
When the electromagnetic coupling in (\ref{SxA}) with the gauge choice
(\ref{derivexpF}) is included, the effect of such a $B$-field term is to shift
the field strength $F_{\mu\nu}$ to
\beq
{\cal F}_{\mu\nu}\equiv2\pi\alpha'\,F_{\mu\nu}-B_{\mu\nu}=2\pi\alpha'\,
(\partial_\mu A_\nu-\partial_\nu A_\mu)-B_{\mu\nu} \ .
\label{Fshift}\eeq
This modification of the $B$-field is actually required to produce an action
which is invariant under the gauge transformations of the antisymmetric tensor
field in (\ref{gBgaugetransf}), which can be absorbed by shifting the vector
potential as
\beq
A_\mu~\longmapsto~A_\mu+\frac1{2\pi\alpha'}\,\Lambda_\mu \ .
\label{Amushift}\eeq
Thus it is the tensor ${\cal F}_{\mu\nu}$ which is the gauge-invariant quantity
in the presence of background supergravity fields, and not the gauge field
strength $F_{\mu\nu}$.

We can now proceed to perform a derivative expansion of the
corresponding disc partition function in exactly the same way we did
in section~\ref{BI}. In the slowly-varying field approximation, the
modification of the Dirac-Born-Infeld action (\ref{DBIaction}) in
arbitrary background supergravity fields may then be computed to
be~\cite{Leigh1}
\beq
S_{\rm DBI}=-T_p\,\int\dd^{p+1}\xi~\e^{-\Phi}~\sqrt{-\det_{a,b}
\left(g_{ab}+B_{ab}+2\pi\alpha'\,F_{ab}\right)} \ ,
\label{SDBISUGRA}\eeq
where the string coupling is generated through the relation
\beq
\frac1{g_s}=\e^{-\Phi} \ ,
\label{gsPhiDBI}\eeq
while $g_{ab}$ and $B_{ab}$ are the pull-backs of the spacetime
supergravity fields to the D$p$-brane worldvolume. In particular, the
induced worldvolume metric is given by
\bea
g_{ab}(\xi)&=&g_{\mu\nu}\Bigl(x(\xi)\Bigr)\,\frac{\partial x^\mu}
{\partial\xi^a}\,\frac{\partial x^\nu}{\partial\xi^b}\nn\\
&=&\eta_{ab}+\partial_ax^\mu\,\partial_bx_\mu+{\cal O}\Bigl((
\partial x)^4\Bigr) \ ,
\label{gpullback}\eea
where the leading terms in (\ref{gpullback}), which coincide with the
induced metric terms in (\ref{DBIaction}), come from setting
$g_{\mu\nu}=\eta_{\mu\nu}$ in static gauge. As before, $F_{ab}$ is the
field strength of the worldvolume $U(1)$ gauge field $A_a$. The
expression (\ref{SDBISUGRA}) is now the correct form of the
worldvolume action which is spacetime gauge-invariant and also reduces
to the appropriate Nambu-Goto type $p$-brane action (\ref{SDBIF0})
when $B=F=0$. It produces an intriguing mixture of gauge theory and
gravity on D-branes.

\subsection{Supersymmetric Yang-Mills Theory \label{SUSYYM}}

Let us now expand the {\it flat} space ($g_{\mu\nu}=\eta_{\mu\nu}$,
$B_{\mu\nu}=0$) action (\ref{DBIaction}) for slowly-varying fields to
order $F^4$, $(\partial x)^4$. This is equivalent to passing to the
field theory limit $\alpha'\to0$, which is defined precisely by
keeping only degrees of freedom of energy $E\ll\frac1{\sqrt{\alpha'}}$,
that are observable at length scales $L\gg\ell_s$. In this limit, the
infinite tower of massive string states decouples, because such states
have masses $m\sim\frac1{\sqrt{\alpha'}}\to\infty$ and are thereby
energetically unfavourable. Using the formula $\det({\cal
  A})=\e^{\Tr\ln({\cal A})}$, the Dirac-Born-Infeld action can be
written as
\beq
S_{\rm DBI}=-\frac{T_p}{g_s}\,V_{p+1}-\frac{T_p\,(2\pi\alpha')^2}
{4g_s}\,\int\dd^{p+1}\xi~\left(F_{ab}F^{ab}+\frac2{(2\pi\alpha')^2}\,
\partial_ax^m\,\partial^ax_m\right)+{\cal O}\left(F^4\right) \ ,
\label{SDBIsmallF}\eeq
where $V_{p+1}$ is the (regulated) $p$-brane worldvolume. This is the
action for a $U(1)$ gauge theory in $p+1$ dimensions with $9-p$ real
scalar fields $x^m$.

But (\ref{SDBIsmallF}) is just the action that would result from the
dimensional reduction of $U(1)$ Yang-Mills gauge theory (electrodynamics) in
ten spacetime dimensions, which is defined by the action
\beq
S_{\rm YM}=-\frac1{4g_{\rm YM}^2}\,\int\dd^{10}x~F_{\mu\nu}F^{\mu\nu}
\ .
\label{SYMU1}\eeq
Indeed, the ten dimensional gauge theory action (\ref{SYMU1}) reduces
to the expansion (\ref{SDBIsmallF}) of the D$p$-brane worldvolume
action (up to an irrelevant constant) if we take the fields $A_a$ and
$A_m=\frac1{2\pi\alpha'}\,x^m$ to depend {\it only} on the $p+1$ brane
coordinates $\xi^a$, and be independent of the transverse coordinates
$x^{p+1},\dots,x^9$. This requires the identification of the
Yang-Mills coupling constant (electric charge) $g^{~}_{\rm YM}$ as
\beq
\begin{tabular}{|c|}\hline\\
$\displaystyle
g_{\rm
  YM}^2=g_s\,T_p^{-1}\,(2\pi\alpha')^{-2}=\frac{g_s}{\sqrt{\alpha'}}
\,\Bigl(2\pi\,\sqrt{\alpha'}\,\Bigr)^{p-2} \ .
$\\\\
\hline\end{tabular}
\label{YMidentify}\eeq

For multiple D-branes, while one can derive a non-abelian extension of
the Dirac-Born-Infeld action~\cite{Myers1,Tseytlin1}, the technical
details would take us beyond the scope of these lectures. Instead, we
will simply make a concise statement about the dynamics of a system of
coinciding D-branes, which naturally generalizes the above
construction and which also incorporates spacetime
supersymmetry~\cite{Witten2}. This will be sufficient for our purposes
here, and will be treated through the following {\it axiom}:
\begin{center}
\begin{tabular}{|c|}\hline\\
\begin{minipage}{15cm}
{\it\baselineskip=12pt The low-energy dynamics of $N$ parallel, coincident
Dirichlet $p$-branes in flat space is described in static gauge by the
  dimensional reduction to $p+1$ dimensions of ${\cal N}=1$
  supersymmetric Yang-Mills theory with gauge group $U(N)$ in ten
  spacetime dimensions.}
\end{minipage}\\\\
\hline\end{tabular}
\end{center}
The ten dimensional action is given by
\beq
\begin{tabular}{|c|}\hline\\
$\displaystyle
S_{\rm YM}=\frac1{4g_{\rm YM}^2}\,\int\dd^{10}x~\Bigl[
\Tr\left(F_{\mu\nu}F^{\mu\nu}\right)+2\,\ii\,\Tr
\left(\,\overline{\psi}\,\Gamma^\mu\,D_\mu\psi\right)\Bigr] \ ,
$\\\\
\hline\end{tabular}
\label{10DSUSYYM}\eeq
where
\beq
F_{\mu\nu}=\partial_\mu A_\nu-\partial_\nu A_\mu-\ii\,[A_\mu,A_\nu]
\label{UNfieldstrength}\eeq
is the non-abelian field strength of the $U(N)$ gauge field $A_\mu$,
and the action of the gauge-covariant derivative $D_\mu$ is defined by
\beq
D_\mu\psi=\partial_\mu\psi-\ii\,[A_\mu,\psi] \ .
\label{gaugecovderiv}\eeq
Again, $g^{~}_{\rm YM}$ is the Yang-Mills coupling constant, $\Gamma^\mu$ are
$16\times16$ Dirac matrices, and the $N\times N$ Hermitian fermion field $\psi$
is a 16-component Majorana-Weyl spinor of the
Lorentz group $SO(1,9)$ which transforms under the adjoint representation of
the $U(N)$ gauge group. The field theory (\ref{10DSUSYYM}) possesses
eight on-shell bosonic, gauge field degrees of freedom, and eight
fermionic degrees of freedom after imposition of the Dirac equation
$\Dirac\psi=\Gamma^\mu\,D_\mu\psi=0$.

\bigskip

\begin{center}
\begin{minipage}{15cm}

\small

{\bf Exercise~8.3.} {\sl\baselineskip=12pt Show that the action
(\ref{10DSUSYYM}) is invariant under the supersymmetry transformations
$$
\begin{array}{rrl}
\delta_\epsilon A_\mu&=&\displaystyle\frac\ii2\,\overline{\epsilon}\,
\Gamma_\mu\psi \ , \nn\\\delta_\epsilon\psi&=&\displaystyle-\frac12\,
F_{\mu\nu}\,[\Gamma^\mu,\Gamma^\nu]\epsilon \ ,
\end{array}
$$
where $\epsilon$ is an infinitesimal Majorana-Weyl spinor.
}

\end{minipage}
\end{center}

\bigskip

Using (\ref{10DSUSYYM}) we can construct a supersymmetric Yang-Mills
gauge theory in $p+1$ dimensions, with 16 independent supercharges, by
dimensional reduction, i.e. we take all fields to be independent of
the coordinates $x^{p+1},\dots,x^9$. Then the ten dimensional gauge
field $A_\mu$ splits into a $p+1$ dimensional $U(N)$ gauge field $A_a$
plus $9-p$ Hermitian scalar fields $\Phi^m=\frac1{2\pi\alpha'}\,x^m$
in the adjoint representation of $U(N)$. The D$p$-brane action is
thereby obtained from the dimensionally reduced field theory as
\beq
\begin{tabular}{|c|}\hline\\
$\begin{array}{rrl}
S_{{\rm D}p}&=&\displaystyle-\frac{T_p\,g_s\,(2\pi\alpha')^2}4\,
\int\dd^{p+1}\xi~\Tr\Biggl(F_{ab}F^{ab}+2D_a\Phi^m\,D^a\Phi_m\Biggr.\nn\\
&&\displaystyle+\left.\sum_{m\neq n}\left[
\Phi^m\,,\,\Phi^n\right]^2+\,{\rm fermions}\right) \ ,
\end{array}$\\\\
\hline\end{tabular}
\label{Dpdimred}\eeq
where $a,b=0,1,\dots,p$, $m,n=p+1,\dots,9$, and for the moment we do not
explicitly display the fermionic contributions. Thus the brane dynamics is
described by a supersymmetric Yang-Mills theory on the D$p$-brane worldvolume,
coupled dynamically to the transverse, adjoint scalar fields $\Phi^m$. This
demonstrates, in particular, how to explicitly write the action for the
collective coordinates $\Phi^m$ representing the fluctuations of the branes
transverse to their worldvolume.

Let us consider the Yang-Mills potential in (\ref{Dpdimred}), which is given by
\beq
V(\Phi)=\sum_{m\neq n}\Tr\left[\Phi^m\,,\,\Phi^n\right]^2
\label{YMpotential}\eeq
and is negative definite because
$[\Phi^m,\Phi^n]^\dag=[\Phi^n,\Phi^m]=-[\Phi^m,\Phi^n]$. A classical vacuum of
the field theory defined by (\ref{Dpdimred}) corresponds to a static solution
of the equations of motion whereby the potential energy of the system is
minimized. It is given by the field configurations which solve simultaneously
the equations
\bea
F_{ab}~=~D_a\Phi^m~=~\psi^\alpha&=&0 \ , \nn\\ V(\Phi)&=&0 \ .
\label{classvacua}\eea
Since (\ref{YMpotential}) is a sum of negative terms, its vanishing is
equivalent to the conditions
\beq
\left[\Phi^m\,,\,\Phi^n\right]=0
\label{potvanish}\eeq
for all $m,n$ and at each point in the $p+1$ dimensional worldvolume of the
branes. This implies that the $N\times N$ Hermitian matrix fields $\Phi^m$ are
simultaneously diagonalizable by a gauge transformation, so that we may write
\beq
\Phi^m=U~\pmatrix{x_1^m& & &0\cr &x_2^m& & \cr & &\ddots& \cr
0& & &x_N^m\cr}~U^{-1} \ ,
\label{Phimdiagsim}\eeq
where the $N\times N$ unitary matrix $U$ is {\it independent} of $m$. The
simultaneous, real eigenvalues $x_i^m$ give the positions of the $N$ distinct
D-branes in the $m$-th transverse direction. It follows that the ``moduli
space'' of classical vacua for the $p+1$ dimensional field theory
(\ref{Dpdimred}) arising from dimensional reduction of supersymmetric
Yang-Mills theory in ten dimensions is the quotient space
$(\real^{9-p})^N/S_N$, where the factors of $\real$ correspond to the positions
of the $N$ D-branes in the $9-p$ dimensional transverse space, and $S_N$ is the
symmetric group acting by permutions of the $N$ coordinates $x_i$. The group
$S_N$ corresponds to the residual Weyl symmetry of the $U(N)$ gauge group
acting in (\ref{Phimdiagsim}), and it represents the permutation symmetry of a
system of $N$ {\it indistinguishable} D-branes.

{}From (\ref{Dpdimred}) one can easily deduce that the masses of the
fields corresponding to the off-diagonal matrix elements are given
precisely by the distances $|x_i-x_j|$ between the corresponding
branes. This description means that an interpretation of the D-brane
configuration in terms of classical geometry is {\it only} possible in
the classical ground state of the system, whereby the matrices
$\Phi^m$ are simultaneously diagonalizable and the positions of the
individual D-branes may be described through their spectrum of
eigenvalues. This gives a simple and natural dynamical mechanism for
the appearence of ``noncommutative geometry'' at short
distances~\cite{CDS1,MavSz1,Witten2}, where the D-branes cease to have
well-defined positions according to classical geometry. Let us now
consider an explicit and important example.

\subsubsection*{Example}

The dynamics of $N$ D0-branes in the low-energy limit in flat ten dimensional
spacetime is the dimensional reduction of ${\cal N}=1$ supersymmetric
Yang-Mills theory in ten dimensions to one time direction $\tau$. The ten
dimensional gauge field $A_\mu$ thereby splits into nine transverse scalars
$\Phi^m(\tau)$ and a one dimensional gauge field $A_0(\tau)$ on the worldline.
By choosing the gauge $A_0=0$, we then get a ``supersymmetric matrix quantum
mechanics'' defined by the action
\bea
S_{{\rm D}0}&=&\frac1{2g_s\,\sqrt{\alpha'}}\,\int\dd\tau~\Tr\left(
\dot\Phi^m\,\dot\Phi_m+\frac1{(2\pi\alpha')^2}\,\sum_{m<n}
\left[\Phi^m\,,\,\Phi^n\right]^2\right.\nn\\&&+\left.\frac1{2\pi\alpha'}
\,\theta^\top\,\ii\,\dot\theta-\frac1{(2\pi\alpha')^2}\,\theta^\top\,
\Gamma_m\,\left[\Phi^m\,,\,\theta\right]\right) \ ,
\label{SD0}\eea
where $\Phi^m$, $m=1,\dots,9$ are $N\times N$ Hermitian matrices (with $N$ the
number of D0-branes), whose superpartners $\theta$ are also $N\times N$
Hermitian matrices which form 16-component spinors under the $SO(9)$ Clifford
algebra generated by the $16\times16$ matrices $\Gamma_m$.

The moduli space of classical vacua of the system of $N$ 0-branes is
$(\real^9)^N/S_N$, which is simply the configuration space of $N$
identical particles moving in nine dimensional space. But for a
general configuration, the matrices only have a geometrical
interpretation in terms of a noncommutative geometry. The Yang-Mills
D0-brane theory is essentially the non-relativistic limit of the
Born-Infeld D0-brane theory (\ref{D0action}), obtained by replacing
the Lagrangian $m\,\sqrt{1-\vec v^{\,2}}$ with its small velocity
limit $-\frac{m\vec v^{\,2}}2$ for $|\vec v\,|\ll1$. The $N\to\infty$
limit of this model is believed to describe the non-perturbative
dynamics of ``M-Theory'' and is known as ``Matrix
Theory''~\cite{BFSS1,Taylor1}.

\subsection{Forces Between D-Branes \label{Forces}}

The final point which we shall address concerning D-brane dynamics
is the nature of the interactions between D-branes. For this, we will
return to the worldsheet formalism and present the string computation
of the static force between two separated D$p$-branes. This will also
introduce another important string theoretical duality, constituting
one of the original dualities that arose in the context of dual
resonance models. We will then compare the results of this formalism
with the Yang-Mills description that we have developed thus far in
this section.

\begin{figure}[htb]
\epsfxsize=2 in
\bigskip
\centerline{\epsffile{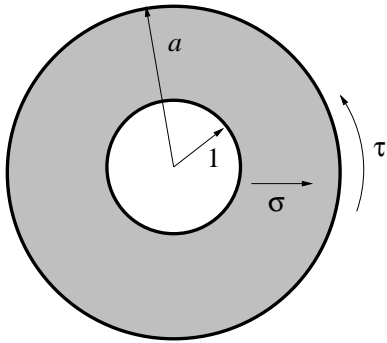}}
\caption{\baselineskip=12pt {\it The annulus diagram, whose local
    coordinates are $(\xi^0,\xi^1)=(\sigma,\tau)$ with
    $0\leq\sigma\leq\pi$, $0\leq\tau\leq2\pi t$ and modulus $0\leq
    t<\infty$. The inner radius of the annulus is set to unity by a
    conformal transformation, while the outer radius is
    $a=\e^{-t}$. Here $\xi^1=\tau$ is the worldsheet time coordinate,
    so that the boundaries represent a pair of D-branes with an open
    string stretched between them.}}
\bigskip
\label{annulusdiag}\end{figure}

\begin{figure}[htb]
\epsfxsize=2 in
\bigskip
\centerline{\epsffile{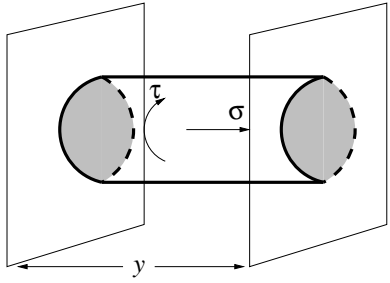}}
\caption{\baselineskip=12pt {\it The cylinder diagram as a modular
    transform of the annulus graph. The separation between the two
    D-branes is $y$. Now $\xi^0=\sigma$ is the worldsheet time
    coordinate and the diagram represents a single closed string
    propagating in the tree channel from one D-brane to the other.}}
\bigskip
\label{cylinderdiag}\end{figure}

We will compute the one-loop open string vacuum amplitude, which is
given by the annulus diagram (fig.~\ref{annulusdiag}). By using an
appropriate modular transformation, this open string graph can be
equivalently expressed as the cylinder diagram obtained by ``pulling
out'' the hole of the annulus (fig.~\ref{cylinderdiag}). In this
latter representation, the worldline of the open string boundary is a
vertex connecting the vacuum to a single closed string state. We have
thereby found that the {\it one-loop open string} Casimir force is
equivalent to a {\it tree-level closed string} exchange between a pair
of D-branes. This equivalence is known as ``open-closed string channel
duality'' and it enables one to straightforwardly identify the
appropriate interaction amplitudes for D-branes~\cite{Pol1}. In particular,
two D$p$-branes interact gravitationally by exchanging closed strings
corresponding, in the massless sector, to graviton and dilaton states.

As is usual in quantum field theory, the one-loop vacuum amplitude
$\cal A$ is given by the logarithm of the partition function $Z_{\rm
  vac}$ determined as the fluctuation determinant of the full theory
arising from integrating out the free worldsheet fields. With $L_0$ the
worldsheet Hamiltonian, we thereby have

\vbox{\bea
{\cal A}&=&\ln(Z_{\rm vac})~=~\ln\left(P_{\rm GSO}\,\frac1{\sqrt{
\det^{~}_{{\rm NS}\oplus{\rm R}}(L_0-a)}}\right)\nn\\
&=&-\frac12\,\Tr^{~}_{{\rm NS}\oplus{\rm R}}
\Bigl[P_{\rm GSO}\,\ln(L_0-a)\Bigr]\nn\\&=&-\frac{V_{p+1}}2\,
\int\frac{\dd^{p+1}k}{(2\pi)^{p+1}}\,\tr^{~}_{{\rm NS}\oplus{\rm R}}
\Bigl[P_{\rm GSO}\,\ln\left(k^2+m^2\right)\Bigr] \ ,
\label{calAtrace}\eea}
\noindent
where $k$ are the zero-mode momenta in the Neumann directions (which
vanish along the Dirichlet directions), the worldvolume factor is
inserted to make the momentum integrations dimensionless, and the open
string mass spectrum is given by (c.f.~(\ref{mijcalN}))
\beq
m^2=\frac1{\alpha'}\,\Bigl(N-a\Bigr)+\left(\frac
  y{2\pi\alpha'}\right)^2
\label{openmassspec}\eeq
with $y^n$, $n=p+1,\dots,9$ the separation of the D$p$-branes. By
using the elementary identity
\beq
-\frac12\,\ln\left(k^2+m^2\right)=\int\limits_0^\infty\frac{\dd t}{2t}~
\e^{-2\pi\alpha'\,(k^2+m^2)\,t} \ ,
\label{lnidentity}\eeq
we may perform the Gaussian integrals over the momenta $k$ in
(\ref{calAtrace}) to get
\beq
{\cal A}=-2V_{p+1}\,\int\limits_0^\infty\frac{\dd t}{2t}~\Bigl(8\pi^2
\alpha'\,t\Bigr)^{-(p+1)/2}~\e^{-y^2\,t/2\pi\alpha'}~
\tr^{~}_{{\rm NS}\oplus{\rm R}}\left(P_{\rm GSO}\,q^{N-a}\,\right) \ ,
\label{calAdt}\eeq
where
\beq
q\equiv\e^{-2\pi t} \ .
\label{qtdef}\eeq

The trace that appears in (\ref{calAdt}) is exactly the {\it same}
sort of trace that we encounterd in the calculation of the one-loop
closed superstring amplitude in section~\ref{1Loop}, with the identification
$t=-\ii\,\tau$ (see (\ref{ZRNSGSO})). In particular, we found there
that the GSO projection, giving the appropriate sum over spin
structures that guarantees modular invariance, yields a {\it
  vanishing} result. We therefore have ${\cal A}=0$, consistent
with the fact that we are computing a vacuum amplitude, and that
the open string spectrum is supersymmetric.

Let us now compare this string result with the static force computed from
quantum field theory (the low-energy limit). Among many others,
the Type~II supergravity action contains the terms
\beq
S_{\rm RR}=-\frac1{2\kappa^2}\,\int\dd^{10}x~
F_{\mu_1\cdots\mu_{p+2}}^{(p+2)}F^{(p+2)\,\mu_1\cdots\mu_{p+2}}
+q_p\int\dd^{p+1}\xi~C_{a_1\cdots a_{p+1}}^{(p+1)}\,
\epsilon^{a_1\cdots a_{p+1}} \ ,
\label{SUGRAaction}\eeq
where $\kappa$ is the gravitational constant, $q_p$ is the charge of
the D$p$-brane under the Ramond-Ramond $p+1$-form potential
$C^{(p+1)}$, and
\beq
C_{a_1\cdots a_{p+1}}^{(p+1)}=C_{\mu_1\cdots\mu_{p+1}}^{(p+1)}\,
\frac{\partial x^{\mu_1}}{\partial\xi^{a_1}}\cdots
\frac{\partial x^{\mu_{p+1}}}{\partial\xi^{a_{p+1}}}
\label{Cpullback}\eeq
is the ``pull-back'' of $C^{(p+1)}$ to the D$p$-brane worldvolume. A
very tedious (but standard) perturbative calculation in the $p+1$-form
quantum field theory defined by the action (\ref{SUGRAaction}) establishes
that the corresponding vacuum energy vanishes {\it provided} we make
the identification~\cite{Pol1}
\beq
\begin{tabular}{|c|}\hline\\
$\displaystyle
T_p=q_p \ .
$\\\\
\hline\end{tabular}
\label{Tpqp}\eeq

This coincidence between the D$p$-brane tension and the
Ramond-Ramond charge is one of the most important results in D-brane
physics, and indeed it was one of the sparks which ignited the second
superstring revolution. It implies that the Ramond-Ramond repulsion
between identical, parallel D$p$-branes cancels exactly their
gravitational and dilaton attraction. The cancellation of the static
force is a consequence of spacetime supersymmetry, and it forces us to
accept the fact that D-branes {\it are} the Type~II R--R charged
states~\cite{Pol1}.

\subsubsection*{BPS States}

The remarkable properties of D-brane interactions that we have
discovered above arise because D-branes describe certain special,
non-perturbative extended states of the Type~II superstring which
carry non-trivial R--R charge. To understand this point better, we
recall that Type~II superstring theory in the bulk (away from
D-branes) possesses ${\cal N}=2$ spacetime supersymmetry. However, the
open string boundary conditions are invariant under only one of these
supersymmetries. In other words, the Type~II vacuum without D-branes
is invariant under ${\cal N}=2$ supersymmetry, but the state
containing a D-brane is invariant only under ${\cal N}=1$~\cite{Pol1}. So a
D-brane may be characterized as a state which preserves only {\it
  half} of the original spacetime supersymmetries. Such a state is
known as a ``Bogomolny-Prasad-Sommerfeld (BPS) state''~\cite{BUSSTEPPJMF}.

Generally, BPS states carry conserved charges which are determined
entirely by their mass in the corresponding (extended) supersymmetry
algebra. In the present case, there is only one set of charges with
the correct Lorentz transformation properties, namely the
antisymmetric Ramond-Ramond charges. So unlike the fundamental string,
a D-brane carries R--R charge, {\it consistent} with the fact that it
is a BPS state. This property is known explicitly from the
realizations of D-branes as solitonic solutions of the classical
supergravity equations of motion~\cite{DKL1}.

Such BPS bound state configurations of D-branes can also be realized
in their low-energy, supersymmetric Yang-Mills theory description. The
corresponding BPS energies of these systems agree with the
supersymmetric Yang-Mills energy given by
\beq
E_{\rm YM}=\frac{\pi^2\,\alpha'\,T_p}{g_s}\,\int\dd^{p+1}\xi~
\Tr\left(F_{ab}F^{ab}\right) \ .
\label{YMenergy}\eeq
The BPS property is protected at the quantum level by supersymmetry,
via the usual non-renormalization theorems. Thus the relation
(\ref{Tpqp}) between the mass and charge of a D-brane cannot be
modified by any perturbative or non-perturbative effects. The fact
that D-branes fill out supermultiplets of the underlying supersymmetry
algebra has been the crucial property in testing the various duality
conjectures that were discussed in section~\ref{History}.

\subsection*{Acknowledgments}

The author is grateful to the participants, tutors and lecturers of
the school for their many questions, suggestions and criticisms which
have all gone into the preparation of these lecture notes. He would
especially like to thank J.~Forshaw for having organised an excellent
school, and for the encouragement to write up these notes. He would
also like to thank J.~Figueroa-O'Farrill and F.~Lizzi for practical
comments on the manuscript. This work was supported in part by an
Advanced Fellowship from the Particle Physics and Astronomy Research
Council~(U.K.).

\end{document}